\RequirePackage{pdf14}

\documentclass[letterpaper]{article}
\usepackage[table]{xcolor}
\usepackage{array,setspace,mathrsfs,amsfonts,amsmath,yfonts,dsfont,bbm,colonequals}
\usepackage{tikz-cd}
\usepackage{jheppub}

\newcommand\be{\beg{equation}}
\newcommand\bea{\beg{eqnarray}}
\newcommand\qq{\mathbbmtt{Q}}
\newcommand\q{\mathbbmtt{Q}_1}
\newcommand\qd{\mathbbmtt{Q}_2}
\newcommand\eea{\end{eqnarray}}
\newcommand\ee{\end{equation}}

\newcommand{\goodchi}{\protect\raisebox{1pt}{$\chi$}}

\newcommand\hb{\bar{h}}
\newcommand\mf{\mathfrak}

\newcommand\nn{\nonumber}
\newcommand\ph{\phantom}
\newcommand\wt{\widetilde}
\newcommand\eg{{\it e.g.}}
\newcommand\ie{{\it i.e.}}
\newcommand\cf{{\it cf.}}

\newcommand\til{\tilde}
\newcommand\pr{\prime}

\newcommand\ve{\varepsilon}

\renewcommand\aa{\alpha}
\newcommand\bb{\beta}
\newcommand\dd{\delta}
\renewcommand\gg{\gamma}
\newcommand\aad{\dot{\aa}}
\newcommand\bbd{\dot{\bb}}
\newcommand\ggd{\dot{\gg}}
\newcommand\ddd{\dot{\dd}}

\newcommand\Cb{\mathbb{C}}
\newcommand\Rb{\mathbb{R}}
\newcommand\Zb{\mathbb{Z}}

\renewcommand\AA{\mathcal{A}}
\renewcommand\SS{\mathcal{S}}

\newcommand\BB{\mathcal{B}}
\newcommand\CC{\mathcal{C}}
\newcommand\DD{\mathcal{D}}
\newcommand\EE{\mathcal{E}}
\newcommand\FF{\mathcal{F}}
\newcommand\HH{\mathcal{H}}
\newcommand\II{\mathcal{I}}
\newcommand\JJ{\mathcal{J}}
\newcommand\KK{\mathcal{K}}
\newcommand\LL{\mathcal{L}}
\newcommand\MM{\mathcal{M}}
\newcommand\NN{\mathcal{N}}
\newcommand\OO{\mathcal{O}}

\newcommand\PP{\mathcal{P}}
\newcommand\QQ{\mathcal{Q}}
\newcommand\RR{\mathcal{R}}
\newcommand\TT{\mathcal{T}}

\newcommand\WW{\mathcal{W}}
\newcommand\XX{\mathcal{X}}

\newcommand\ZZ{\mathcal{Z}}

\newcommand\Qm{\mathcal{Q}}

\newcommand\Sm{\mathcal{S}}

\newcommand\IIh{\hat{\II}}
\newcommand\JJh{\hat{\JJ}}

\newcommand\zb{{\bar z}}
\newcommand\Lt{{\tilde L}}
\newcommand\Lb{{\bar L}}

\newcommand\dce{h^{\vee}}
\renewcommand\sl[1]{\mf{sl}(#1)}

\newcommand{\beg}[1]{\begin{#1}} %This is a hack for my compiler -- doesn't matter in general (CJB)

\def\QQ{\mathcal{Q}}
\def\q{\mathbbmtt{Q}\,_{1}}
\def\qd{\mathbbmtt{Q}\,_{2}}

\def\cal{\mathcal}

\def\ad{\dot{\alpha}}
\def\bd{\dot{\beta}}

\def\Tr{\mbox{Tr}}

\def\a{\alpha}
\def\b{\beta}

\def\ph{\phantom}

\def\QQ{\mathcal{Q}}

\def\SO{\mathrm{SO}}
\def\SU{\mathrm{SU}}
\def\USp{\mathrm{USp}}

\def\hf{\frac{1}{2}}

\def\del{\partial}

\newcommand\restr[2]{{% we make the whole thing an ordinary symbol
  \left.\kern-\nulldelimiterspace % automatically resize the bar with \right
  #1 % the function
  \vphantom{\big|} % pretend it's a little taller at normal size
  \right|_{#2} % this is the delimiter
  }}
\newcommand\fverb{\setbox\fverbbox=\hbox\bgroup\verb}
\newcommand\fverbdo{\egroup\medskip\noindent%
			\fbox{\unhbox\fverbbox}\ }
\newcommand\fverbit{\egroup\item[\fbox{\unhbox\fverbbox}]}
\newbox\fverbbox

\title{Infinite Chiral Symmetry in Four Dimensions}

\author[\!\!1,2]{Christopher Beem,}
\author[\!\!3]{Madalena Lemos,}
\author[\!\!3,4]{Pedro Liendo,}
\author[\!\!3]{Wolfger Peelaers,}
\author[\!\!1,3]{Leonardo Rastelli,}
\author[3,5]{and Balt C. van Rees}

\affiliation[1]{Institute for Advanced Study, Einstein Dr., Princeton, NJ 08540, USA}
\affiliation[2]{Simons Center for Geometry and Physics, Stony Brook University, Stony Brook, NY 11794-3636, USA}
\affiliation[3]{C.~N.~Yang Institute for Theoretical Physics, Stony Brook University, Stony Brook, NY 11794-3840, USA}
\affiliation[4]{IMIP, 
Humboldt-Universit\"at zu Berlin,
IRIS Adlershof, Zum Gro\ss en Windkanal 6, 12489 Berlin, Germany}
\affiliation[5]{Theory Group, Physics Department, CERN, CH-1211 Geneva 23, Switzerland}

\preprint{YITP-SB-13-45,  CERN-PH-TH/2013-311,  HU-EP-13/78}
\bigskip
\abstract{We describe a new correspondence between four-dimensional conformal field theories with extended supersymmetry and two-dimensional chiral algebras. The meromorphic correlators of the chiral algebra compute correlators in a protected sector of the four-dimensional theory. Infinite chiral symmetry has far-reaching consequences for the spectral data, correlation functions, and central charges of any four-dimensional theory with $\NN=2$ superconformal symmetry.}


\begin{document}
\setcounter{tocdepth}{2}
\maketitle

\section{Introduction}

It has long been recognized that supersymmetric quantum field theories enjoy many special properties that make them particularly useful testing grounds for more general ideas about quantum field theory. This is largely a consequence of the fact that many observables in such theories are ``protected'', in the sense of being determined by a semiclassical calculation with a finite number of corrections taken into account, or alternatively by some related ``finite-dimensional'' problem that admits the type of closed-form solution that is uncharacteristic of interacting quantum field theories. In most circumstances, these techniques have a semiclassical flavor to them. For example, in cases where supersymmetric partition functions can be computed by localization, the calculation is generally performed starting from a weakly coupled Lagrangian description of the theory.

A notable omission from the currently available techniques is a way to directly access the interacting superconformal phases of theories that do not admit a Lagrangian formulation. By now, there exists a veritable menagerie of models in various dimensions that exhibit conformal phases with varying amounts of supersymmetry, but only in the nicest cases do such models belong to families that include free theories as special points, allowing for properties of the interacting theory to be studied semiclassically. Even for those Lagrangian models, the standard supersymmetric toolkit does not seem to exploit some of the most powerful structures of conformal field theory, such as the existence of a state/operator map and of a well-controlled and convergent operator product expansion.
 
Meanwhile, recent years have witnessed a surprising resurgence of progress centering around precisely these aspects of conformal field theory in the form of the conformal bootstrap \cite{Polyakov:1974gs,Ferrara:1973yt}. In large part, this progress has been inspired by the development of numerical techniques for extracting constraints on the defining data of a CFT using unitarity and crossing symmetry \cite{Rattazzi:2008pe,ElShowk:2012ht}. Generally speaking, these techniques are equally applicable to theories with and without supersymmetry, and despite promising early results \cite{Poland:2010wg, Poland:2011ey, Beem:2013qxa, Alday:2013opa}, it has not been entirely clear the extent to which supersymmetry improves the situation. 
Nevertheless, the possibility that supersymmetry may act as a crucible in which exact results can be forged even for strongly interacting CFTs is irresistible, and we are led to ask the question:
%%%
\medskip
\begin{quote}
\emph{Do the conformal bootstrap equations in dimension $d>2$ admit a solvable truncation in the case of superconformal field theories?}
\end{quote}
\medskip
%%%
Having formulated the question, it is worth pausing to consider in what sense the answer could be ``yes''. The most natural possibilities correspond to known situations in which bootstrap-type equations are rendered solvable. There are two primary scenarios in which the constraints of crossing symmetry are nontrivial, yet solvable:
%%%
\begin{quote}
\begin{description}
\item[(I)\ph{I}] Meromorphic (and rational) conformal field theories in two dimensions.
\item[(II)] Topological quantum field theories.
\end{description}
\end{quote}
%%%
The subject of this paper is the realization of the first option in the context of $\NN\geqslant2$ superconformal field theories in four dimensions. The same option is in fact viable for $(2,0)$ superconformal theories in six dimensions. That subject is elaborated upon in a separate article \cite{WIP_6d}. Although we will not discuss the subject at any length in the present work, the second option can also be realized using similar techniques to those discussed herein.

The primary hint that such an embedding should be possible was already observed in \cite{Dolan:2006ec,Beem:2013qxa}, building upon the work of \cite{Eden:2000bk,Eden:2001ec,Dolan:2001tt,Heslop:2002hp,Dolan:2004mu,Nirschl:2004pa}. In a remarkable series of papers \cite{Eden:2000bk,Eden:2001ec,Dolan:2001tt,Heslop:2002hp,Dolan:2004mu,Nirschl:2004pa,Dolan:2006ec}, the constraints of superconformal symmetry on four-point functions of half-BPS operators in $\NN=2$ and $\NN=4$ superconformal field theories were studied in detail. This analysis revealed that the superconformal Ward identities obeyed by these correlators can be conveniently solved in terms of a set of arbitrary real-analytic functions of the two conformal cross ratios $(z,\bar z)$, along with a set of \emph{meromorphic} functions of $z$ alone. In a decomposition of the four-point function as an infinite sum of conformal blocks, these meromorphic functions capture the contribution to the double operator product expansion of intermediate ``protected'' operators belonging to shortened representations. The real surprise arises when these results are combined with the constraints of crossing symmetry. One then finds \cite{Beem:2013qxa,Dolan:2006ec} that the meromorphic functions obey a \emph{decoupled} set of crossing equations, whose general solution can be parametrized in terms of a finite number of coefficients. For example, in the important case of the four-point function of stress-tensor multiplets in an $\NN=4$ theory, there is a one-parameter family of solutions, where the parameter has a direct physical interpretation as the central charge (conformal anomaly) of the theory. The upshot is that the protected part of this correlator is entirely determined by abstract symmetry considerations, with no reference to a free-field description of the theory. 

\bigskip

In this paper we establish a conceptual framework that explains and vastly generalizes this observation. For a general $\NN=2$ superconformal field theory, we define a protected subsector by passing to the cohomology of a certain nilpotent supercharge $\qq\,$. This is a familiar strategy -- for example, the definition of the chiral ring in an $\NN=1$ theory follows the same pattern -- but our version of this maneuver will be slightly unconventional, in that we take $\qq = \QQ + \SS$ to be a linear combination of a Poincar\'e and a conformal supercharge. In order to be in the cohomology of $\qq\,$, local operators must lie in a fixed plane $\mathbb{R}^2 \subset \mathbb{R}^4$. Crucially, their correlators can be shown to be non-trivial \emph{meromorphic} functions of their positions. This is in contrast to correlators of $\NN=1$ chiral operators, which are purely topological in a general $\NN=1$ model, and strictly vanish in an $\NN=1$ \emph{conformal} theory due to $R$-charge conservation.

The meromorphic correlators identified by this cohomological construction are precisely the ingredients that define a two-dimensional \emph{chiral algebra}.\footnote{We have settled on the expression ``chiral algebra'' as it is the most common in the physics literature. We consider it to be synonymous with ``vertex operator algebra'', though in the mathematical literature some authors make a distinction between the two notions. We trust no confusion will arise with the overloading of the word ``chiral'' due to its unavoidable use in the four-dimensional context, \eg, ``chiral and anti-chiral $4d$ supercharges'', ``the $\NN=1$ chiral ring'', etc.} Our main result is thus the definition of a map $\goodchi$ from the space of four-dimensional $\NN=2$ superconformal field theories to the space of two-dimensional chiral algebras,
%%%%%%
$$
\goodchi~:~\text{4d $\NN = 2$ SCFT}~\longrightarrow~\text{2d Chiral Algebra}.
$$
%%%%%%
In concrete terms, the chiral algebra computes correlation functions of certain operators in the four-dimensional theory, which are restricted to be coplanar and further given an \emph{explicit} space-time dependence correlating their $SU(2)_R$ orientation with their positions, see \eqref{eq:twistedtranslate}. For the case of four-point functions of half-BPS operators, assigning the external operators this ``twisted'' space-time dependence accomplishes precisely the task of projecting the full correlator onto the meromorphic functions appearing in the solution to the superconformal Ward identities. To recapitulate, those mysterious meromorphic functions are given a direct interpretation as correlators in the associated chiral algebra, and turn out to be special instances of a much more general structure.

The explicit space-time dependence of the four-dimensional operators is instrumental in making sure that they are annihilated by a common supercharge $\qq\,$ for any insertion point on the plane. From this viewpoint, our construction is in the same general spirit as \cite{Drukker:2009sf} (see also \cite{deMedeiros:2001kx}). These authors considered particular examples of correlators in $\NN=4$ super Yang-Mills theory that are invariant under supercharges of the same schematic form $\QQ + \SS$. Their choices of supercharges are inequivalent to ours, and do not lead to meromorphic correlators.
 
The operators captured by the chiral algebra are precisely the operators that contribute to the Schur limit of the superconformal index \cite{Kinney:2005ej,Gadde:2011ik,Gadde:2011uv}, and we will refer to them as \emph{Schur operators}. Important examples are the half-BPS operators that are charged under $SU(2)_R$ but neutral under $U(1)_r$, whose vacuum expectation values parameterize the Higgs branch of the theory, and the $SU(2)_R$ Noether current. The class of Schur operators is much larger, though, and encompasses a variety of supermultiplets obeying less familiar semi-shortening conditions. Operators associated to the Coulomb branch of the theory (such as the half-BPS operators charged under $U(1)_r$ but neutral under $SU(2)_R$) are \emph{not} of Schur type. In a pithy summary, the cohomology of $\qq\,$ provides a ``categorification'' of the Schur index. It is a surprising and useful fact that this vector space naturally possesses the additional structure of a chiral algebra. 

Chiral algebras are rigid structures. 
Associativity of their operator algebra translates into  strong constraints on the spectrum and OPE coefficients of Schur operators in the parent four-dimensional theory. We have already pointed out that this leads to a unique determination of the protected part of four-point function of stress-tensor multiplets in the $\NN=4$ context \cite{Beem:2013qxa}. Another canonical example is the four-point function of ``moment map'' operators in a general $\NN=2$ superconformal field theory. The moment map $M$ is the lowest component of the supermultiplet that contains the conserved flavor current of the theory, and as such it transforms in the adjoint representation of the flavor group $G$. We find that the associated two-dimensional meromorphic operator $J (z)\colonequals\goodchi [M]$ is the dimension-one generating current of an affine Lie algebra $\hat{\mf{g}}_{k_{2d}}$, with level $k_{2d}$ fixed in terms of the four-dimensional flavor central charge. As the four-point function of affine currents is uniquely fixed, this relation completely determines the protected part of the moment map four-point function. In turn, this information serves as essential input to the full-fledged bootstrap equations that govern the contributions from generic long multiplets in the conformal block decomposition of these four-point functions. These equations can be studied numerically to derive interesting bounds on non-protected quantities, following the approach of \cite{Beem:2013qxa}. We will present numerical bounds that arise for various choices of $G$ in a separate publication \cite{WIP_N2_Numerics}. It is worth emphasizing that the protected part of the four-point function receives contributions from an \emph{infinite} tower of intermediate shortened multiplets, and without knowledge of its precise form the numerical bootstrap program would never get off the ground. In theories that admit a Lagrangian description, one could appeal to non-renormalization theorems and derive the same protected information in the free field limit; the chiral algebra then just serves as a powerful organizing principle to help obtain the same result. However, the abstract chiral algebra approach seems indispensable for the analysis of non-Lagrangian theories -- for example, when $G$ is an exceptional group.
 
As a byproduct of a detailed study of the moment map four-point function, we are able to derive new unitarity bounds that must be obeyed by the central charges of any interacting $\NN=2$ superconformal field theory. By exploiting the relation between the two- and four-dimensional perspectives, we are able to express certain coefficients of the four-dimensional conformal block decomposition of the four-point function in terms of central charges; the new bounds arise because those coefficients must be non-negative in a unitary theory. Saturation of the bounds signals special properties of the Higgs branch chiral ring. This is a particular instance of a more general encoding of four-dimensional physics in the chiral algebra, the surface of which we have only barely scratched. One notable aspect of this correspondence is the interplay between the geometry of the Higgs branch and the representation theory of the chiral algebra; for example, null vectors that appear at special values of the affine level imply Higgs branch relations.
   
We describe several structural properties of the map $\goodchi$. Two universal features are the affine enhancement of the global flavor symmetry, and the Virasoro enhancement of the global conformal symmetry. The affine level in the chiral algebra is related to the flavor central charge in four dimensions as $k_{2d}=-\frac12 k_{4d}$, while the Virasoro central charge is proportional to the four-dimensional conformal anomaly coefficient,\footnote{There are two tensorial structures in the four-dimensional trace anomaly, whose coefficients are conventionally denoted $a$ and $c$. It is the $c$ anomaly that is relevant for us, in contrast to the better studied $a$ anomaly, which decreases monotonically under RG flow \cite{Cardy:1988cwa,Komargodski:2011vj}.} $c_{2d}=-12c_{4d}$. A perhaps surprising feature of these relations is that the two-dimensional central charges and affine levels must be negative. Another universal aspect of the correspondence is a general prescription to derive the chiral algebra associated to a gauge theory whenever the chiral algebra of the original theory whose global symmetry is being gauged is known.
 
Turning to concrete examples, we start with the SCFTs of free hypermultiplets and free vector multiplets, which are associated to free chiral algebras. With the help of the general gauging prescription, we can combine these ingredients to find the chiral algebra associated to an arbitrary Lagrangian SCFT. We also sketch the structure of the chiral algebras associated to SCFTs of class $\SS$, which are generally non-Lagrangian. In several concrete examples, we present evidence that the chiral algebra has an economical presentation as a $\WW$-algebra, \ie, as a chiral algebra with a finite set of generators \cite{Bouwknegt:1992wg}. We do not know whether all chiral algebras associated to SCFTs are finitely generated, or how to identify the complete set of generators in the general case. Indeed, an important open problem is to give a more precise characterization of the class of chiral algebras that can arise from physical four-dimensional theories. Ideally the distinguishing features of this class could be codified in a set of additional axioms. Since chiral algebras are on sounder mathematical footing than four-dimensional quantum field theories, it is imaginable that this could lead to a well-defined algebraic classification problem. If successful, this approach would represent concrete progress towards the loftier goal of classifying all possible $\NN=2$ SCFTs.

On a more formal note, four-dimensional intuition leads us to formulate a number of new conjectures about chiral algebras that may be of interest in their own right. The conjectures generally take the form of an ansatz for the cohomology of a BRST complex, and include new free-field realizations of affine Lie algebras at special values of the level and new examples of quantum Drinfeld-Sokolov reduction for nontrivial modules. We present evidence for our conjectures obtained from a low-brow, level-by-level analysis, but we suspect that more powerful algebraic tools may lead to rigorous proofs. 
 
\bigskip

The organization of this paper is as follows. In \S\ref{sec:section2} we review the arguments behind the appearance of infinite-dimensional chiral symmetry algebras in the context of two-dimensional conformal field theories. We explain how the same structure can be recovered in the context of $\NN=2$ superconformal field theories in four dimensions by studying observables that are well-defined after passing to the cohomology of a particular nilpotent supercharge in the superconformal algebra. This leads to the immediate conclusion that chiral symmetry algebras will control the structure of this subclass of observables. In \S\ref{sec:new2d4d}, we describe in greater detail the resulting correspondence between $\NN=2$ superconformal models in four dimensions and their associated two-dimensional chiral algebras. We outline some of the universal features of the correspondence. We further describe an algorithm that defines the chiral algebra for any four-dimensional SCFT with a Lagrangian description in terms of a BRST complex. In \S\ref{sec:4d_consequences}, we describe the immediate consequences of this structure for more conventional observables of the original theory. It turns out that superconformal Ward identities that have previously derived for four-point functions of BPS operators are a natural outcome from our point of view. We further derive new unitarity bounds for the anomaly coefficients of conformal and global symmetries, many of which are saturated by interesting superconformal models. We point out that the state space of the chiral algebra provides a categorification of the Schur limit of the superconformal index. In \S\ref{sec:lagrangian_examples}, we detail the construction and analysis of the chiral algebras associated to some simple Lagrangian SCFTs. We also make a number of conjectures as to how to describe these chiral algebras as $\WW$-algebras. In \S\ref{sec:classS} we provide a sketch of the class of chiral algebras that are associated to four-dimensional theories of class $\SS$. We conclude in \S\ref{sec:conclusions} by listing a number of interesting lines of inquiry that are opened up by the results reported here. Several appendices are included that review relevant material concerning the superconformal algebras and representation theory used in our constructions.

\section{Chiral symmetry algebras in four dimensions}
\label{sec:section2}

The purpose of this section is to establish the existence of infinite chiral symmetry algebras acting on a restricted class of observables in any $\NN=2$ superconformal field theory in four dimensions. This is accomplished in two steps. First, working purely in terms of the relevant spacetime symmetry algebras, we identify a particular two-dimensional conformal subalgebra of the four-dimensional superconformal algebra,\footnote{In this section, we adopt the convention of specifying the complexified versions of symmetry algebras. This will turn out to be particularly natural in the discussion of \S\ref{subsec:twisted_subalgebra}. We generally attempt to select bases for the complexified algebras that are appropriate for a convenient real form. Our basic constructions are insensitive to the signature of spacetime, though in places we explicitly impose constraints that follow from unitarity in Lorentzian signature.}
%%%%%%
$$
\mf{sl}(2)\times\widehat{\mf{sl}(2)}\subset\mf{sl}(4\,|\,2)~,
$$
%%%%%%
with the property that the holomorphic factor $\mf{sl}(2)$ commutes with a nilpotent supercharge, $\qq\,$, while the antiholomorphic factor $\widehat{\mf{sl}(2)}$ is exact with respect to the same supercharge. We then characterize the local operators that represent nontrivial $\qq\,$-cohomology classes. The only local operators for which this is the case are restricted to lie in a plane $\Rb^2\subset\Rb^4$ that is singled out by the choice of conformal subalgebra. The correlation functions of these operators are meromorphic functions of the insertion points, and thereby define a chiral algebra. As a preliminary aside, we first recall the basic story of infinite chiral symmetry in two dimensions in order to distill the essential ingredients that need to be reproduced in four dimensions. The reader who is familiar with chiral algebras in two-dimensional conformal field theory may safely proceed directly to \S\ref{subsec:twisted_subalgebra}.

\subsection{A brief review of chiral symmetry in two dimensions}
\label{subsec:chiral_symmetry_review}

Let us take as our starting point a two-dimensional quantum field theory that is invariant under the global conformal group $SL(2,\Cb)$. The complexification of the Lie algebra of infinitesimal transformations factorizes into holomorphic and anti-holomorphic generators,
%%%%%%
\be
\begin{split}
&L_{-1}=-\partial_z~,\qquad L_{0}=-z\partial_z~,\qquad L_{+1}=-z^2\partial_z~,\\ 
&\Lb_{-1}=-\partial_{\zb}~,\qquad \Lb_0=-\zb\partial_{\zb}~,\qquad \Lb_{+1}=-\zb^2\partial_{\zb}~,
\end{split}
\ee
%%%%%%
which obey the usual $\mf{sl}(2)\times\mf{sl}(2)$ commutation relations,
%%%%%%
\be
\begin{split}
&[L_{+1},L_{-1}]=2L_0~,\qquad [L_0,L_{\pm1}]=\mp L_{\pm1}~,\\
&[\Lb_{+1},\Lb_{-1}]=2L_0~,\qquad [\Lb_0,\Lb_{\pm1}]=\mp \Lb_{\pm1}~.
\end{split}
\ee
%%%%%%
We need not assume that the theory is unitary, but for simplicity we will assume that the space of local operators decomposes into a direct sum of irreducible highest weight representations of the global conformal group. Such representations are labelled by holomorphic and anti-holomorphic scaling dimensions $h$ and $\bar h$ of the highest weight state,
%%%%%
\be
L_0|\psi\rangle_{h.w.}=h|\psi\rangle_{h.w.}~,\qquad \Lb_0|\psi\rangle_{h.w.}=\bar h|\psi\rangle_{h.w.}~,
\ee
%%%%%
and we further assume that $h$ and $\bar h$ are not equal to negative half-integers (in which case one would find finite-dimensional representations of $\mf{sl}(2)$).

Chiral symmetry arises as a consequence of the existence of any local operator $\OO(z,\bar z)$ which obeys a meromorphicity condition of the form
%%%%%%
\be\label{eqn:2d_holomorphicity}
\partial_{\bar z}\OO(z,\bar z)=0 \implies \OO(z,\bar z)\colonequals\OO(z)~.
\ee
%%%%%%
Under the present assumptions, such an operator will transform in the trivial representation of the anti-holomorphic part of the symmetry algebra and by locality will have $h$ equal to an integer or half-integer. Meromorphicity implies the existence of infinitely many conserved charges (and their associated Ward identities) defined by integrating the meromorphic operator against an arbitrary monomial in $z$,
%%%%%%
\be\label{eqn:charges_from_operators}
\OO_n:=\oint \frac{dz}{2\pi i}\,z^{n+h-1}\,\OO(z)~.
\ee
%%%%%%
The operator product expansion of two meromorphic operators contains only meromorphic operators, and the singular terms determine the commutation relations among the associated charges, \cf\ \cite{Bouwknegt:1992wg}. This is the power of meromorphy in two dimensions: an infinite dimensional symmetry algebra organizes the space of local operators into much larger representations, and the associated Ward identities strongly constrain the correlation functions of the theory. 

Some examples of this structure are ubiquitous in two-dimensional conformal field theory. The energy-momentum tensor in a two-dimensional CFT is conserved and traceless in flat space, $\partial^\mu T_{\mu\nu}=T_\mu^{~\mu}=0$, leading to two independent conservation equations
%%%%%%%
\be\label{eqn:2d_stress}
\begin{split}
\partial_{\bar z}T_{zz}(z,\bar z)=0\implies T_{zz}(z,\bar z)\colonequals T(z)~,\\
\partial_{z}T_{\bar z\bar z}(z,\bar z)=0\implies T_{\bar z\bar z}(z,\bar z)\colonequals \overline T(\bar z)~.
\end{split}
\ee
%%%%%%
The holomorphic stress tensor $T(z)$ is a meromorphic operator with $(h,\bar h)=(2,0)$, and its self-OPE is fixed by conformal symmetry to take the form
%%%%%%
\be
\label{eq:2d_TT_OPE}
T(z)T(w)\sim\frac{c/2}{(z-w)^4}+\frac{2T(w)}{(z-w)^2}+\frac{\partial T(w)}{(z-w)}~,
\ee
%%%%%%
which implies that the associated conserved charges obey the commutation relations of the Virasoro algebra with central charge $c$,
%%%%%%
\be
\label{eq:virasoro}
L_n\colonequals\oint \frac{dz}{2\pi i}z^{n+1}T(z)~,\qquad [L_m,L_n]=(m-n)L_{m+n}+\frac{c}{12}m(m^2-1)\delta_{m+n,0}~.
\ee
%%%%%%
Similarly, global symmetries can give rise to conserved holomorphic currents $J^A_z(z,\bar z) \equalscolon J^A(z)$ with $(h,\bar h)=(1,0)$. The self-OPEs of such currents are fixed to take the form
%%%%%%
\be
\label{eq:2d_JJ_OPE}
J^A(z)J^B(w)\sim\frac{k\, \delta^{AB}}{(z-w)^2}+\sum_C i f^{ABC}\frac{J^C(w)}{(z-w)}~,
\ee
%%%%%%
with the structure constants $f^{ABC}$ those of the Lie algebra of the global symmetry. The conserved charges in this case obey the commutation relations of an affine Lie algebra at level $k$,
%%%%%%
\be
\label{eq:2d_AKM}
J^A_n\colonequals\oint\frac{dz}{2\pi i}\,z^{n}\,J^A(z)~,\qquad [J^A_m,J^B_n]=\sum_c if^{ABC}J^C_{m+n}+mk\,\delta^{AB}\delta_{m+n,0}~.
\ee
%%%%%%
The algebra of all meromorphic operators, or alternatively the algebra of their corresponding charges, constitutes the \emph{chiral algebra} of a two-dimensional conformal field theory. 

In most physics applications, the spectrum of a CFT will include non-meromorphic operators that reside in modules of the chiral algebra of the theory. In the generic case in which the chiral algebra is the Virasoro algebra, this just means that there are Virasoro primary operators with $\bar h\neq0$. Nevertheless, the correlation functions of the meromorphic operators can be taken in and of themselves to define a certain meromorphic theory. Such theories are referred to by various authors as chiral algebras, vertex operator algebras, or meromorphic conformal field theories. Though some of these names are occasionally assigned to structures that possess some extra nice properties, such as modular invariant partition functions, we will be discussing the most basic version. Henceforth, by chiral algebra we will mean the operator product algebra of a set of meromorphic operators in the plane.\footnote{In a preview of later discussions, we mention that by $\WW$-algebra we will mean a chiral algebra for which the space of local operators is generated by a finite number of operators via the operations of taking derivatives and normal-ordered multiplication.} So defined, a chiral algebra is strongly constrained by the requirements of crossing symmetry. In what follows, we show that any $\NN=2$ superconformal field theory in four dimensions possesses a class of observables that define a chiral algebra in this sense.

\subsection{Twisted conformal subalgebras}
\label{subsec:twisted_subalgebra}

Chiral algebras are ordinarily thought to be a special feature of conformal-invariant models in two dimensions. Indeed, the appearance of an infinite number of conserved charges as defined in \eqref{eqn:charges_from_operators} follows from the interaction of two different ingredients that are special to two dimensions. Firstly, the operators that give rise to the chiral symmetry charges are invariant under (say) the anti-holomorphic factor of the two-dimensional conformal algebra, while transforming in a nontrivial representation of the holomorphic factor, so they are nontrivial holomorphic operators on the plane. The powerful machinery of complex analysis in a single variable then produces the infinity of conserved charges in \eqref{eqn:charges_from_operators}.\footnote{From another point of view, one can hardly hope to find a meromorphic sector in a higher dimensional CFT due to Hartogs' theorem, which implies the absence of singularities of codimension greater than one in a meromorphic function of several variables. This has been overcome in, \eg, \cite{Johansen:1994ud,Kapustin:2006hi} by considering extended operators that intersect in codimension one. The problem, then, is that the meromorphic structure does not impose constraints on the natural objects in the original theory -- the local operators.}

In dimension $d>2$, it is the first of these conditions that fails the most dramatically, while the latter seems more superficial. Indeed, correlation functions in a conformal field theory in higher dimensions can be restricted so that all operators lie on a plane $\Rb^2\subset\Rb^d$, and the resulting observables will transform covariantly under the subalgebra of the $d$-dimensional conformal algebra that leaves the $\Rb^2$ in question fixed,
%%%%%%
\be
\label{eq:2d_conformal_subalgebra}
\mf{sl}(2)\times\overline{\mf{sl}(2)}\subset\mf{so}(d+2)~.
\ee
%%%%%%
These correlation functions will be largely indistinguishable from those of an authentic two-dimensional CFT, and if one could locate operators that were chiral with respect to this subalgebra, then the arguments of \S\ref{subsec:chiral_symmetry_review} would go through unhindered and a chiral symmetry algebra could be constructed that would act on $\Rb^2$-restricted correlation functions. However, a local operator that transforms in the trivial representation of either copy of $\mf{sl}(2)$ in \eqref{eq:2d_conformal_subalgebra} will necessarily be trivial with respect to \emph{all} of $\mf{so}(d+2)$. As such, the only ``meromorphic'' operator on the plane in a higher dimensional theory is the identity operator, and no chiral symmetry algebra can be constructed. This is ultimately a consequence of the simple fact that the higher dimensional conformal algebras do not factorize into a holomorphic and anti-holomorphic part: any two $\mf{sl}(2)$ subalgebras will be related by conjugation.

The brief arguments given above are common knowledge, and essentially spell the end to any hopes of recovering chiral symmetry algebras in a general higher-dimensional conformal field theory. We have reproduced them here to clarify the mechanism by which they will be evaded in the coming discussion. In particular, we will see that the additional tools at our disposal in the case of \emph{super}conformal field theories are sufficient to give life to chiral algebras in four dimensions. Before describing the construction, let us recall a simple example which illustrates the mechanism that will be used.

\subsubsection{Intermezzo: translation invariance from cohomology}
\label{subsubsec:intermezzo}

In a quantum field theory with $\NN=1$ supersymmetry in four dimensions, there exists a special class of operators known as \emph{chiral operators} (not to be confused with the meromorphic operators of \S\ref{subsec:chiral_symmetry_review}, which are chiral in a different sense) that lie in short representations of the supersymmetry algebra and satisfy a shortening condition in terms of a chiral half of the supercharges,
%%%%%%
\be
\label{eq:Poincare_chiral_operator}
\{Q_{\alpha},\OO(x)]=0~,\qquad\alpha=\pm\,.
\ee
%%%%%%
The translation generators in $\Rb^4$ are exact with respect to the chiral supercharges,
%%%%%%
\be
\label{eq:Poincare_translation_exact}
P_{\alpha\dot\alpha}=\{Q_{\alpha},\wt Q_{\dot\alpha}\}~,
\ee
%%%%%%
and consequently, via the Jacobi identity, the derivative of a chiral operator is also exact,
%%%%%%
\be
[ P_{\alpha\dot\alpha},\OO(x) ]=\{Q_{\alpha},\OO^\prime(x)]~.
\ee
%%%%%%
Because the chiral supercharges are nilpotent and anti-commute, the cohomology classes of chiral operators with respect to the supercharges $Q_\alpha$ are well-defined and independent of the insertion point of the operator. Schematically, one can write
%%%%%%
\be
[\OO_i(x)]_{Q_\alpha}\colonequals\OO_i~.
\ee 
%%%%%%
Products of chiral operators are then free of short distance singularities and form a ring at the level of cohomology. Correlation functions of chiral operators have the excellent property of being independent of the positions of the operators,
%%%%%%
\be
\label{eq:Poincare_position_independent}
\langle\OO_1(x_1)\OO_2(x_2)\ldots\OO_n(x_n)\rangle=\langle[\OO_1(x_1)][\OO_2(x_2)]\ldots[\OO_n(x_n)]\rangle=\langle\OO_1\OO_2\ldots\OO_n\rangle~.
\ee
%%%%%%
A suggestive way of phrasing this well-known feature of the chiral ring is that although chiral operators transform in a nontrivial representation of the four-dimensional translation group, their cohomology classes with respect to the chiral supercharges transform in the trivial representation. The passage from local operators to their cohomology classes modifies the transformation properties of these local operators under the spacetime symmetry algebra, in this case rendering them trivial. 

\subsubsection{Holomorphy from cohomology}

To recover chiral algebras in four dimensions, we adopt the same philosophy just illustrated in the example of the chiral ring. We will find a nilpotent supercharge with the property that \emph{cohomology classes} of local operators with respect to said supercharge transform in a chiral representation of an $\mf{sl}(2)\times\widehat{\mf{sl}(2)}$ subalgebra of the full superconformal algebra, and as such behave as meromorphic operators. Such local operators will then necessarily constitute a chiral algebra as described in \S\ref{subsec:chiral_symmetry_review}.

The first task that presents itself is an algebraic one. To realize chiral symmetry at the level of cohomology classes, we identify a two-dimensional conformal subalgebra of the four-dimensional superconformal algebra,
$$
\mf{sl}(2)\times\widehat{\mf{sl}(2)}\subset\mf{sl}(4\,|\,2)~,
$$ 
along with a privileged supercharge $\qq\,$ for which the following criteria are satisfied:
%%%%%%
\begin{enumerate}
\item[$\bullet$]{The supercharge is nilpotent: $\qq\,^2=0$.}
\item[$\bullet$]{$\mf{sl}(2)$ and $\widehat{\mf{sl}(2)}$ act as the generators of holomorphic and anti-holomorphic M\"obius transformations on a complex plane $\Cb\subset\Rb^4$.}
\item[$\bullet$]{The holomorphic generators spanning $\mf{sl}(2)$ commute with $\qq\,$.}
\item[$\bullet$]{The anti-holomorphic generators spanning $\widehat{\mf{sl}(2)}$ are $\qq\,$ commutators.}
\end{enumerate}
%%%%%%

In searching for such a subalgebra, we can first restrict our attention to subalgebras of $\mf{sl}(4|2)$ that keep the plane fixed set-wise. There are two inequivalent maximal subalgebras of this kind: $\mf{sl}(2|1)\times \mf{sl}(2|1)$, which is the symmetry algebra of an $\NN=(2,2)$ SCFT in two dimensions, and $\mf{sl}(2)\times\mf{sl}(2|2)$, which is the symmetry algebra of an $\NN=(0,4)$ SCFT in two dimensions. One easily determines that the first subalgebra cannot produce the desired structure; we proceed directly to consider the second case.

The four-dimensional $\NN=2$ superconformal algebra and the two-dimensional $\NN=(0,4)$ superconformal algebra are summarized in Appendix \ref{app:SCAs}. In embedding the latter into the former, we take the fixed two-dimensional subspace to be the one that is fixed pointwise by the rotation generator
%%%%%%
\be
\MM^\perp\colonequals\MM^{\ph{+}+}_{+}-{\MM}^{\dot+}_{\ph{+}\dot+}~.
\ee
%%%%%%
The generator of rotations acting within the fixed plane is the orthogonal combination,
%%%%%%
\be
\MM\colonequals\MM^{\ph{+}+}_{+}+{\MM}^{\dot+}_{\ph{+}\dot+}~.
\ee
%%%%%%
In more conventional terms, we are picking out the plane with $x_1=x_2=0$. Introducing complex coordinates $z\colonequals x_3+ix_4$, $\bar z\colonequals x_3-i x_4$, the two-dimensional conformal symmetry generators in $\mf{sl}(2)\times\mf{sl}(2|2)$ can be expressed in terms of the four-dimensional ones as
%%%%%%
\be\label{eq:conformal_embedding}
\begin{split}
L_{-1}&=\PP_{+\dot+}~,\qquad L_{+1}=\KK^{\dot++}~,\qquad 2L_0=\HH+\MM~,\\
\Lb_{-1}&=\PP_{-\dot-}~,\qquad \Lb_{+1}=\KK^{\dot--}~,\qquad 2\Lb_0=\HH-\MM~.
\end{split}
\ee
%%%%%%
The fermionic generators of $\mf{sl}(2)\times\mf{sl}(2|2)$ are obviously all anti-holomorphic, and upon embedding are identified with four-dimensional supercharges according to
%%%%%%
\be\label{eq:supercharge_embedding}
\QQ^\II=\QQ^\II_-~,\qquad\wt\QQ_\II=\wt\QQ_{\II\dot-}~,\qquad\SS_\II=\SS_\II^{-}~,\qquad\wt\SS^\II=\wt\SS^{\II\dot-}~,
\ee
%%%%%%
where $\II=1,2$ is an $\mf{sl}(2)_R$ index. Finally, the $\mf{sl}(2|2)$ superalgebra has a central element $\ZZ$, which upon embedding is given in terms of four-dimensional symmetry generators as
%%%%%%
\be\label{eqn:2d_4d_central}
\ZZ = r + \MM^\perp~,
\ee
%%%%%%
where $r$ is the generator of $U(1)_r$.

Amongst the supercharges listed in \eqref{eq:supercharge_embedding}, one finds a variety of nilpotent operators. Any such operator will necessarily commute with the generators $L_{\pm1}$ and $L_0$ in \eqref{eq:conformal_embedding} since all of the supercharges do so. The requirement of $\qq\,$-exact anti-holomorphic M\"obius transformations is harder to satisfy. In fact, up to similarity transformation using generators of the bosonic symmetry algebra, there are only two possible choices:
%%%%%
\be
\begin{split}
\qq\,_1 \colonequals \QQ^1 +  \tilde \SS^2 		~ &,  \quad \qq\,_2 \colonequals \SS_1 - \tilde \QQ_2 ~,\\
\qq\,_1^\dagger \colonequals  \SS_1 + \tilde \QQ_2 ~ &, \quad \qq\,_2^\dagger \colonequals \QQ^1 -  \tilde \SS^2 \,.
\end{split}
\ee
%%%%%
Interestingly, $\qq\,_1$ and $\qq\,_2$ give rise to \emph{the same} $\qq\,$-exact generators of an anti-holomorphic $\widehat{\mf{sl}(2)}$ algebra,
%%%%%%
\be\label{qqcomm}
\begin{alignedat}{7}
&\{ \q \, , \tilde \QQ_1 \} &~~=~~& \{ \qd \, , -\QQ^2        \}  &~~=~~& \ph{2}\bar L_{-1} +  \RR^- &~~\equalscolon~~& \ph{2} \widehat L_{-1} ~,\\
&\{ \q \, , \SS_2        \} &~~=~~& \{ \qd \, , \tilde \SS^1 \}  &~~=~~& \ph{2}\bar L_{+1} -  \RR^+ &~~\equalscolon~~& \ph{2} \widehat L_{+1} ~,\\
&\{ \q \, , \q^\dagger   \} &~~=~~& \{ \qd \, , \qd^\dagger  \}  &~~=~~&     2(\bar L_{ 0} -  \RR)  &~~\equalscolon~~&     2  \widehat L_{ 0} ~.
\end{alignedat}
\ee
%%%%%%
In addition, the central element of $\mf{sl}(2|2)$ is exact with respect to both supercharges,
%%%%%%
\be\label{eq:central_element_exact}
\{ \q \, , \qd \} = -\ZZ~.
\ee
%%%%%%
Note that while $\widehat{\mf{sl}(2)}$ \emph{does} act on the plane by anti-holomorphic conformal transformations, it is not simply a subalgebra of the original global conformal algebra. Rather, it is an $\mf{sl}(2)_R$ twist of $\overline{\mf{sl}(2)}$.\!\footnote{In light of this, we may understand the absence of a similar construction using the $\mf{sl}(2|1)\times\mf{sl}(2|1)$ algebra as a consequence of there being no $\mf{sl}(2)_R$ with which to twist. Similarly, our construction does not extend to $\mathcal N = 1$ superconformal theories since they only have an abelian R-symmetry.} Because the relevant real forms of the $\overline{\mf{sl}(2)}$ conformal algebra and $\mf{sl}(2)_R$ are different, the generators of $\widehat{\mf{sl}}(2)$ do not enjoy any reasonable hermiticity properties when acting on the Hilbert space of the four-dimensional theory. In particular, we can immediately see that $\widehat{L}_{\pm1}^\dagger\neq\widehat{L}_{\mp1}$. This would complicate matters considerably if our intention was to study operators that transform in nontrivial representations of this twisted algebra. Fortunately, our plan is precisely the opposite: chiral algebras can appear after passing to $\qq\,$-cohomology, at which point all of the objects of interest will effectively be invariant under the action of $\widehat{\mf{sl}(2)}$. Consequently, reality/hermiticity conditions will play no role in the structure of the ``physical'' operators/observables defined at the level of cohomology.

\subsection{The cohomology classes of local operators}
\label{subsec:local_op_cohomology}

Our next task is to study the properties of operators that define non-trivial $\qq\,_i$-cohomology classes. For the purposes of the present paper, we are restricting our attention to \emph{local operators} in four dimensions; the inclusion of non-local operators, such as line or surface operators, is an interesting extension that will be addressed in future work.

We begin by identifying the requirements for an operator inserted at the origin to define a nontrivial $\qq\,_i$-cohomology class. In particular, we will derive the conditions under which an operator $\OO(x)$ obeys
%%%%%%
\be\label{eq:qq_closed}
\{\qq\,_i,\OO(0)]=0~,\qquad\OO(0)\neq\{\qq\,_i,\OO^\prime(0)]~,
\ee
%%%%%%
for $i=1$ or $i=2$. Because both $\qq\,_i$ commute with $\widehat L_0$ and $\ZZ$, we lose no generality in restricting to definite eigenspaces of these charges. A standard cohomological argument then implies that since $\widehat L_0$ and $\ZZ$ are actually $\qq\,_i$-\emph{exact}, an operator satisfying \eqref{eq:qq_closed} must lie in the zero eigenspace of both charges.
%%%%%%
In terms of four-dimensional quantum numbers, this means that such an operator must obey\footnote{In fact, the second relation in \eqref{macrelations} follows from the first as a consequence of unitarity and the four-dimensional superconformal algebra (see \S\ref{subsec:schur}). We list it separately here since it is an algebraically independent constraint at the level of the quantum numbers.}
%%%%%%
\be \label{macrelations}
\tfrac 12(E-(j_1+j_2)) - R = 0~,\quad r + (j_1-j_2) = 0~,
\ee
%%%%%%
where $E$ is the conformal dimension/eigenvalue of $\HH$, $j_1$ and $j_2$ are $\mf{sl}(2)_1$ and $\mf{sl}(2)_2$ Lorentz quantum numbers/eigenvalues of $\MM_+^{\ph{+}+}$ and $\MM^{\dot+}_{\ph{+}\dot+}$, and $R$ is the $\mf{sl}(2)_R$ charge/eigenvalue of $\RR$. As long as the four-dimensional SCFT is unitary, the last line of \eqref{qqcomm} implies that any operator with zero eigenvalue under $\widehat L_0$ must be annihilated by $\qq\,_i$ and $\qq\,_i^\dagger$ for \emph{both} $i=1$ and $i=2$. The relations in \eqref{macrelations} therefore characterize the harmonic representatives of $\qq\,_i$-cohomology classes of operators at the origin, and we see that the two supercharges actually define the \emph{same} cohomology. Notably, these relations are known to characterize the operators that contribute to the Schur (and Macdonald) limits of the superconformal index in four dimensions \cite{Gadde:2011uv}, suggesting that the cohomology will be non-empty in any nontrivial $\NN=2$ SCFT. We will refer to the class of local operators obeying \eqref{macrelations} as the {\it Schur operators} of the SCFT. We will have more to say about the features of these operators in \S\ref{sec:new2d4d}.

Note that in contrast to the case of ordinary chiral operators in a supersymmetric theory, which are annihilated by a given Poincar\'e supercharge regardless of the insertion point, for operators to be annihilated by the $\qq\,_i$ when inserted away from the origin requires that they acquire a more intricate dependence on their position in $\Rb^4$. This is a consequence of the fact that the translation generators do not commute with the superconformal charges $\SS^-_1$ and $\tilde \SS^{2\dot{-}}$ appearing in the definitions of the $\qq\,_i$. Indeed, there is no way to define the translation of a Schur operator from the origin to a point outside of the $(z,\zb)$ plane so that it continues to represent a $\qq\,_i$-cohomology class. Within the plane, though, we can accomplish this task using the $\qq\,_i$-exact, twisted $\widehat{\mf{sl}(2)}$ of the previous subsection. In particular, because the twisted anti-holomorphic translation generator $\widehat L_{-1}$ is a $\qq\,_i$ anti-commutator and the holomorphic translation generator $L_{-1}$ is $\qq\,_i$-closed, we can define the \emph{twisted-translated} operators
%%%%%%
\be\label{eq:twistedtranslate}
\OO(z,\bar z)=e^{zL_{-1} + \bar z\widehat L_{-1}}\,\OO(0)\,e^{- zL_{-1} - \bar z\widehat L_{-1}}~,
\ee
%%%%%%
where $\OO(0)$ is a Schur operator. One way of thinking about this prescription for the translation of local operators is as the consequence of introducing a constant, complex background gauge field for the $\mf{sl}(2)_R$ symmetry that is proportional to the $\mf{sl}(2)$ raising operator. By construction, the twisted-translated operator is $\qq\,_i$ closed for both $i=1,2$, and the cohomology class of this operator is well-defined and depends on the insertion point holomorphically,
%%%%%%
\be
[\OO(z,\bar z)]_\qq~~~\Longrightarrow~~~\OO(z)~.
\ee
%%%%%%
What does such an operator look like in terms of a more standard basis of local operators at the point $(z,\zb)$? To answer this, we must first note that Schur operators at the origin occupy the highest-weight states of their respective $\mf{sl}(2)_R$ representation (this fact will be explained in greater detail in \S\ref{sec:new2d4d}). If we denote the whole spin $k$ representation of $\mf{sl}(2)_R$ as $\OO^{\II_1\II_2\cdots\II_{2k}}$ with $\II_i=1,2$, then the Schur operator at the origin is $\OO^{11\cdots1}(0)$, and the twisted-translated operator at any other point is defined as
%%%%%%
\be\label{displaced}
\OO(z, \bar z) \colonequals u_{\II_1}(\bar z)\,\cdots\,u_{\II_{2k} }(\bar z) \; \OO^{\II_1 \dots \II_{2k}} (z, \bar z) \,, \qquad\quad u_{\II} (\bar z) \colonequals (1, \bar z) \,.
\ee
%%%%%%
At any given point $(z,\zb)$, this is a particular complex-linear combination of the different elements of the $\mf{sl}(2)_R$ representation of the corresponding Schur operator. The precise combination depends on the insertion point as indicated. What we have discovered is that the correlation functions of these operators are determined at the level of their $\qq\,_i$-cohomology classes, and are therefore meromorphic functions of the insertion points.\footnote{For $\NN = 4$ SYM, a similar contraction of the $SU(4)_R$ indices with position-dependent vectors was studied in \cite{Drukker:2009sf}. The twists considered in that paper are different, and do not give rise to meromorphic operators and chiral algebras.}

\subsection{A chiral operator product expansion}
\label{subsec:chiral_OPE}

The most efficient language for describing chiral algebras is that of the operator product expansion. Let us therefore study the structure of the operator product expansion of the twisted-translated Schur operators in order to see the emergence of meromorphic OPEs befitting a chiral algebra.

Consider two operators: $\OO_1(z,\zb)$ is the twisted translation of a Schur operator from the origin to $(z,\zb)$, and $\OO_2(0,0)$ is a Schur operator inserted at the origin. Given the general expression for the twisted-translated operator given in \eqref{displaced}, the OPE of these two operators should take the form
%%%%%%
\be\label{eq:chiral_ope_general}
\OO_1(z,\zb)\OO_2(0)=\sum_k\lambda_{12k}\frac{\bar z^{R_1+R_2-R_k}}{z^{h_1+h_2-h_k}\zb^{\hb_1+\hb_2-\hb_k}}\OO_k(0)~,
\ee
%%%%%%
where the $\bar z^{R_1+R_2-R_k}$ in the numerator comes from the explicit factors of $\zb$ appearing in \eqref{displaced}, and $R_k$ is the $R$-charge of the operator $\OO_k$. This form of the OPE is so far a consequence of two-dimensional conformal invariance and conservation of $R$-charge under multiplication. We have introduced the two-dimensional quantum numbers $h$ and $\hb$, which are expressible in terms of four-dimensional quantum numbers as
%%%%%%
\be
h = \frac{E+(j_1+j_2)}{2}~,\qquad \hb = \frac{E-(j_1+j_2)}{2}~.
\ee
%%%%%%
Though the OPE does not look meromorphic yet, we are already well on our way. The left hand side of \eqref{eq:chiral_ope_general} is $\qq\,_i$-closed for any $(z,\zb)$, with the $\zb$ dependence being $\qq\,_i$-exact. As a result, each individual term on the right hand side must be $\qq\,_i$-closed, and the sum should be reorganized into two groups. The first group will consist of the terms in which the operator $\OO_k(0)$ is a Schur operator, while the second will consist of the remaining terms, for which the operator $\OO_k(0)$ is $\qq\,_i$-exact. Recalling that the quantum numbers of Schur operators obey $\hb=R$, we immediately see that for those terms in the OPE the $\zb$ dependence cancels between the denominator and the numerator, thus providing the desired meromorphicity result:
%%%%%%
\be
\OO_1 (z,\zb)\,\OO_2(0,0)\, = \sum_{k_{\text{Schur}}}  \frac{\lambda_{12k}}{z^{h_1 + h_2 - h_k}}\, \OO_k(0) \, + \{\qq\,, \dots ]~.
\ee
%%%%%%
From the four-dimensional construction, we expect this OPE to be single-valued, which implies that $h_1+h_2-h_k$ should be an integer. Indeed, this integrality follows from the fact that $h$ is a sum of $SU(2)$ Cartans after applying $SU(2)$ selection rules. Clearly, in passing to $\qq\,_i$-cohomology classes the OPE stays well-defined and the $\qq\,_i$-exact piece can be set to zero. Thus at the level of cohomology, the twisted-translated operators can be reinterpreted as two-dimensional meromorphic operators with interesting singular OPEs.

It may be instructive to see how this meromorphic OPE plays out in a simple example. An extremely simple case, to which we shall return in \S\ref{sec:new2d4d}, is that of free hypermultiplets in four dimensions. The scalar squarks $Q$ and $\tilde Q$ of the hypermultiplet are Schur operators, and the corresponding twisted-translated operators take the form
%%%%%%
\be\label{eq:freehyperexpansion}
q(z)\colonequals[Q(z,\bar z)+\bar z \tilde{Q}^*(z,\bar z)]_\qq~,\qquad 
\tilde{q}(z)\colonequals[\tilde Q(z,\bar z)-\bar z Q^*(z,\bar z)]_\qq~.
\ee
%%%%%%
The singular OPE of these twisted operators can be easily worked out using the free OPE in four dimensions; we have
%%%%%%
\be \label{qOPE}
\begin{alignedat}{2}
q(z)q(w)		&\sim&~ {\rm regular}~,\qquad \tilde q(z)\tilde q(w)&\sim~{\rm regular}~,\\
q(z)\tilde q(w) &\sim& \frac{1}{z-w}~,\qquad \tilde q(z)q(w) &\sim-\frac{1}{z-w}~.
\end{alignedat}
\ee
%%%%%%
This is example is in some respects deceptively simple, in that the terms appearing in the singular part of the OPE are meromorphic on the nose. In more complicated theories, there will be cohomologically trivial terms appearing in the singular part of the OPE, and meromorphicity will depend on a more detailed knowledge of the action of the nilpotent supercharges.

Let us briefly point out one difference between the structure observed here and that of a more conventional cohomological subalgebra. The chiral ring in the free hypermultiplet theory is generated by the operators $q(x)$ and $\tilde q(x)$. Because these operators both have $R=1/2$, there can be no nonzero correlation functions in the chiral ring. The existence of nontrivial correlation functions in the chiral algebra described here follows precisely from the presence of subleading terms in the $\zb$ expansion \eqref{eq:freehyperexpansion} with $SU(2)_R$ quantum numbers of opposite sign relative to the leading term.

Having established existence of nontrivial $\qq$-cohomology classes with meromorphic OPEs and correlators, we now take some time to develop the dictionary between four-dimensional SCFT structures and their two-dimensional counterparts.

\section{The SCFT/chiral algebra correspondence}
\label{sec:new2d4d}

For any four-dimensional $\NN=2$ superconformal field theory, we have identified a subsector of operators whose correlation functions are meromorphic when they are restricted to be coplanar. This sector thus defines a map from four-dimensional SCFTs to two-dimensional chiral algebras:
%%%%%%
$$
\goodchi~:~\text{4d SCFT}~\longrightarrow~\text{2d Chiral Algebra}.
$$
%%%%%%
The aim of this section is to elaborate on the structure of this correspondence, focusing primarily on its more universal aspects. We begin with a short preview of some of the more prominent features of the correspondence.

Our first main result is the generic enhancement of the global $\mf{sl}(2)$ conformal symmetry algebra to a full fledged Virasoro algebra. In other words, for any SCFT $\TT$, we find that $\goodchi[\,\TT\,]$ contains a meromorphic stress tensor. The two-dimensional central charge turns out to have a simple relationship to the four-dimensional conformal anomaly coefficient, 
%%%%%%
$$
c_{2d} = - 12 c_{4d}~.
$$
%%%%%%
In particular, this implies that when $\TT$ is unitary (which we always take to be the case), $\goodchi[\,\TT\,]$ is necessarily non-unitary. In a similar vein, we find that global symmetries of $\TT$ are always enhanced into affine symmetries of $\goodchi[\,\TT\,]$, and the respective central charges of these flavor symmetries enjoy another simple relationship,
%%%%%%
$$
k_{2d} = -\frac12 k_{4d}~.
$$
%%%%%%

It is often helpful to think of a chiral algebra in terms of its generators. In the chiral algebra sense of the word, generators are those operators that cannot be expressed as the conformally normal-ordered products of derivatives of other operators. While we do not find a complete characterization of the generators of our chiral algebras, we do identify certain operators in four dimensions whose corresponding chiral operator will necessarily be generators. In particular, operators that are $\NN=1$ chiral \emph{and} satisfy the Schur shortening condition form a ring which is a consistent truncation of the $\NN=1$ chiral ring, to which we refer as the Hall-Littlewood (HL) chiral ring. We find that every generator of the HL chiral ring necessarily leads to a generator of the associated chiral algebra. There may be additional generators of the chiral algebra beyond the stress tensor and the operators associated to generators of the HL chiral ring. We will find such additional generators in the example of \S\ref{subsec:genustwo}.

For the special case of free SCFTs we completely characterize the associated chiral algebras. Unsurprisingly, free SCFTs give rise to free chiral algebras. In particular, free hypermultiplets correspond to the chiral algebra of dimension $1/2$ symplectic bosons, while free vector multiplets correspond to the small algebra of a $(b,c)$ ghost system of dimension $(1,0)$.

Finally, we describe the two-dimensional counterpart of gauging a flavor symmetry $G$ in some general SCFT $\TT_G$. Assuming that the chiral algebra associated to the ungauged SCFT is known, the prescription to find the chiral algebra of the new theory is as follows. The direct product of the original chiral algebra $\goodchi[\,\TT_G\,]$ with a $(b,c)$ system in the adjoint representation of $G$ admits a nilpotent BRST operator precisely when the beta function for the four-dimensional gauge coupling vanishes. The chiral algebra of the gauged theory is then obtained by restricting to the BRST coholomogy. We find that this BRST operator precisely captures the one-loop correction to a certain four-dimensional supercharge, so that restricting to its cohomology is equivalent to the requirement that one should only retain those states that remain in their original short representations once one-loop corrections are taken into account.

\subsection{Schur operators}
\label{subsec:schur}

As a first order of business, we pursue a more concrete characterization of the four-dimensional operators whose correlation functions are captured by the chiral algebra. Let us first reiterate the basic facts about these operators that were derived in \S\ref{sec:section2}. The chiral algebra computes correlation functions of operators that define nontrivial cohomology classes of the nilpotent supercharges $\qq\,_i$. Such operators are obtained by \emph{twisted translations} \eqref{displaced} of Schur operators from the origin to an arbitrary point $(z,\zb)$ on the plane. A Schur operator is any operator that satisfies
%%%%%%
\bea
\makebox[1in][l]{$[\widehat L_0, \OO] = 0 $} &\Longleftrightarrow& \makebox[2in][l]{\quad\quad$\tfrac12\left(E-(j_1 + j_2)\right) - R = 0 ~,$}\label{mac2}\\ 
\makebox[1in][l]{$[\ZZ,\OO] = 0$} &\Longleftrightarrow& \makebox[2in][l]{\quad\quad$r + j_1 - j_2 = 0~.$}\label{mac2b}
\eea
%%%%%%
If $\TT$ is unitary, then these conditions can be equivalently formulated as the requirement that when inserted \emph{at the origin}, an operator is annihilated by the two Poincar\'e and the two conformal supercharges that enter in the definition of the $\qq\,_i$, \ie,
%%%%%%
\be  \label{schurQS}
[\QQ\,^1_-,\OO(0)]=[\widetilde\QQ\,_{2\dot-},\OO(0)] = [\SS_1^-,\OO(0)] =  [\widetilde\SS^{2\dot-},\OO(0)] =0~.
\ee
%%%%%%
This follows from the hermiticity conditions $\Qm^{1\dagger}_-\colonequals\Sm_{1}^-$ and $\Qm_{2 \dot -}^\dagger \colonequals \widetilde \Sm^{2 \dot -}$ in conjunction with the relevant anticommutators from Appendix \ref{app:SCAs},
%%%%%%
\be \label{relevantanti}
\{ \Qm^1_-\, ,  \Qm^{1\dagger}_-  \} = \widehat L_0 - \frac12{\ZZ}    \, , \quad     \{  \widetilde \Qm_{2 \dot -}  \, ,  \widetilde \Qm_{2 \dot -}^\dagger    \}  =  \widehat L_0  + \frac12{\ZZ} \,.
\ee
%%%%%%
It follows immediately that the state  $\OO(0) | 0 \rangle$ is annihilated by all four supercharges if and only if its quantum numbers obey \eqref{mac2} and \eqref{mac2b}. Actually, \eqref{relevantanti} implies the additional inequality
%%%%%%
\be
\widehat L_0 \geqslant   \frac{| {\ZZ} |}{2}~,
\ee
%%%%%%
from which we may conclude that imposing only \eqref{mac2} is a necessary and sufficient condition to define a Schur operator. We further note that Schur operators are necessarily the highest-weight states of their respective $SU(2)_R$ representations, and so carry the maximum eigenvalue $R$ of the Cartan generator. If this were not the case, states with greater $R$ would have negative $\widehat L_0$ eigenvalues, in contradiction with unitarity. Similarly, Schur operators are necessarily the highest weight states of their $SU(2)_1\times SU(2)_2$ Lorentz symmetry representation, carrying the largest eigenvalues for $j_1$ and $j_2$. The index structure of a Schur operator is therefore of the form $\OO^{1\dots1}_{+\dots+\,\dot+\dots\dot+}$.

From the definition of $L_0$ in \eqref{eq:conformal_embedding} and \eqref{mac2} we find that the holomorphic dimension $h$ of a Schur operator is non-zero and fixed in terms of its quantum numbers,
%%%%%%
\be\label{eq:schur_left_dim}
h = \tfrac12\left(E + j_1 + j_2\right) = R + j_1 + j_2~.
\ee
%%%%%%
This is always a half integer, since $R$, $j_1$ and $j_2$ are all $SU(2)$ Cartans. It follows from \eqref{mac2b} and \eqref{eq:schur_left_dim}, in conjunction with the non-negativity of $j_1$ and $j_2$, that the holomorphic dimension of a Schur operator is bounded from below in terms of its four-dimensional $R$-charges,
%%%%%%
\be \label{L_0bound}
h=R+j_1+j_2\geqslant R+|j_1-j_2| = R+|r|~.
\ee
%%%%%%
The inequality is saturated if and only if $j_1$ or $j_2$ is zero.

\renewcommand{\arraystretch}{1.5}
\begin{table}
\centering
\begin{tabular}{|l|l|l|l|l|}
\hline \hline
Multiplet  & $\OO_{\rm Schur}$  & $h$ & $r$  & Lagrangian  ``letters''  \\ 
\hline 
$\hat \BB_R$  &  $\Psi^{11\dots 1}$   &    $R$ &  $0$ & $Q$, $\tilde Q$ \\ 
\hline
$\DD_{R (0, j_2 )}$  &    $ \widetilde {\QQ}^1_{\dot +} \Psi^{11\dots 1}_{\dot  + \dots \dot  + }$ &   $R+ j_2 +1$  & $j_2 + \frac{1}{2}$  &  $Q$, $\tilde Q$, $\tilde \lambda^1_{\dot +}$ \\
\hline
$\bar \DD_{R (j_1, 0 )}$  & $ {\QQ}^1_{ +} \Psi^{11\dots 1}_{+   \dots +}$ &     $R+ j_1 +1$  & $-j_1 - \frac{1}{2}$  & $Q$, $\tilde Q$,  $\lambda^1_{+}$ \\
\hline
$\hat \CC_{R (j_1, j_2) }$ &   ${\QQ}^1_{+} \widetilde  {\QQ}^1_{\dot +} \Psi^{11\dots 1}_{+   \dots + \, \dot  + \dots \dot  + }$&   
 $R+ j_1 + j_2 +2$  & $j_2 - j_1$  &
 $D_{+ \dot +}^n Q$,  $D_{+ \dot +}^n \tilde Q$,  $D_{+ \dot +}^n \lambda^1_{+}$,
  $D_{+ \dot +}^n  \tilde \lambda^1_{\dot +}$ \\
\hline
\end{tabular}
\caption{\label{schurTable} This table summarizes the manner in which Schur operators fit into short multiplets of the $\NN=2$ superconformal algebra. For each supermultiplet, we denote by $\Psi$ the superconformal primary. There is then a single conformal primary Schur operator ${\OO}_{\rm Schur}$, which in general is obtained by the action of some Poincar\'e supercharges on $\Psi$. We list the holomorphic dimension $h$ and $U(1)_r$ charge $r$ of ${\OO}_{\rm Schur}$ in terms of the quantum numbers $(R,j_1,j_2)$ that label the shortened multiplet (left-most column). We also indicate the schematic form that ${\OO}_{\rm Schur}$ can take in a Lagrangian theory by enumerating the elementary ``letters'' from which the operator may be built. We denote by $Q$ and $\tilde Q$ the complex scalar fields of a hypermultiplet, by $\lambda_{\alpha}^\II$  and $\tilde \lambda_{\dot \alpha}^\II$ the left- and right-moving fermions of a vector multiplet, and by $D_{\alpha \dot \alpha}$ the gauge-covariant derivatives.
} 
\end{table}

\subsubsection{Supermultiplets of Schur type}
\label{subsubsec:schur_supermultiplets}

Schur operators belong to shortened representations of the $\NN =2$ superconformal algebra. The complete list of possible shortening conditions is reviewed in Appendix \ref{app:shortening}. In the notations of \cite{Dolan:2002zh}, the superconformal multiplets that contain Schur operators are the following,
%%%%%%
\be \label{schursuper}
\hat\BB_R~,\quad \DD_{R(0,j_2)}~,\quad \bar\DD_{R(j_1,0)}~,\quad\hat\CC_{R(j_1,j_2)}~.
\ee
%%%%%%
For the purpose of enumeration, it is sufficient to focus on those Schur operators that are conformal primaries. Given such a \emph{primary Schur operator}, there is a tower of descendant Schur operators that are obtained by the action $L_{-1} =  P_{+ \dot +} = - \partial_{+ \dot +}$. It turns out that each of the supermultiplets listed in \eqref{schursuper} contains exactly one conformal primary Schur operator. In the case of $\hat\BB_R$, this is also the \emph{super}conformal primary of the multiplet, whereas in the other cases it is a superconformal descendant. This representation-theoretic information is summarized in Table \ref{schurTable}, where we also provide the schematic form taken by each type of operator in a Lagrangian theory.

The shortening conditions obeyed by the Schur operators make crucial use of the extended $\NN=2$ supersymmetry. Indeed, the hallmark of a Schur operator is that it is annihilated by two Poincar\'e supercharges of \emph{opposite} chiralities ($\Qm_-^1$ and  $\widetilde \Qm_{2 \dot -}$ in our conventions). This defines a consistent shortening condition because the supercharges have the same $SU(2)_R$ weight, and thus anticommute with each other. No analogous shortening condition exists in an $\NN=1$ supersymmetric theory, because the anticommutator of opposite-chirality supercharges necessarily yields a momentum operator, which annihilates only the identity.

Although the most general Schur operators, which are those belonging to  $\hat \CC_{R(j_1,j_2)}$ multiplets, may seem somewhat exotic, the Schur operators of type $\hat \BB_R$,  $\DD_{R (0, j_2 )}$ and $\bar \DD_{R (j_1, 0 )}$ are relatively familiar. Indeed, they can be understood as special cases of conventional $\NN=1$ chiral or anti-chiral operators. Let us focus for the moment on the ${\NN}=1$ Poincar\'e subalgebra that contains the supercharges
%%%%%%
\be \label{N1sub}
\Qm_\alpha^2~,\quad \widetilde \Qm_{2 \dot \alpha}~.
\ee
%%%%%%
We then ask what subset of Schur operators are also elements of the chiral ring for this $\NN=1$ subalgebra. In particular, such operators will be annihilated by \emph{both} spinorial components of the anti-chiral supercharge $\widetilde \Qm_{2 \dot \alpha}$, $\dot \alpha = \dot \pm$. These operators have $j_2 = 0$, and a quick glance at Table \ref{schurTable} tells us that they are Schur operators of types $\hat\BB_R$ and $\bar\DD_{R(j_1,0)}$. These operators saturate the inequality \eqref{L_0bound}, with $r = -j_1 < 0$ for $\bar \DD_{R (j_1,0)}$ and $r = 0$ for the $\hat\BB_R$. As these are precisely the operators that contribute to the Hall-Littlewood (HL) limit of the superconformal index, we refer to them as \emph{Hall-Littlewood operators}. They form a ring, the \emph{Hall-Littlewood chiral ring}, which is a consistent truncation of the full $\NN=1$ chiral ring.

In a Lagrangian theory, the $\hat \BB_R$ type Schur operators are gauge-invariant combinations of $Q$ and $\tilde Q$, the complex hypermultiplet scalars that are bottom components of $\NN=1$ chiral superfields (we are suppressing color and flavor indices). Schur operators of type $\bar {\DD}_{R (j_1, 0 )}$ are obtained by further allowing as possible letters the gauginos $\lambda^1_+$, which are the bottom components of the field strength chiral superfield ${W}_+$. In the full $\NN=1$ chiral ring, one also has the other Lorentz component $W_-$ of the field strength, as well as the ${\NN}=1$ chiral superfield belonging to the ${\NN}=2$ vector multiplet. Operators that contain those letters are, however, \emph{not} a part of the HL chiral ring.

In complete analogy, we may also define a Hall-Littlewood anti-chiral ring,  which contains the Schur operators of type $\hat \BB_R$ and $\DD_{R(0,j_2)}$. These operators are annihilated by chiral supercharges $\Qm_\alpha^1$, $\alpha = \pm$, and are thus  $\NN=1$ anti-chiral with respect to the $\NN=1$ subalgebra that is orthogonal to \eqref{N1sub}. Schur operators of type $\hat \BB_R$ belong to both HL rings -- these are half-BPS operators that are annihilated by both $\Qm_\alpha^1$ and $\widetilde \Qm_{2 \dot \alpha}$. They form a further truncation of the $\NN=1$ chiral ring to the \emph{Higgs chiral ring}, and their vacuum expectation values parametrize the Higgs branch of the theory. We note that in Lagrangian theories that are represented by acyclic quiver diagrams, all $\DD$-type multiplets recombine and are lifted from the $\NN=1$ chiral ring at one-loop order \cite{Gadde:2011uv}. In such cases, the HL chiral ring will coincide with the more restricted Higgs branch chiral ring.

Let us now look in greater detail at some Schur-type shortened multiplets of particular physical interest:
%%%%%%
\begin{itemize}
  \item $\hat \CC_{0 (0, 0) }$: Stress-tensor multiplet. The superconformal primary is a scalar operator of dimension two that is a singlet under the $SU(2)_R \times U(1)_r$. The $SU(2)_R$ and $U(1)_r$ conserved currents, the supercurrents, and the stress tensor all lie in this multiplet. The Schur operator is the  highest weight component of the $SU(2)_R$ current: $J_{ + \dot +}^{1 1}$ of the $SU(2)_R$.
  
  \item $\hat \CC_{0 (j_1, j_2) }$: Higher-spin currents multiplets.  These generalize the stress-tensor multiplet and contain conserved currents of spin higher than two. If any such multiplets are present, the SCFT must contain a decoupled free sector \cite{Maldacena:2011jn}. Requiring the absence of these higher spin multiplets will lead to interesting unitarity bounds for the central charge of interacting SCFTs in \S\ref{sec:4d_consequences}.
  
  \item $\hat \BB_{\frac12}$: This is the superconformal multiplet of free hypermultiplets.
  
  \item $\hat \BB_{1}$: Flavor-current multiplet. The superconformal primary is the ``moment map'' operator $M^{\II\JJ}$, which is a scalar operator of dimension two that is an $SU(2)_R$ triplet, is $U(1)_r$ neutral, and transforms in the adjoint representation of the flavor group $G_F$. The highest weight state of the moment map -- $M^{1 1}$ -- is the Schur operator. The claim to fame of $\hat{\BB}_{1}$ multiplets is that they harbor the conserved currents $J^F_{\alpha \dot \alpha}$ that generate the continuous ``flavor'' symmetry group $G_F$ of the SCFT, that is, the symmetry group that commutes with the superconformal group. Because $\hat \BB_{1}$ multiplets do not appear in any of the recombination rules for short multiplets listed in Appendix \ref{app:shortening}, it is absolutely protected: $J^F_{\alpha \dot \alpha}$ remains conserved on the entire  conformal manifold of the SCFT.\footnote{The only other supermultiplet that contains a global flavor symmetry current is $\hat {\CC}_{0(\frac12, \frac12)}$. However, that multiplet also contains higher-spin currents, thus showing that the only points on a conformal manifold at which the flavor symmetry enhances are the points where the SCFT develops a free decoupled subsector.}

  \item $\DD_{0 (0, 0 )} \oplus \bar \DD_{0 (0, 0 )}$: This is the superconformal multiplet of free $\NN=2$ vector multiplets.
  
  \item $\DD_{\frac12 (0, 0 )} \oplus \bar \DD_{\frac12 (0, 0 )}$: ``Extra'' supercurrent multiplets.  The top components of these multiplets are spin $3/2$ conserved currents of dimension $\Delta = 7/2$ ($J_{\alpha \dot \alpha \dot \beta}$ and $J_{\alpha \beta \dot \alpha}$). They generate \emph{additional} supersymmetry transformations beyond the $\NN=2$ superalgebra in question. In particular, in the $\NN=2$ description of an $\NN=4$ SCFT, one finds two copies of each of these multiplets transforming as a doublet of the ``flavor'' $SU(2)_F \subset SU(4)_R$  that commutes with  $SU(2)_R \times U(1)_r \subset  SU(4)_R$. The Schur operators have $\Delta= 5/2$, and have index structure $\OO^{11}_{\dot +}$ and  $\OO^{11}_{+}$. In ${\NN}=4$ supersymmetric Yang-Mills theory, these are the operators ${\rm Tr}\,  q^{1}_i \tilde \lambda^1_{\dot +}$ and ${\rm Tr} \, q^{1}_i \lambda^1_{+}$, where $i = 1, 2$ is the $SU(2)_F$ index.
\end{itemize}

\subsection{Notable elements of the chiral algebra}
\label{subsec:notable}

Armed with a working knowledge of the relevant four-dimensional operators, we now proceed to derive some universal entries in the $4d/2d$ dictionary. We first recall from \S\ref{subsec:local_op_cohomology} the process by which a meromorphic operator in two dimensions is obtained from an appropriate protected operator in four dimensions. Starting with a Schur operator in four dimensions, we obtain a two-dimensional chiral operator via the following series of specializations:
%%%%%%
$$
\begin{tikzcd}[row sep = small]
{}\OO^{1\cdots1}_{+\cdots+\dot+\cdots\dot+}(x)\dar{} & \text{Schur operator}\\
{}\OO(z,\zb)\cong u_{\II_1}(\zb)\cdots u_{\II_{2R}}(\bar z)\OO^{(\II_1\cdots\II_{2R})}(z,\zb)\dar{} & \text{Twisted-translated Schur operator}\\
{}[\OO(z,\zb)]_{\qq\,} \dar{}& \text{Chiral cohomology class}\\
{}\OO(z)& \text{Two-dimensional chiral operator}\\
\end{tikzcd}
$$
%%%%%%
In general we will refer to this associated chiral operator via the following notation:
%%%%%%
$$
\OO(z)=\goodchi[\OO^{1\cdots1}_{+\cdots+\dot+\cdots\dot+}]~,
$$
where sometimes we will be lax about the argument of the $\goodchi$ map and allow $\OO^{1\cdots1}_{+\cdots+\dot+\cdots\dot+}$ to be replaced by the more generic form of the operator $\OO^{\II_1\cdots\II_{2R}}_{\alpha_1\cdots\alpha_{2j_1}\dot\alpha_1\cdots\dot\alpha_{2j_2}}$. Our first task will be to understand the chiral operators that are related to certain characteristic Schur operators of a four-dimensional theory.
In doing so we will discover some interesting and generic features of this correspondence.

\subsubsection{Virasoro enhancement of the $\mf{sl}(2)$ symmetry}

The holomorphic $\mf{sl}(2)$ algebra generated by $\{ L_{-1}, L_0, L_1 \}$ is a manifest symmetry of the chiral algebra. Remarkably, this global conformal symmetry is enhanced to the full Virasoro algebra. The Virasoro algebra is generated by the modes $L_n$, $n \in \Zb$, of a holomorphic stress tensor of dimension two $T(z)$. Surveying Table \ref{schurTable}, we find a suitable candidate that is present in any theory $\TT$: the Schur operator belonging to stress tensor multiplet $\hat\CC_{0(0,0)}$. One should note that the Schur operator in this multiplet is \emph{not} the four-dimensional stress tensor, but rather the component $J_{+ \dot +}^{11}$ of the $SU(2)_R$ current $J_{\alpha\dot\alpha}^{\II\JJ}$.

The corresponding twisted-translated operator is defined as follows,
%%%%%%
\be\label{ttJ}
\JJ_R(z,\bar z)\colonequals u_{\II}(\bar z)~u_{\JJ}(\bar z)~J_{+\dot+}^{\II\JJ}(z, \bar z)~.
\ee
%%%%%%
Per the discussion of \S\ref{sec:section2}, we identify the cohomology class $[\JJ_R(z, \bar z)]_{\qq\,_i}$ with a dimension two meromorphic operator in the chiral algebra $\goodchi[\,\TT\,]$,
%%%%%%
\be\label{T_J}
T_{\JJ}(z)\colonequals \kappa~[{\JJ}_R(z,\bar z)]_{\qq\,_i}~.
\ee
%%%%%%
We provisionally include the subscript $\JJ$ as a reminder of the definition \eqref{T_J}; we still need to establish that the OPEs of $T_{\JJ}(z)$ with itself and with other operators in the chiral algebra take the standard forms appropriate to a two-dimensional stress tensor. With this in mind, we have also included a normalization factor $\kappa$, to be fixed momentarily in order to recover the canonical $TT$ OPE.

The two- and three-point functions of the $R$-symmetry current with itself are fixed by $\NN=2$ superconformal invariance in terms of a single parameter $c_{4d}$, which is one of the two conformal anomaly coefficients (the other being $a_{4d}$). Starting from the OPE of two $SU(2)_R$ currents \cite{Argyres:2007cn},
%%%%%%
\be 
J^{\II\JJ }_\mu (x) J^{\KK \LL}_\nu(0)\sim\frac{3c_{4d}}{4\pi^4} \epsilon^{\KK(\II} \epsilon^{ \JJ)\LL}\frac{x^2 g_{\mu \nu}-2 x_\mu x_\nu}{x^8}+\frac{2 i}{\pi^2} \frac{x_\mu x_\nu x\cdot J^{(\KK(\II}\epsilon^{\JJ)\LL)}}{x^6}+\cdots~,
\ee
%%%%%%
we find the following OPE of twisted-translated Schur operators,
%%%%%%
\bea\label{eq:TjTjOPE}
\JJ_R(z, \bar z) \JJ_R(0,0) & \sim & -\frac{3 c_{4d}}{2\pi^4 z^4} -
\frac{1}{\pi^2} \frac{ \JJ_R(0,0) }{ z^2 }   \nn\\&& 
-\frac{1}{\pi^2} \bar z \frac{ u_\II u_\JJ J^{\II \JJ}_{-\dot{-}}(0) }{ z^3 } 
+ \frac{i}{\pi^2} \bar z \frac{ J^{2 1}_{+\dot{+}}(0)}{z^2}+\frac{i}{\pi^2}\bar z^2 \frac{J^{2 1}_{-\dot{-}}(0)}{z^3}+\cdots~.
\eea
%%%%%%
Because the last three terms have non-zero $\widehat{L}_0$ eigenvalue, they are guaranteed to be $\qq\,_i$-exact. Upon setting $\kappa = -2\pi^2$, we find the following meromorphic OPE for $T_\JJ$,\footnote{The term corresponding to the simple pole does not immediately follow from the OPE given in \eqref{eq:TjTjOPE}. In particular, though the presence of $\partial T_\JJ(0)$ is guaranteed as a consequence of the double pole, we may worry that an additional quasiprimary (in the two-dimensional sense) may also appear. Such a quasiprimary $\OO$ would have to be a boson of holomorphic dimension $h=3$ and have nonzero three point function $\langle T_\JJ T_\JJ \OO\rangle$. This is forbidden by Bose symmetry.}
%%%%%%
\be\label{eq:TJ_OPE}
T_\JJ(z)~ T_\JJ(0) \sim  \frac{ - 6\,c_{4d}  }{z^4} + \frac{2 ~ T_\JJ(0)}{z^2} +  \frac{ \partial T_\JJ(0)}{z} ~ .
\ee
%%%%%%
Happily, we recognize in \eqref{eq:TJ_OPE} the familiar two-dimensional $TT$ OPE with central charge $c_{2d}$ given by
%%%%%%
\be \label{c2c4}
c_{2d} = -12\,c_{4d}~.
\ee
%%%%%%
This is the first major entry in our dictionary. Note that unitarity of the four-dimensional theory requires $c_{4d} > 0$, so the chiral algebra will have negative central charge and will therefore necessarily be non-unitary.

It is not immediately clear from the arguments presented thus far that $T_\JJ(z)$ will have the correct OPE with operators of the chiral algebra. In other words, the assertion that $T_\JJ$ acts as the stress tensor of the chiral algebra means that the ``geometric'' $\mf{sl}(2)$ generators $\{ L_{-1}, L_0, L_{+1} \}$ defined by the embedding \eqref{eq:conformal_embedding} of the two-dimensional conformal algebra into the four-dimensional one should coincide in cohomology with the generators $\{L^\JJ_{-1}, L^\JJ_0, L^\JJ_{+1}\}$ defined by the mode expansion of $T_\JJ(z)$. It would be sufficient to verify that this is the case for quasiprimary operators, by which we mean operators $\OO(z)$ that, when inserted at the origin, are annihilated by the holomorphic special conformal generator
%%%%%%
\be \label{quasiprimary}
[L_{+1}, \OO(0)]= 0~.
\ee
%%%%%%
In our construction, such an $\OO(z)$ arises as the cohomology class of a twisted-translated primary Schur operator. The assertion is then that in the chiral algebra (\ie, up to $\qq\,_i$-exact terms), the $T_\JJ$ OPEs take the form
%%%%%%
\be \label{TJclaim}
T_\JJ(z)\OO(0)\sim\cdots+ \frac{0}{z^3} + \frac{h~\OO(0)}{z^2} + \frac{\partial\OO(0)}{z}~,
\ee
%%%%%%
where $h$ is the holomorphic dimension of $\OO$ and the dots indicate possible poles of order four or higher. Though we have not been able to find a general proof, we believe \eqref{TJclaim} to be a universal consequence of superconformal Ward identities. It is thanks to the relation for the conformal dimension $h=R+j_1+j_2$ that the $SU(2)_R$ current can reproduce the appropriate scaling dimension, and the absence of additional operators should be excluded by selection rules for three-point functions of Schur-type superconformal multiplets. In practice, we have been able to give an abstract argument that this OPE holds only for the case where $\OO$ is a scalar operator. For non-scalar operators in the abstract setting, we leave the structure of these OPEs as a conjecture. Later in this section, the OPE \eqref{TJclaim} will be shown to hold in full generality in the theories of free hypermultiplets and free vector multiplets. The abstract claim would follow if the most general solution of the requisite Ward identity is expressible as a linear combination of structures corresponding to free field models, which is empirically the case in all analogous situations with which the authors are familiar.

\subsubsection{Affine enhancement of the flavor symmetry}

We next turn to the role played by the flavor symmetries of $\TT$ in the associated chiral algebra. When $\TT$ enjoys a flavor symmetry $G_F$, the corresponding conserved current $J_{\alpha \dot \alpha}$ is an element of a $\hat \BB_1$ supermultiplet, which additionally contains as its Schur primary the moment map operator $M^{11}$ described in the list at the end of \S\ref{subsubsec:schur_supermultiplets}. We expect the presence of $G_F$ symmetry to make itself known via the chiral operator associated to the moment map. Following the now-familiar procedure, we define a $\qq\,_i$-closed operator $M(z,\bar z)$ via twisted translations of the Schur moment-map operator from the origin, and identify the corresponding cohomology class as a meromorphic operator in the chiral algebra,
%%%%%%
\be
M(z,\bar z) \colonequals u_{\II}(\bar z)u_{\JJ}(\bar z)\,M^{\II\JJ}(z,\bar z)~,\qquad J(z)\colonequals\kappa[M(z,\bar z)]_{\qq\,_i}~.
\label{twistedB1}
\ee
%%%%%%
The normalization constant $\kappa$ will be determined momentarily. The meromorphic operator $J(z)$ has holomorphic dimension $h=1$. We have suppressed flavor indices up to this point, but these operators all transform in the adjoint representation of the flavor symmetry group, and so we actually find $\dim G_F$ dimension one currents $J^A(z)$ in the chiral algebra. It is natural to suspect that these operators will behave as affine currents for the flavor symmetry. Indeed, a little calculation bears out this expectation. First, recall that the central charge $k_{4d}$ of the flavor symmetry is defined in terms of the self-OPE of the conserved flavor symmetry current as follows,
%%%%%%
\be \label{k4d}
 J^A_\mu(x) ~ J^B_\nu ( 0)  \sim  \frac{3 k_{4d}}{4 \pi^4} \delta^{AB} \frac{x^2 g_{\mu \nu} - 2 x_\mu x_\nu}{x^8} + \frac{2}{\pi^2} \frac{x_\mu x_\nu f^{ABC} x \cdot J^C(0)}{x^6}  + \cdots~.
\ee
%%%%%%
Here $A,B,C=1,\ldots,\dim G_F$ are adjoint flavor indices, and we are using normalizations such that long roots of a Lie algebra have length $\sqrt{2}$ as in \cite{Argyres:2007cn}.
In the same conventions, the OPE of two moment maps reads
%%%%%%
\be
M^{A\, \II\JJ}(x)M^{B\, \KK\LL}(0)\sim
- \frac {3 k_{4d} }{ 48 \pi^4 } \frac{\epsilon^{\KK(\II}\epsilon^{\JJ)\LL} \delta^{AB}}{x^4}
-\frac{ \sqrt{2} }{ 4 \pi^2 } ~ \frac{f^{ABC}M^{C\,(\II(\KK}\epsilon^{\LL)\JJ)}}{x^2}+\cdots~ .
\ee
%%%%%%
The OPE for the corresponding twisted-translated operators follows directly,
%%%%%%
\be 
M^A(z,\bar z)M^B(0,0)\sim-\frac{3k_{4d}}{48\pi^4}\frac{\delta^{AB}}{z^2}
+\frac{\sqrt{2}}{4\pi^2}~i~\frac{f^{ABC}M^C(0,0)}{z}
+\frac{\sqrt{2}}{4\pi^2}~f^{ABC} M^{C\,21}(0)\frac{\bar z}{z}+\cdots~,
\ee
%%%%%%
where the last term is $\qq\,_i$-exact. Setting $\kappa = 2 \sqrt{2} \pi^2$, we recognize the canonical current algebra OPE,\footnote{In two dimensions it is standard to define a convention-independent affine level $k_{2d}$ as $k_{2d}\colonequals\frac{2\tilde{k}_{2d}}{\theta^2}$, where $\tilde{k}_{2d}$ is the level when the length of the long roots are normalized to be $\theta$. In our conventions $\theta^2=2$ and so $\tilde{k}_{2d}=k_{2d}$.}
%%%%%%
\be
 J^A(z) J^B(w) \sim k_{2d} \frac{\delta^{AB} }{ (z-w)^2 } + \sum_C i f^{ABC}\frac{J^C (w)}{z-w}~,
\ee
%%%%%%
where the two-dimensional affine level $k_{2d}$ is related to the four-dimensional flavor central charge $k_{4d}$ by
%%%%%%
\be \label{kdictionary}
k_{2d} = - \frac{k_{4d}}{2}~.
\ee
%%%%%%
This is the second important entry in the dictionary.

\subsubsection{The Hall-Littlewood chiral ring and chiral algebra generators}

An interesting problem that will be of particular concern in \S\ref{sec:lagrangian_examples} is that of giving a simple description of the chiral algebra $\goodchi[\,\TT\,]$ associated to a given $\TT$ in terms of a set of generating currents. Generators of a chiral algebra are by definition those $\mf{sl}(2)$ primary operators $\{\OO_j\}$ for which the normal ordered products of their descendants, \ie, operators of the form $\partial^{n_1} \OO_1\partial^{n_2}\OO_2 \dots \partial^{n_k}\OO_k$, span the whole algebra.\!\footnote{We are adopting the normal ordering conventions of \cite{Thielemans:1991uw}, in which a sequence of chiral operators represents left-nesting of conformally normal-ordered products:
%%%%%%
\be
\OO_1\OO_2\cdots\OO_{n-1}\OO_n\colonequals(\OO_1(\OO_2(\cdots(\OO_{n-1}\OO_n))))~.
\ee
%%%%%%
The algebra of operators so-defined is non-commutative and non-associative.} When the chiral algebra has only a finite number of generators, it is customary to refer to it as a $\WW$-algebra.

While we have given a clear set of rules that identifies the spectrum of the chiral algebra given the spectrum of the four-dimensional theory $\TT$, these rules have little to say about the question of what operators are \emph{generators} of $\goodchi[\,\TT\,]$. There turns out to be a subset of generators that is always relatively easy to identify. Recall from \S\ref{subsubsec:schur_supermultiplets} that the HL chiral and anti-chiral rings are consistent truncations of the $\NN=1$ chiral and anti-chiral rings of $\TT$, respectively. As such, they are commutative rings, and it is often possible to give them presentations in terms of generators and relations. What we show now is that the meromorphic operators associated to the generators of the HL chiral and antichiral rings are in fact generators of $\goodchi[\,\TT\,]$ in the chiral algebra sense.

Given the shortening conditions they obey, one finds that the chiral algebra operators associated to HL operators have holomorphic dimension $h = R + |r|$. In order to establish the claim made above, we will show that an HL operator can never arise as a normal ordered product of other operators that are not themselves of HL type. Let $\OO_1 (z, \bar z)$ and $\OO_2(z, \bar z)$ be two generic twisted-translated Schur operators, and let us assume that their OPE contains an HL operator $\OO_3^{\rm HL}$,
%%%%%%
\be
\OO_1 (z, \bar z) \, \OO_2(0,0) \sim  \frac{1}{z^{h_1 + h_2 - h_3} }\OO_3^{\rm HL} (0,0) + \dots
\ee
%%%%%%
By assumption, $h_3 = R_3 + | r_3|$, while \eqref{L_0bound} implies that $h_1 \geqslant R_1 + |r_1|$, $h_2 \geqslant R_2 + |r_2|$. The $U(1)_r$ charge is conserved, so $r_3 = r_1 + r_2$ and $|r_3| \leqslant |r_1 | + |r_2|$. Furthermore, $SU(2)_R$ selection rules imply the triangular inequality $R_3 \leqslant R_1 + R_2$. Combining these (in)equalities, we find that $h_3 \leqslant h_1+h_2$, which implies that an HL operator may only appear on the right hand side as a singular term (if $h_3  < h_1 + h_2$) or as the leading non-singular term (if $h_3 = h_1 + h_2$). The latter possibility requires that $\OO_1$ and $\OO_2$ saturate the respective bounds \eqref{L_0bound} for $h_1$ and $h_2$, which is to say that they themselves must be HL operators. This argument establishes that HL operators cannot be generated as normal ordered products of non-HL operators, and so the generators of the HL chiral and antichiral rings must necessarily be generators of the chiral algebra. 

\subsubsection{The Hall-Littlewood chiral ring and Virasoro primaries}

A further interesting feature of the HL chiral ring operators is that their corresponding meromorphic operators are always Virasoro primaries. For the generators of the HL chiral ring, this is already clear since the generators of any chiral algebra that includes a stress tensor are necessarily primaries of the Virasoro subalgebra. For other HL operators, though, this is a useful result that will help organize our thinking about some of the examples studied in \S\ref{sec:lagrangian_examples}.

The statement follows from a relatively straightforward analysis of the OPE of the meromorphic stress tensor with an arbitrary HL operator. In particular, let $\OO_1(z)$ be the meromorphic operator associated to an HL operator in four dimensions. The quantum numbers of $\OO_1$ obey the HL relation
\be
h_1=R_1+|r_1|~.
\ee
%%%%%%
Now the crucial observation from which our result follows is this: from a four-dimensional perspective, the meromorphic stress tensor is a $\zb$-dependent linear combination of operators with $r=0$ and $R=0,\pm1$. Consequently, in the OPE of the meromorphic stress tensor with $\OO_1(0)$, the only operators that may appear will have $R=R_1\pm1$ or $R=R_1$ and $r=r_1$. With what power of $z$ can such an operator appear in the OPE? A Schur operator $\OO_\gamma(0)$ with $R=R_1+\gamma$ and $\MM=|r_1|+2\text{min}(j_1,j_2)$ will appear in the OPE as
%%%%%%
\be
T(z)\OO_1(0)\supset \frac{\OO_\gamma(0)}{z^{2+R_1+|r_1|-R-\MM}}=\frac{\OO_\gamma(0)}{z^{2-\gamma-2\text{min}(j_1,j_2)}}~.
\ee
%%%%%%
This is at most a pole of order three (when $\gamma=-1$ and $j_1=0$ or $j_2=0$), but such a pole cannot appear because HL operators are always $\mf{sl}(2)$ primaries -- thus the most singular term possible is a pole of order two. This is precisely the property that characterizes Virasoro primary operators, and so we have our result.

\subsection{The chiral algebras of free theories}
\label{subsec:freetheories}

The simplest $\NN=2$ SCFTs are the theories of a free hypermultiplet and that of a free vector multiplet. For these special cases, we give a complete description of the associated chiral algebras. These chiral algebras are useful as the building blocks of interacting Lagrangian theories, some of which are discussed in \S\ref{sec:4d_consequences}. We describe in turn the cases of hypermultiplets and vector multiplets.

\subsubsection{Free hypermultiplets}

Let us consider the field theory of a single free hypermultiplet. The hypermultiplet itself lies in the short supermultiplet $\BB_{\frac{1}{2}}$, in which the primary Schur operators are the scalars $Q$ and $\tilde Q$. These are the highest weight states in a pair of $SU(2)_R$ doublets, 
%%%%%%
\be
Q^\II=\left(\begin{array}{c} Q \\ \tilde Q^* \end{array}\right)~,\qquad \tilde Q^\II=\left(\begin{array}{c} \tilde Q \\ -Q^* \end{array}\right)~.
\ee
%%%%%%
The single free hypermultiplet enjoys an $SU(2)_F$ flavor symmetry, under which $Q^\II$ and $\tilde Q^\II$ transform as a doublet. To work covariantly in terms of this $SU(2)_F$, we can introduce the following tensor,
%%%%%%
\be
Q^ {\II }_{\hat \II} \colonequals \left(\begin{array}{cc} Q & \tilde Q \\ \tilde Q^* &  -Q^*  \end{array}\right)~,
\ee
%%%%%%
where $\hat \II=1,2$ is the newly minted $SU(2)_F$ index.

The Schur operators in this free theory are all the ``words'' that can be constructed out of the ``letters'' $\{Q, \tilde Q, \partial_{+\dot+}\}$. As there are no singularities in the products of ($\del_{+\dot+}$ derivatives of) $Q$ and $\tilde Q$, the operator associated to any given word is well-defined and the Schur operators in this theory form a commutative ring. The set of all meromorphic operators in the free hypermultiplet chiral algebra are therefore precisely the $\qq\,_i$ cohomology classes of the twisted-translated versions of these words. This chiral algebra is itself a free chiral theory in two dimensions. Let us see how this works.

The twisted-translated operators and the associated meromorphic operators for the hypermultiplet scalars themselves are defined as follows,
%%%%%%
\be\label{eq:chiral_hyper_def}
Q_{\hat\II}(z,\zb)  \colonequals u_{\II}(\zb)\,Q^\II_{\hat\II}(z,\zb)~,\qquad 
q_{\hat\II}(z)      \colonequals [Q_{\hat\II}(z,\zb)]_{\qq\,_i}~.
\ee
%%%%%%
The relation to the operators defined in \S\ref{subsec:chiral_OPE} is $q_{\hat\II}(z) = (q(z),\tilde q(z))$.
%%%%%%
This is an $SU(2)_F$ doublet of dimension $1/2$ meromorphic fields, the OPE of which can be computed using the free-field OPE in four dimensions and the definition of the twisted translated operators in \eqref{eq:chiral_hyper_def},
%%%%%%
\be\label{eq:full_hyper_OPE}
q_{\hat\II}(z)\,q_{\hat\JJ}(w)\sim\,\frac{\varepsilon_{\hat\II\hat\JJ}}{z-w}~.
\ee
%%%%%%
It is reasonably easy to see that the entire spectrum of the chiral algebra of four-dimensional hypermultiplets is obtained by taking normal ordered products of the $q_{\hat\II}(z)$ and their descendants. In particular, one can show that the following diagram commutes,\!\footnote{We will see when we come to consider interacting theories in \S\ref{sec:lagrangian_examples} that product structures on Schur operators do not always translate so simply into those of the chiral algebra. 
}
%%%%%%
\be\label{eq:commutative_diagram}
\begin{tikzcd}
\{\OO_i~,\OO_j\} \arrow{r}{\times_{4d}} \arrow[swap]{d}{\qq\,_i} & \OO_i\OO_j \arrow{d}{\qq\,_i} \\
\{[\OO_i]~,[\OO_j]\}  \arrow{r}{\times_{::}} & :[\OO_i][\OO_j]:
\end{tikzcd}~,
\ee
%%%%%%
where the top row represents multiplication in the ring of Schur operators, the bottom row represents \emph{creation/annihilation} normal ordered products of chiral vertex operators, and the vertical arrows represent the identification of a Schur operator with its meromorphic counterpart in the chiral algebra. It follows that the meromorphic operator associated to any given word in ($\partial_{+\dot+}$derivatives of) $Q$ and $\tilde Q$ is simply the corresponding creation/annihilation normal ordered product of (holomorphic derivatives of) $q$ and $\tilde q$.

The chiral algebra of the free hypermultiplet is thus none other than the free symplectic boson algebra (\cf\ \cite{Goddard:1987td}). This simple example serves to illustrate some of the general points made in the previous subsections. The symplectic boson theory has a canonical stress tensor,
%%%%%%
\be \label{Tsympl}
T(z) = \frac12\ve^{ \hat \II  \hat \JJ}q_{ \hat \II} \partial q_{ \hat \JJ}(z)~,
\ee
%%%%%%
and it is easy to check that the modes $\{L_{+1},L_0,L_{-1}\}$ appearing in Laurent expansion of \eqref{Tsympl} reproduce the action of the holomorphic $\mf{sl}(2)$ symmetry inherited from four dimensions. Thus the holomorphic $\mf{sl}(2)$ is indeed enhanced to Virasoro symmetry. Moreover, we observe that given the form of the $SU(2)_R$ current in four dimensions
%%%%%%
\be
\JJ_{\mu}^{\II\JJ}(x) \sim \ve^{\IIh\JJh}Q_{\IIh}^{(\II} \partial_\mu Q_{\JJh}^{\JJ)}(x)~,
\ee
%%%%%%
The corresponding meromorphic operator $T_\JJ(z)$ will be equivalent to the canonical stress tensor,
%%%%%%
\be
T(z)=T_\JJ(z)~.
\ee
%%%%%%
From the $TT$ OPE we read off the central charge $c_{2d} = -1$. Recalling that the conformal anomaly coefficient of a free hypermultiplet is $c_{4d} = 1/12$, this result is in agreement the universal relation $c_{2d} = -1 2 c_{4d}$. The symplectic boson theory is like the theory of a complex free fermion (which of course has $c_{2d} = 1$), but with opposite statistics, hence the opposite value of the central charge.

Finally we mention a minor generalization of the above story for hypermultiplets. Gauge theories with $\NN=2$ supersymmetry are often described in terms of half-hypermultiplets instead of whole hypermultiplets. The generalization of the chiral algebra to the half-hypermultiplet conventions is straightforward. Let us consider half-hypermultiplets transforming in a pseudo-real representation $R$ of some symmetry group $G$ (at the moment we are working at zero coupling, so $G$ is just a global symmetry group). The corresponding chiral algebra will be generated by $\dim R$ meromorphic fields,
%%%%%%
\be
q_i~,\quad i=1,\ldots,\dim R~,
\ee
%%%%%%
and the singular OPE of these operators will be given by
%%%%%%
\be
q_i(z)q_j(w)\sim \frac{\Omega_{ij}}{z-w}~.
\ee
%%%%%%
Here $\Omega_{ij}$ is the anti-linear involution that maps the representation $R$ to its conjugate and squares to minus one. The description of the single full hypermultiplet in \eqref{eq:full_hyper_OPE} actually fits into this framework with $G=SU(2)_F$.

\subsubsection{Free vector multiplet}

The other key ingredient in Lagrangian SCFTs is the theory of free vector multiplets. Free vectors lie in the short supermultiplet $\bar\DD_{0(0,0)}$ and its conjugate $\DD_{0(0,0)}$, whose superconformal primaries are the complex scalar $\phi$ and its conjugate $\bar\phi$, respectively. The primary Schur operators in these multiplets are the fermions $\lambda^1_{+}$ and $\tilde \lambda^1_{\dot +}$, and as in the case of hypermultiplets, the entire set of Schur operators in this theory is comprised of the words built out of the letters $\lambda^1_+$, $\tilde\lambda^1_{\dot+}$, and $\partial_{+\dot+}$.

The twisted-translated operators associated to the vector multiplet fermions are defined as follows,
%%%%%%
\be
\lambda(z,\bar z)\colonequals u_{\II}(\bar z)\,\lambda_+^\II (z,\bar z)~, \quad \tilde\lambda(z,\bar z)\colonequals u_{\II}(\bar z)\tilde\lambda_{\dot +}^\II (z,\bar z)~,
\ee
%%%%%%
and the $\qq\,_i$-cohomology classes of these operators are Grassmann-odd, holomorphic fields of dimension $h=1$,
%%%%%%
\be
\lambda(z) \colonequals [\lambda(z,\bar z)]_{\qq\,_i}~,\quad\tilde\lambda(z) \colonequals [\tilde\lambda(z,\bar z)]_{\qq\,_i}~.
\ee
%%%%%%
Using the four-dimensional free field OPEs, it is easy to derive the OPEs of these holomorphic fields. They are again the OPEs of a free chiral algebra:
%%%%%%
\be
\tilde\lambda(z) \lambda(0) \sim \frac{1}{z^2}~, \quad \lambda(z)\tilde\lambda(0) \sim -\frac{1}{z^2}~.
\ee
%%%%%%
Indeed, the free-field form of these OPEs leads to an analogous commutative diagram to \eqref{eq:commutative_diagram}, which ensures that all the meromorphic operators in this theory are generated by $\lambda(z)$ and $\tilde\lambda(z)$ in the chiral algebra sense. We can recognize this chiral algebra as the $(b,c)$ ghost system of weight $(1, 0)$,\footnote{Recall that the derivative of a dimension zero conformal primary field -- $c(z)$ in this case -- is again a conformal primary.}
%%%%%%
\be \label{identifications}
\tilde\lambda\colonequals b(z)~,\quad\lambda(z)\colonequals \partial c(z)~.
\ee
%%%%%%
In making this identification, we have introduced an extra spurious mode -- the zero mode $c_0$ of $c(z)$ -- which is of absent in the algebra generated by $\lambda(z)$ and $\tilde \lambda(z)$. Thus, the more precise statement is that the chiral algebra associated to the vector multiplet is the so-called ``small algebra''  of the $(b,c)$ system, which is by definition the algebra generated by $b(z)$ and $\partial c(z)$ (\cf\ \cite{Friedan:1985ge,Kausch:1995py}). In other words, the Fock space of the small algebra is the subspace of the $(b,c)$ Fock space that does \emph{not} contain $c_0$, or equivalently, the subspace annihilated by $b_0$,
%%%%%%
\be
\FF_{\text{small}} \colonequals \{\psi\in\FF_{bc} \;|\; b_0\psi = 0\}~.
\ee
%%%%%%
The small algebra enjoys a global $SL(2,\Rb)$ symmetry under which $\lambda(z)$ and $\tilde\lambda(z)$ transform as a doublet. We can make this symmetry manifest by introducing the notation $\rho^\alpha$ with $\alpha = \pm$, where $\rho^+\colonequals\tilde \lambda$ and $\rho^-\colonequals\lambda$. Note that the Cartan generator of this symmetry acts as the $U(1)_r$ charge. In the language of the small algebra, the OPE can be put in a covariant form,
%%%%%%
\be
\rho^\alpha(z)\,\rho^\beta(0) \sim \frac{\varepsilon^{\alpha\beta}}{z^2}~.
\ee
%%%%%%

As in the hypermultiplet case, the action of the $\{L_{+1},L_0,L_{-1}\}$ modes of the canonical ghost stress tensor can easily be seen to match the action of the geometric $\mf{sl}(2)$ action inherited from the four-dimensional conformal algebra. Furthermore, given the $SU(2)_R$ current of the free vector theory,
%%%%%%
\be
\JJ_{\alpha \dot\alpha}^{\II\JJ}(x) \sim \lambda^{(\II}_\alpha \tilde\lambda^{\JJ)}_{\dot\alpha}(x)~,
\ee
%%%%%%
we see that the canonical stress tensor coincides precisely with the dimension two current $T_\JJ$ obtained from the $R$-symmetry current by the usual map,
%%%%%%
\be
T(z) =- \frac{1}{2} \ve_{\alpha \beta}\rho^\alpha \rho^\beta(z) = T_\JJ(z)~.
\ee
%%%%%%
The central charge of the $(b,c)$ ghost system/small algebra is $c_{2d} = -2$, which can be seen to agree with the relation \eqref{c2c4} upon recalling that $c_{4d} = \frac{1}{6}$ for a free vector multiplet.

\subsection{Gauging prescription}

The natural next step is to consider interacting SCFTs. Lagrangian $\NN=2$ SCFTs can be described using hypermultiplets and vector multiplets as elementary building blocks (see \cite{Bhardwaj:2013qia} for a recent classification of all possibilities). In particular, such an SCFT consists of vector multiplets transforming in the adjoint representation of a semisimple gauge group $G = G_1 \times G_2 \dots \times G_k$, along with a collection of (half)hypermultiplets transforming in some representation $R$ of the gauge group such that the one-loop beta functions for all gauge couplings vanish. Supersymmetry ensures that the theory remains conformal at the full quantum level. The building blocks of the corresponding chiral algebra are a collection of symplectic bosons $\{q\,,\tilde q\}$ in the representation $R$, and a collection of $(b\,,c)$ ghost small algebras in the adjoint representation of $G$. When the gauge couplings are strictly zero, the chiral algebra is simply obtained by imposing the Gauss law constraint, \ie, by restricting to the gauge-invariant operators of the free chiral algebra of symplectic bosons and ghosts. Our next step will be to determine what happens as we turn on the gauge couplings.
  
In fact, as Lagrangian theories are a small subset of all possible $\NN=2$ SCFTs, it is worthwhile to put the discussion in a more general context. Given a general superconformal field theory $\TT$ with $G_F$ flavor symmetry, a new SCFT is obtained by gauging a subgroup $G\subset G_F$ provided the gauge coupling beta function vanishes. We will denote the gauged theory with a nonzero gauge coupling $g$ as $\TT_G$.\footnote{More precisely, there is one independent gauge coupling for each simple factor of the gauge group. To avoid clutter we focus on the procedure for gauging one simple factor at the time, so $G$ will taken to be a simple group in what follows.} Though $\TT$ may be strongly coupled, the gauging procedure can be described in semi-Lagrangian language. By assumption, $\TT$ possesses a conserved flavor symmetry current $J^A_{\alpha \dot \alpha}$, where $A = 1, \dots \dim\,G$, which by $\NN=2$ supersymmetry is the top component of the moment map supermultiplet $\hat\BB_1$. The gauged theory $\TT_G$ is described by minimally coupling an $\NN=2$ vector multiplet to $\hat\BB_1$. Of particular importance is the addition to the action, in $\NN=1$ notation, of the superpotential coupling
%%%%%%
\be \label{univsuper}
g\int d^2\theta\,\Phi^A\,M^{11,A}+\,{h.c.}\,, 
\ee
%%%%%%
where $\Phi$ is the $\NN=1$ chiral superfield in the $\NN=2$ vector multiplet, and $M^{11}$ is the $\NN=1$ chiral superfield whose bottom component is the complex moment map $M^{11}$; both transform in the adjoint representation of $G$.

Let us assume that the chiral algebra $\goodchi[{\TT}]$ is known. It will suffice to work abstractly, in the sense that the only features of $\goodchi[\TT]$ that we will use follow directly from the existence of the global $G$ symmetry. In particular, there will be an affine current $J^A(z)$ at level $k_{2d}=-\frac12 k_{4d}$ (\cf\ \S\ref{subsec:notable}). As we mentioned above, at zero gauge coupling the chiral algebra of the gauged theory is obtained by imposing the Gauss law constraint on the tensor product algebra of $\goodchi[{\TT}]$ with the $G$-ghost small algebra $(\rho^+,\rho^-)$. In fact, it will be more useful to introduce the full $(b,c)$ system and restrict to the small algebra by imposing the auxiliary condition $b^A_0\psi = 0$ for any state $\psi$.

The affine current associated to the $G$ symmetry in the ghost sector is
%%%%%%
\be
J_{\rm gh}^A \colonequals -i\,f^{ABC} \,(c^Bb^C)~.
\ee
%%%%%%
The Gauss law, or gauge-invariance, constraint requires that all physical states should have vanishing total gauge charge, which is measured by the zero mode of the total gauge symmetry current,
%%%%%%
\be
J_{\rm tot}^A(z) \colonequals J^A(z) + J_{\rm gh}^A(z)~.
\ee
%%%%%%
Symbolically, we can therefore define the chiral algebra at zero gauge coupling as follows:
%%%%%%
\be
\goodchi[\TT_G^{(0)}] = \{ \psi \in \goodchi[\TT] \otimes (b^A , c^A)  \, | \, b_0^A \psi = J_{{\rm tot} \, 0}^A \psi = 0 \}\,.
\ee
%%%%%%
We are now ready to address the problem of identifying the chiral algebra for $\TT_G$ with $g\neq0$.

\subsubsection{BRST reduction of the chiral algebra}

On general grounds, we expect that the chiral algebra of the interacting gauge theory will contain fewer operators than the non-interacting version, because some of the short multiplets containing Schur operators that are present at zero coupling will recombine into long multiplets and acquire anomalous dimensions. Ideally, we would like to describe this phenomenon using only the general algebraic ingredients that we have introduced so far. A crucial hint comes from phrasing the condition of conformal invariance of the gauge theory more abstractly. The vanishing of the one-loop beta function amounts to the requirement that in the ungauged theory, the flavor symmetry central charge is given by
%%%%%%
\be\label{eq:4d_marginal_central_charge}
k_{4d} = 4 h^\vee~, 
\ee
%%%%%%
where $h^\vee$ is the dual Coxeter number of the gauge group. This means that in two-dimensional language, the corresponding symmetry in $\goodchi[\TT]$ must have its affine level given by
%%%%%%
\be
k_{2d } = - 2 h^\vee~.
\ee
%%%%%%
The affine level of the ghost-sector flavor currents $J_{\rm gh}$ is easily calculated to be $2 h^\vee$, so the requirement of conformal invariance translates into the condition that the level of the total affine current $J_{\rm tot}^A$ be zero. Precisely in this case, it is possible to construct a nilpotent BRST operator in the chiral algebra. Imitating a construction familiar from coset conformal field theory \cite{Karabali:1989dk}, we define
%%%%%%
\be\label{QBRST}
Q_{\rm BRST} \colonequals \oint \frac{dz}{2 \pi i} \, j_{\rm BRST} (z) \, , \quad  j_{\rm BRST} \colonequals c_A \left[ J^A + \frac{1}{2} J_{\rm gh}^A \right ] \, .
\ee
%%%%%%
Our contention is that the chiral algebra corresponding to the gauged theory at finite coupling is obtained by passing to the cohomology of $Q_{\rm BRST}$ \emph{relative to} the ghost zero modes $b_0^A$,\footnote{In other terms, the BRST cohomology is being defined entirely in the small algebra: two $Q_{\rm BRST}$-closed states belong to the same cohomology class if and only if they differ by an exact state $Q_{\rm BRST} \lambda$, where $\lambda$ is also in the small algebra.}
%%%%%%
\be\label{gaugingconjecture}
\goodchi[\TT_G] = \HH^*_{\rm BRST} [\psi\in\goodchi[\TT]\otimes(b^A,c^A)~\big|~ b_0^A \psi = 0 ] \,.
\ee
%%%%%%
Apart from its elegance, there are compelling physical arguments behind this claim. We will show that states of the chiral algebra that define nontrivial cohomology classes of $Q_{\rm BRST}$ correspond to the four-dimensional Schur states that survive in the interacting theory. By construction, all states of $\goodchi[\TT_G^{(0)}]$ are annihilated by the four supercharges in \eqref{schurQS}. As we turn on the gauge coupling, those supercharges receive quantum corrections, and only a subset of states remains supersymmetric. We will see that $Q_{\rm BRST}$ precisely implements the $O(g)$ correction to one of the Poincar\'e supercharges, which will justify our conjecture under the assumption that higher order corrections do not remove any additional states.

A preliminary remark is that the Gauss law constraint is imposed automatically. Because 
%%%%%%
\be \label{gaugingclaim}
\{b_0^A\,,Q_{\rm BRST}\} = J^A_{{\rm tot}\,0}~,
\ee
%%%%%%
states in the small algebra that are $Q_{\rm BRST}$-closed are automatically gauge invariant. Consequently, we have the simpler expression,
%%%%%%
\be
\goodchi[\TT_G]= \HH^*_{\rm BRST}[\goodchi[\TT_G^{(0)}]]~.
\ee
%%%%%%
We can rewrite $Q_{\rm BRST}$ and separate out the ghost zero modes,
%%%%%%
\be
Q_{\rm BRST} = c_0^A J^A_{{\rm tot} \,0} + b_0^A X^A + Q^-~,
\ee
%%%%%%
where we have defined
\be 
X^A \colonequals -\frac{i}{2} f^{ABC} \left( \sum_{n \neq 0} :c^B_{-n} c^C_n :  - c^B_0 c^C_0 
\right)\, ,
\ee
while $Q^-$ anticommutes with both $c^A_0$ and $b^A_0$ and can thus be expressed purely in terms of $(\rho^{+ A} , \rho^{-A} )$,
%%%%%%
\be \label{Q-}
Q^- \colonequals \sum_{n \neq 0} \frac{1}{n} : \rho^{- A}_{-n} J^A_n : + \frac{i}{2} f^{ABC} \sum_{\substack{n \neq 0 \\ m \neq 0 \\ m \neq n}} \frac{1}{n m} :\rho^{- A}_{-n} \rho^{- B}_m \rho^{+ C}_{n-m}:\, .
\ee
%%%%%%
The operator $Q^-$ fails to be nilpotent by a term proportional to $J^A_{tot \,0}$, so it \emph{is} nilpotent when acting on gauge-invariant states. It follows that \eqref{gaugingclaim} can be equivalently written as
%%%%%%
\be\label{eq:corrected_chiral_algebra}
\goodchi[\TT_G] 
 = \HH^*_{Q^-} [ \psi \in \goodchi[\TT] \otimes (\rho^{+ A} , \rho^{-A} ) \, , {\rm with}\; J^A_{{\rm tot} \,0} \psi = 0 ] \,.
\ee
%%%%%%
This is the form of our conjecture that makes more immediate contact with four-dimensional physics. We will show that the action of $Q^-$ precisely matches to the action of $\widetilde\QQ^{(1)}_{2\dot-}$, the $O(g)$ term in the expansion of the supercharge $\widetilde \QQ_{2 \dot -}$,
%%%%%%
\be
\widetilde \QQ_{2 \dot -} = \widetilde \QQ^{(0)}_{2 \dot -} + g \, \widetilde \QQ^{(1)}_{2 \dot -} + O(g^2) \,.
\ee
%%%%%%
In fact, $Q^-$ is the lowest component of an $SL(2,\Rb)$ doublet of operators $Q^\alpha$, with 
%%%%%%
\be\label{Q+}
Q^+ \colonequals \sum_{n \neq 0} \frac{1}{n} : \rho^{+ A}_{-n} J^A_n : + \frac{i}{2} f^{ABC} \sum_{\substack{n \neq 0 \\ m \neq 0 \\ m \neq n}} \frac{1}{m n} :\rho^{+ A}_{-n} \rho^{+ B}_m \rho^{- C}_{n-m}:\,.
\ee
%%%%%%
In complete analogy, the action of $Q^+$ will be shown to be isomorphic to that of $\QQ^{1 (1)}_{-}$, the $O(g)$ term in the expansion of $\QQ^{1}_{-}$. The two Poincar\'e supercharges $\QQ^{1}_{-}$ and $\widetilde \QQ_{2 \dot -}$ play a completely symmetric role in the definition of Schur operators. The fact that $Q_{\rm BRST}$ contains $Q^-$ rather than $Q^+$ is a consequence of our choice \eqref{identifications}, which treated $\lambda$ and $\tilde \lambda$ in a slightly asymmetric fashion.

Fortunately, to leading order in the gauge coupling the action of the relevant supercharges takes a universal form in the subspace of operators that obey the tree-level Schur condition. Such operators are obtained by forming gauge-invariant combinations of more elementary building blocks, namely the conformal primaries of the ``matter'' SCFT $\TT$, the gauge-covariant derivative $D_{+ \dot +}$, and the gauginos $\tilde\lambda^1_{\dot +}$ and $\lambda^1_{+}$. The supersymmetry variation of a gauge-invariant ``word'' is found by using the Leibniz rule to act on each elementary ``letter''.\footnote{For the special case of $\NN=2$ superconformal QCD, a very explicit description of the action of $\QQ^{1(1)}_{-}$ in the subsector of tree-level Schur operators can be found in Section 5 of \cite{Liendo:2011xb}. } It is then sufficient to specify the SUSY variations of the letters:
%%%%%%
\begin{enumerate}
	\item $\QQ^{1}_{-}$ and $\widetilde \QQ_{2 \dot -}$ (anti)commute with the conformal primary operators in the matter sector $\TT$.

	\item For the gauge-covariant derivative $D_{+ \dot +} \colonequals \partial_{+ \dot +} + g A_{+ \dot +}$,
	%%%%%%
	\be
		[\QQ^{1}_{-}, D_{+ \dot + }] = g \tilde \lambda^1_{\dot +} \, , \quad [\widetilde \QQ_{2 \dot-}, D_{+ \dot + }] = g \lambda^1_{+}~,
	\ee
	%%%%%%
	where we have just used the tree-level variation of the gauge field, times the explicit factor of $g$.

	\item Finally the variations of the gauginos can be deduced from the non-linear classical equations of motions of the vector multiplet, minimally coupled to the moment map supermultiplet $\hat \BB_1$,
	%%%%%%
	\begin{eqnarray}\label{eq:HL_one_loop_action}
		\{ \widetilde \QQ_{2 \dot -} , \tilde \lambda^{1}_{\dot +}  \} & = & \{\QQ^{1}_{-}  , \lambda^{1}_{+} \} = F^{11} = g M^{11} \\
		\{ \widetilde \QQ_{2 \dot -} , \lambda^{1}_{+} \} & = & \{\QQ^{1}_{-}  ,  \tilde \lambda^{1}_{\dot +} \} = 0 \, ,\nonumber
	\end{eqnarray}
	%%%%%%
	where $F^{11}$ is the highest-weight of the $SU(2)_R$ triplet of auxiliary fields in the $\NN=2$ vector multiplet.\footnote{In an $\NN=1$ description of the $\NN=2$ vector multiplet, $F^{11} = \bar F$, where $F$ is the top component of chiral superfield $\phi$, whose superpotential coupling with the moment map is given in \eqref{univsuper}.} 
\end{enumerate}
%%%%%%
If a Schur operator in the free theory is to retain its Schur status at $O(g)$, then when inserted at the origin it must be annihilated by the one-loop corrections to the four relevant supercharges,
$\{\widetilde \QQ^{(1)}_{2\dot-}, (\widetilde\QQ^{(1)}_{2\dot-})^\dagger, \QQ^{1(1)}_{-},(\QQ^{1(1)}_{-})^\dagger\}$.
Equivalently, it must define a nontrivial cohomology class with respect to $\widetilde \QQ^{(1)}_{2 \dot -}$ and $\QQ^{1 (1)}_{-}$. Conveniently, the recombination rules for shortened multiplets of Schur type (\cf\ Appendix \ref{app:shortening}) are such that in any such recombination, the Schur operators of $\TT^{(0)}$ are lifted in quartets that are related by the action of these two supercharges in the manner indicated in the following diagram:
%%%%%%
\be\label{eq:quartet}
\begin{tikzcd}[row sep = scriptsize]
{}					& \hat \CC_{R + \frac{1}{2} (j_1-\frac{1}{2}, j_2)}  \drar{ \widetilde \QQ^{(1)}_{2 \dot -} }	& {} \\
\hat \CC_{R (j_1, j_2)} \urar{ \QQ^{1 (1)} _-  } \drar[swap]{ \widetilde \QQ^{(1)}_{2 \dot -}} 	&  	& \hat \CC_{R+1 (j_1-\frac{1}{2}, j_2- \frac{1}{2})}  \\
{}					&  \hat \CC_{R+ \frac{1}{2} (j_1, j_2- \frac{1}{2})}  \urar[swap]{ \QQ^{1 (1)} _-}	& {}
\end{tikzcd}
\ee
%%%%%%
In the diagram, we are labeling Schur operators by the name of the supermultiplet to which they belong.\footnote{To include all possible recombinations, we must formally allow $j_1$ and $j_2$ to take the value $-\frac{1}{2}$ as well, and re-interpret a $\hat \CC$ multiplet with negative spins as a $\hat\BB$, $\DD$ or $\bar\DD$ multiplet, according to the rules: \\
%%%%%%
$\hat \CC_{R (j_1, -\frac{1}{2}) } \colonequals \bar\DD_{R + \frac{1}{2} (j_1, 0)}$, $\hat \CC_{R (-\frac{1}{2}, j_2) } \colonequals \DD_{R + \frac{1}{2} (0, j_2)}$, 
$\hat \CC_{R (-\frac{1}{2}, - \frac{1}{2}) } \colonequals \hat \BB_{R + 1}$.
%%%%%%
}
Consequently, if an operator remains in the cohomology of \emph{either} supercharge, it necessarily remains in the cohomology of both, and so stays a Schur operator at one-loop order. For example, if an operator becomes $\QQ^{1 (1)} _-$ exact then it is either at the right or at the top of the diagram and it follows that it is either $\widetilde \QQ^{(1)}_{2 \dot -}$ exact or not $\widetilde \QQ^{(1)}_{2 \dot -}$ closed, respectively. The other cases can be treated analogously.

Under the $4d/2d$ identifications
%%%%%%
\be
\widetilde \QQ^{(1)}_{2 \dot -} \to Q^- \, , \quad \QQ^{1 (1)} _- \to Q^+\, ,\quad D_{+ \dot +} \to \partial \, , \quad \lambda^1_+ \to \rho^- \, , \quad \tilde \lambda^1_{\dot +} \to \rho^+\, ,
\ee
%%%%%%
one easily checks that \eqref{Q-} and \eqref{Q+} have precisely the right form to reproduce the action of the $O(g)$ correction to the four-dimensional supercharges. Thus, the BRST cohomology specified in \eqref{gaugingconjecture} is just the right thing to project out states whose corresponding Schur operators are lifted at one-loop order.

It is  of some interest to note that this story of one-loop corrections to the spectrum of Schur operators admits a simple truncation to the case of HL chiral ring operators. The tree-level HL operators will be gauge-invariant combinations of the HL operators of $\TT$ and the gaugino $\lambda^1_+$. The operators that are lifted from the spectrum at one-loop will be those that are related by the corrected supercharge $\wt\QQ^{(1)}_{2\dot-}$, whose action in this sector is completely determined by \eqref{eq:HL_one_loop_action}. The problem of finding the HL operators in the spectrum of the interacting theory thus becomes a miniature ``HL-cohomology'' problem. In examples, it is sometimes useful to solve this problem as a first step in order to determine some important operators that will necessarily make an appearance in the chiral algebra.

Finally, a caveat is in order. We have assumed that the Schur operators that persist at infinitesimal coupling will remain protected at any finite value of the coupling. In some concrete cases, it can be demonstrated that no further recombination of shortened multiplets is possible. Moreover, in the examples of \S\ref{sec:lagrangian_examples} we will propose simple economical descriptions for the chiral algebras defined by this cohomological recipe, and demonstrate that they have the symmetries expected at finite coupling from S-duality, giving strong evidence for our proposal, at least in those examples. 

\subsubsection{Non-renormalization of three-point couplings}

So far, we have studied how the spectrum of operators is modified when the coupling is turned on, but we have said nothing about the OPE coefficients of the remaining physical operators in the gauged theory. Our implicit assumption has been that the OPE coefficients of operators that remain protected at finite coupling are actually independent of the coupling. From a two-dimensional perspective, it seems unlikely that the OPE coefficients could change due to the extremely rigid structure of chiral algebras, and we expect a corresponding non-renormalization statement to hold in four dimensions. Indeed, such a non-renormalization theorem directly follows from the methods and results of \cite{Baggio:2012rr}. Let us consider the four-point function of three Schur-type operators and of the exactly marginal operator $\OO_\tau$ responsible for changing the complexified gauge coupling,
%%%%%%
\be \label{mustvanish}
\langle \, \OO^{\II_1}_1 (x_1) \, \OO^{\II_2}_2 (x_2) \, \OO^{\II_3}_3 (x_3) \, \OO_\tau (x_4) \, \rangle\, ,
\ee
%%%%%%
where $\II=(\II^{(1)} \dots \II^{(k)})$ with $\II^{(i)}=1, 2$ are $SU(2)_R$ multi-indices and we have suppressed Lorentz indices. Non-renormalization of the appropriate three-point function of Schur-type operators will follow at once if we can argue that the above four-point function vanishes for any $x_4$ when $x_{1,2,3}$ all lie on the plane. By a conformal transformation, we can always take the fourth operator to lie on the same plane, and then focus on the $SU(1,1|2)$ subalgebra of $SU(2,2|2)$ defined by the embedding \eqref{eq:supercharge_embedding}. The Schur-type operators are chiral primaries of this subalgebra. The marginal operator $\OO_\tau $, being the top component of an $\bar \EE_2$ multiplet of $SU(2,2|2)$, is of the form $\OO_\tau =\{\QQ^1,[\QQ^2,\dots]\}$ where $\QQ^\II \colonequals \QQ^\II_-$ are supercharges of $SU(1,1|2)$.\footnote{Similarly, the conjugate operator $\bar \OO_\tau$  is the top component of an $\EE_2$ and can be written as $\{\widetilde \QQ_1,[\widetilde\QQ_2,\dots]\}$. An entirely analogous argument holds for the four-point function containing $\bar \OO_\tau$.} All the properties exploited in \cite{Baggio:2012rr} to show the vanishing of the four-point function \eqref{mustvanish} are satisfied. The authors of \cite{Baggio:2012rr} interpreted this result as a non-renormalization theorem for three-point functions of chiral primaries of two-dimensional $(0,4)$ theories, but exactly the same argument applies to our case as well. 

\bigskip

We close this section by pointing out a curious aspect of the gauging prescription given here. Given a chiral CFT $\goodchi[\TT]$ with affine $G$ symmetry, one can introduce a two-dimensional vector field $A_{\bar z}$ and gauge $G$. Following standard arguments (for example, see \cite{Gawedzki:1988nj,Karabali:1989dk}), a change of variables in the path integral eliminates the gauge field in favor of an extra $G$ current algebra at level $-(2h^\vee+k_{2d})$ and an adjoint-valued $(b,c)$ ghost system. One must also impose invariance under the standard BRST operator associated to the gauge symmetry. In our case, $2h^\vee+k_{2d}= 0$ so the extra current algebra is trivial, and the BRST operator associated to the two-dimensional gauging takes precisely the form of \eqref{QBRST}. In some sense, we have found that ``$4d$ gauging = $2d$ gauging''. We find it plausible that a localization-style argument may shed light on this correspondence.

\section{Consequences for four-dimensional physics}
\label{sec:4d_consequences}

The chiral symmetry algebras that we have uncovered have extensive consequences for the spectrum and structure constants of any $\NN=2$ SCFT. To give a simple example, Virasoro symmetry implies that any Higgs branch half-BPS supermultiplet $\hat {\cal B}_R$ is accompanied by an entire module of semi-short $\hat\CC_{R^\prime(j,j)}$ multiplets with $R^\prime=R-1,R,R+1$. In the four-dimensional theory, the descendant operators arise by taking repeated normal ordered products with certain components of the $SU(2)_R$ current, but the chiral algebra perspective makes this structure much more transparent.
   
In this section we elaborate on the relationship between the observables associated to the chiral algebra (\ie, its correlation functions and torus partition function) and those of the parent four-dimensional theory. We first point out that the superconformal Ward identities for four-point functions of $\hat {\cal B}_R$ operators \cite{Dolan:2004mu,Nirschl:2004pa} are a simple consequence of our cohomological construction. This new perspective makes it clear that analogous Ward identities must hold for four-point functions of general Schur operators. The presence of meromorphic functions in the solution of the Ward identities of \cite{Eden:2000bk, Dolan:2004mu,Nirschl:2004pa} was one of the initial clues that led to our work. We now have a neat conceptual interpretation for them: they are nothing but the correlation functions of the associated chiral algebra. By exploiting the relationship between the two-dimensional and four-dimensional perspectives we are able to derive new unitarity bounds that must be satisfied by the conformal and flavor anomalies of a general \emph{interacting} $\NN=2$ SCFT. Finally, we delineate the relationship between the torus partition function of the chiral algebra and the superconformal index of the parent four-dimensional theory.

\subsection{Conformal twisting and superconformal Ward identities}

By construction, for a given SCFT $\TT$, the correlation functions of $\goodchi[\TT]$ are equal to certain correlation functions of physical operators in $\TT$ restricted to lie on the plane. From the four-dimensional point of view these are somewhat unnatural correlators to study, as they have explicit space-time dependence built into the operators. On the other hand, each correlation function of $\goodchi[\TT]$ is canonically associated to a family of more natural correlation functions of $\TT$ that are obtained by replacing the twisted-translated operators with the corresponding untwisted operators at the same points in $\Rb^2$. 

Let us consider such a correlator now. For simplicity, we specialize to a four-point function, in which case there is actually no loss of generality in restricting the operators to be coplanar. We denote the untwisted operators as $\OO^{\II}(z,\zb)$, with $SU(2)_R$ multi-indices $\II=(\II^{(1)},\ldots, \II^{(k)})$ where $\II^{(i)}=1,2$. The components of the multi-index are symmetrized; the operator transforms in the spin $k/2$ representation of $SU(2)_R$. Recall that in our conventions, the Schur operator in this $SU(2)_R$ multiplet is the highest-weight state $\OO^{1\ldots1}(z,\zb)$. We represent the four-point function of such operators as
%%%%%%
\be
\label{eq:phys_four_point}
\FF^{\II_1 \II_2 \II_3 \II_4}(z_i,\zb_i)=\langle\;
\OO_1^{\II_1}(z_1,\zb_1)\,
\OO_2^{\II_2}(z_2,\zb_2)\,
\OO_3^{\II_3}(z_3,\zb_3)\,
\OO_4^{\II_4}(z_4,\zb_4)\;
\rangle~.
\ee
%%%%%%
This is actually a collection of four-point functions labelled by the different possible assignments for the $R$-symmetry indices. The full collection of four-point functions can be conveniently packaged by introducing two-component $SU(2)_R$ vectors $u(y_i)=(1,y_i)$ and defining contracted operators that depend on the auxiliary variable $y$ as follows \cite{Dolan:2004mu,Nirschl:2004pa}
%%%%%%
\be\label{eq:contracted_ops}
\OO_i(z_i,\zb_i;y_i)=u_{I_1}(y_i)\cdots u_{\II_{k_i}}(y_i)\OO^{(\II_1\cdots \II_{k_i})}_i(z_i,\zb_i)~.
\ee
%%%%%%
A single function of $x_i$ and $y_i$ can be defined that encodes the full content of the collection of correlation functions in \eqref{eq:phys_four_point},
%%%%%%
\be\label{eq:r-symmetry_compact}
\FF(z_i,\zb_i;y_i)=\langle\;
\OO_1(z_1,\zb_1;y_1)\,
\OO_2(z_2,\zb_2;y_2)\,
\OO_3(z_3,\zb_3;y_3)\,
\OO_4(z_4,\zb_4;y_4)\;
\rangle~.
\ee
%%%%%%
Charge conservation ensures that this function is homogeneous in the auxiliary $y_i$ with weight $\frac12\sum k_i$, and the correlation function for a given choice of external $R$-symmetry indices can be read off by selecting the coefficient of the appropriate monomial in the $y_i$ variables.

This repackaging makes it simple to state the relationship with correlation functions of $\goodchi[\TT]$. The twisted chiral operators defined in \S\ref{subsec:twisted_subalgebra} are the specialization of the repackaged operators in \eqref{eq:contracted_ops} to $y_i=\zb_i$. So if the related four-point function of meromorphic operators $\OO_i(z)=\goodchi[\OO_i(z,\zb)]$ is defined as
%%%%%%
\be
\label{eq:chiral_four_point}
f(z_1,z_2,z_3,z_4)=\langle\OO_1(z_1)\OO_2(z_2)\OO_3(z_3)\OO_4(z_4)\rangle~,
\ee
%%%%%%
then the correlation functions are related according to
%%%%%%
\be\label{eq:meromorphic_two_ways}
f(z_i)=\FF(z_i,\zb_i;y_i)\restr{}{y_i\rightarrow \zb_i}~.
\ee
%%%%%%
The fact that the left-hand side of this equation is a meromorphic function of the operator insertion points is a consequence of the cohomological arguments of the previous sections, but it is also \emph{precisely} the final form of the superconformal Ward identities for such a correlation function \cite{Eden:2000bk,Eden:2001ec,Dolan:2001tt,Heslop:2002hp,Dolan:2004mu,Nirschl:2004pa}.

This is a rather wonderful result: the entirety of the constraints imposed by superconformal Ward identities on the four-point function of half-BPS operators are captured by the existence of the twist of \S\ref{subsec:twisted_subalgebra}. It is worth noting that while the Ward identities of \cite{Dolan:2004mu} were derived specifically for half-BPS operators in $\hat\BB_R$ multiplets, here we see that the same type of Ward identities holds more generally for any Schur-type operators.

\subsection{Four-dimensional unitarity and central charge bounds}
\label{subsec:unitarity_bounds}

The natural inner product on the Hilbert space of the radially quantized four-dimensional theory $\TT$ does not survive the passage to $\qq\,$ cohomology. This is an immediate consequence of the fact that $\qq\,$ is not hermitian. Hence, unitarity in four dimensions does not imply unitarity in the chiral algebra. 
In fact, we have seen that a unitary theory $\TT$ always gives rise to a chiral algebra $\goodchi[\TT]$ with negative central charge, which is necessarily non-unitary. Nevertheless, there is an interesting interplay between the structure of the chiral algebra and four-dimensional unitarity. 
This leads to new unitarity bounds for the anomaly coefficients of any four-dimensional SCFT. 
In this section, we explore an elementary example that provides us with such bounds. It is possible that more extensive analysis could lead to further constraints; we leave such an analysis for future study.

The origin of nontrivial consistency conditions can be found in the fact that, as summarized in \eqref{eq:meromorphic_two_ways}, the meromorphic correlator $f(z_i)$ can be computed in two different ways that must agree. The first computation is the two-dimensional one: once the singular OPEs of the meromorphic operators appearing in the correlator are known, the full correlation function is completely fixed by meromorphy. The meromorphic correlator further admits a unique decomposition into $\sl{2}$ conformal blocks,\footnote{The result could also be expanded in Virasoro conformal blocks, but this is less natural for comparison to four-dimensional quantities.} leading to an expression of the form
%%%%%%
\be\label{eq:4pt_conf_block_2d}
f(z_i)=\left(\frac{z_{24}}{z_{14}}\right)^{h_{12}}\!\!\!\left(\frac{z_{14}}{z_{13}}\right)^{h_{34}}\!\!\!\!\frac{1}{z_{12}^{h_1+h_2}z_{34}^{h_3+h_4}}\sum_{\ell=0}^{\infty}(-1)^{\ell}\,a_{\ell}\,g_\ell(z)~,\quad
g_{\ell}(z)\colonequals(-\tfrac12z)^{\ell-1}z\;{}_2F_1(\ell,\ell;2\ell;z)~,
\ee
%%%%%%
where we have adopted the standard notation $z_{ij}\colonequals z_i-z_j$ and $z\colonequals\frac{z_{12}z_{34}}{z_{13}z_{24}}$. Additionally, $h_i$ is the holomorphic scaling dimension of the $i$'th operator, and we have defined $h_{ij}=h_i-h_j$.

The second computation is the four-dimensional one. The correlator in \eqref{eq:phys_four_point} admits a decomposition into $\mf{su}(2,2|2)$ superconformal blocks that each represent the contribution of a given superconformal multiplet to the four-point function. The contribution of each superconformal block to the meromorphic part of the amplitude defined by \eqref{eq:meromorphic_two_ways} is fixed up to the three-point coefficients. Thus for a given theory $\TT$, the spectrum and three-point coefficients of BPS operators appearing in the conformal block expansion of a given correlation function can be determined directly from the correlation functions of $\goodchi[\TT]$. Non-trivial constraints arise when we require that the three-point coefficients determined in this manner be consistent with unitarity.

Let us now turn to a specific example to study in detail. We consider the four-point function of superconformal primary operators in $\hat\BB_1$ multiplets. As was explained in \S\ref{sec:new2d4d}, these multiplets contain the spin one conserved currents that generate the global (non-$R$) symmetry of the theory, and the superconformal primaries are scalar moment map operators $M^A$. Consequently the results derived from this example will be relevant to any theory with non-trivial flavor symmetry. The moment map operators have dimension two and transform in the adjoint representations of both the flavor group $G_F$ and $SU(2)_R$. The four-point function of such operators can be expanded in channels corresponding to each irreducible representation ${\cal R}$ of $G_F$ in which the exchanged operators in the conformal block expansion may transform,
%%%%%%
\be
\langle M^{A}(z_1,\bar z_1;y_1)M^B(z_2, \bar z_2;y_2)M^C(z_3, \bar z_3;y_3)M^D(z_4, \bar z_4;y_4)\rangle=
\!\!\sum_{ {\cal R} \in\otimes^2\bf{adj}}\!\!P_{\cal R}^{ABCD}\FF_{\cal R}(z_i,\bar z_i;y_i)~,
\ee
%%%%%%
where $P_{\cal R}^{ABCD}$ is the projector onto the irreducible representation denoted by ${\cal R}$. The projectors for the various groups can be obtained following the procedures described in \cite{Cvitanovic:2008zz}.

Per the discussion of \S\ref{subsec:notable}, the chiral operators $J^A=\goodchi[M^A]$ are affine currents, and the mermorphic correlators that emerge in the limit $y_i\rightarrow\bar z_i$ are equal to the four-point functions in the corresponding chiral algebra,
%%%%%%
\be 
z_{12}^2z_{34}^2\langle J^A(z_1) J^B(z_2) J^C(z_3) J^D(z_4)\rangle=f^{ABCD}(z)=\sum_{\cal R}P_{\cal R}^{ABCD}f_{\cal R}(z)\,.
\ee
%%%%%%
Each such function can be examined independently as a potential source of nontrivial consistency conditions. In \S\ref{sec:new2d4d} we found that the level of the affine Lie algebra symmetry generated by these currents is $k_{2d}=-\frac12 k_{4d}$, so this meromorphic four-point function is completely fixed in terms of the structure constants of the associated non-affine Lie algebra and the flavor central charge,\footnote{Here we have rescaled the currents in such a way that the identity operator appears with unit normalization in the current-current OPE.}
%%%%%%
\be\label{ffunction}
f^{ABCD}(z)=\delta^{AB} \delta^{CD}+z^2\delta^{AC} \delta^{BD}+\frac{z^2}{(1-z)^2} \delta^{AD} \delta^{CB}-\frac{z}{k_{2d}}f^{ACE}f^{BDE}-\frac{z}{k_{2d}(z-1)}f^{ADE}f^{BCE}\,.
\ee
%%%%%%
This correlator can be decomposed into $G_F$ channels, each of which can be expanded in $\mf{sl}(2)$ conformal blocks as in \eqref{eq:4pt_conf_block_2d}. For example, for the singlet channel ${\cal R}={\bf 1}$, the above correlator gives
%%%%%%
\be\label{eq:singlet_channel_expansion}
\begin{split}
f_{{\cal R} ={\bf 1}}&=\dim{G_F} + z^2 \left(1+\frac{1}{(1-z)^2}\right) + \frac{4 z^2 h^{\vee}}{k_{2d}(z-1)} \\
&= \dim{G_F}-\sum_{\ell=0,2,\cdots} \frac{2^{\ell } (\ell +1) (\ell !)^2 \left(2 (\ell +1) (\ell +2) k_{2d} - 8\, h^{\vee} \right)}{k_{2d}(2 \ell +1)!} g_{\ell+2}(z)~,
\end{split}
\ee
%%%%%%
where $\dce$ is the dual Coxeter number.

This operator product expansion can be compared with that of the full four-point function in four dimensions. The superconformal block decomposition of such a four-point function has been worked out in \cite{Dolan:2001tt}. In particular, operators that can potentially appear in the intermediate channel must belong to one of the following superconformal multiplets:
%%%%%%
\begin{enumerate}
\item[$\bullet$]$\AA_{\Delta(j,j)}$: Long multiplets that are $SU(2)_R$ singlets with $j_1=j_2=j$.
\item[$\bullet$]$\hat\CC_{0(j,j)}$: Semishort multiplets with $j_1=j_2=j$ that contain conserved currents of spin $2j+2$.
\item[$\bullet$]$\hat\CC_{1(j,j)}$: Semishort multiplets with $j_1=j_2=j$.
\item[$\bullet$]$\hat\BB_{1}$: Half-BPS multiplets containing Higgs branch moment map operators.
\item[$\bullet$]$\hat\BB_{2}$: Half-BPS multiplets containing Higgs branch chiral ring operators of dimension four.
\item[$\bullet$]$\II$: The identity operator.
\end{enumerate}
%%%%%%
The contribution of each such multiplet to the full four-point function is fixed up to a single coefficient corresponding to the three-point coupling (squared), and unitarity requires that this coefficient be real and positive. The contribution of each multiplet to the meromorphic functions $f_{\cal R}(z)$ appearing in the superconformal Ward identities has also been determined in \cite{Dolan:2001tt}. The results are summarized as follows:
%%%%%%%%%%
\be
\label{eq:CB_short_contribution}
\begin{alignedat}{3}
&{\AA_{\Delta(\frac\ell2,\frac\ell2)}}&	&~~~:~~~&	&0~,\\
&{\hat\CC_{0(\frac\ell2,\frac\ell2)}}&	&~~~:~~~&	&\lambda_{\hat\CC_{0(\frac\ell2,\frac\ell2)}}^2 g_{\ell+2}(z)~,\\
&{\hat\CC_{1(\frac\ell2,\frac\ell2)}}&	&~~~:~~~&-2 &\lambda_{\hat\CC_{1(\frac\ell2,\frac\ell2)}}^2 g_{\ell+3}(z)~,\\
&{\hat\BB_1}&							&~~~:~~~&	&\lambda_{\hat\BB_1}^2 g_1(z)~,\\
&{\hat\BB_2}&							&~~~:~~~&-2 &\lambda_{\hat\BB_2}^2 g_{2}(z)~,\\
&\rm{Id}&								&~~~:~~~&	&\lambda^2_{\mathrm{Id}}~.
\end{alignedat}
\ee
%%%%%%%%%%
The coefficient $\lambda_{\bullet}^2$ of each contribution is required by unitarity to be non-negative. 

Some of the coefficients appearing in \eqref{eq:CB_short_contribution} can be completely fixed by symmetry. For example, the identity operator can only appear in the singlet channel $f_{{\cal R}={\bf 1}}(z)$, where the corresponding coefficient is necessarily given by
%%%%%%
\be
\lambda^2_{\mathrm{Id}}=\dim{G_F}~.
\ee
%%%%%%
The multiplet $\hat\CC_{0(0,0)}$ contains a spin two conserved current, \ie, the stress tensor. There can only be one such multiplet, and it contributes to the meromorphic part of the four point function only in the singlet channel. The three-point coupling is fixed in terms of the four-dimensional central charge. In particular, one finds that in $f_{{\cal R}={\bf 1}}(z)$,
%%%%%%
\be
\label{eq:stress_tensor_coeff}
\lambda^2_{\hat\CC_{0(0,0)}}=\frac{\mathrm{dim}\,G_F}{3c_{4d}}~.
\ee
%%%%%%
Finally, multiplets of type $\hat\BB_{1}$ can contributes only to the adjoint channel, and the corresponding three-point coupling in $f_{\rm adj}(z)$ is fixed to be
%%%%%%
\be 
\lambda_{\hat{\BB}_{1}}^2=\frac{4 h^{\vee} }{k_{4d}}\,.
\ee
%%%%%%
As far as we know, these are the only contributions to this four-point function that are fixed by symmetry in terms of anomaly coefficients. Additionally, the multiplets $\hat\CC_{0(\frac\ell2,\frac\ell2)}$ for $\ell\neq0$ necessarily contain conserved currents of spin greater than two, and so are expected to be absent in interacting theories \cite{Maldacena:2011jn}. We will take this to be the case in the following analysis.

\begin{table}[t] 
\centering
\begin{tabular}{lll||lll}
\hline\hline
$G_F$ 	~~~~~~~~~~&			 $\dce$			~~~~~~~~~~		 & $\mathrm{dim}\,G_F$  ~~~~~~~~~~&
$G_F$ 	~~~~~~~~~~&			 $\dce$			~~~~~~~~~~		 & $\mathrm{dim}\,G_F$			\\[0.5ex] 
\hline 
$ \SU(N)$ 					&$N$							&$N^2-1$
& $E_6$ 					&$12$   			         	&$78$								\\[.3ex]
$ \SO(N)$ 			 		&$N-2$ 							&$\frac{N(N-1)}{2}$
&$E_7$						&$18$				  			&$133$								\\[.3ex]
$\USp(2N)$			 		&$N+1$							&$N(2N+1)$
&$E_8$						&$30$  							&$248$								\\[.3ex]
$G_2$ 						&$4$  							&$14$
&$F_4$ 						&$9$  							&$52$								\\[1ex]
\hline
\end{tabular} 
\caption{Dual Coxeter number and dimensions for simple Lie groups.\label{tab:groupprops} }
\end{table}

We can determine the three-point coefficients in, say, the ${\cal R}={\bf 1}$ channel by comparing with the expansion of the $\goodchi[\TT]$ four-point function in \eqref{eq:singlet_channel_expansion}. In particular, we find
%%%%%%
\be
\label{eq:2d4d_OPE_coef_relation}
\begin{split}
\lambda^2_{\mathrm{Id}}&=\mathrm{dim}\,G_F~,\\
\lambda_{\hat\CC_{0(0,0)}}^2-2\lambda_{\hat\BB_2}^2&=\frac{8\dce}{k_{4d}}-4~,\\
\lambda_{\hat\CC_{1(\frac{\ell}{2},\frac{\ell}{2})}}^2&=\frac{2^{\ell+1}(\ell+2)((\ell+1)!)^2}{k_{4d}(2 \ell + 3 ) !}\left((\ell+2)(\ell+3)k_{4d}-4\dce\right)~,
\end{split}
\ee
%%%%%%
where in the last line only odd $\ell$ may appear. The second line of \eqref{eq:2d4d_OPE_coef_relation}, after substituting the contribution of the stress tensor multiplet from \eqref{eq:stress_tensor_coeff}, implies a nontrivial bound that must be satisfied in order for the contribution of the $\hat\BB_2$ multiplet to be consistent with unitarity,
%%%%%%
\be
\label{centralchargebound}
\frac{\dim{G_F} }{ c_{4d} }\geqslant \frac{24 h^{\vee} }{ k_{4d} }-12~.
\ee
%%%%%%
For reference, the dimensions and dual Coxeter numbers of the semi-simple Lie algebras are displayed in Table~\ref{tab:groupprops}. Similarly, the positivity of the last line in \eqref{eq:2d4d_OPE_coef_relation} for $\ell=1$ implies the bound
%%%%%%
\be
\label{singletkbound}
k_{4d} \geqslant \frac{\dce}{3}~.
\ee
%%%%%%

\begin{table}[t] 
\centering
\begin{tabular}{lllc}
\hline\hline
$G_F$ &~~~~~~~~~~~~~~~~~~~~& Bound~~~~~~~~~~~~~~~~ & Representation \\[0.5ex] 
\hline 
$\SU(N)$ & $N \geqslant 3$ 		& $k_{4d}\geqslant N$				& $\mathbf{N^2-1}_{\mathrm{symm}}$			\\[.3ex]
$\SO(N)$ & $N = 4,\ldots,8$	& $k_{4d}\geqslant 4$ 	  	 		& $\mathbf{\frac{1}{24} N(N-1)(N-2)(N-3)}$	\\[.3ex]
$\SO(N)$ & $N \geqslant 8$ 		& $k_{4d}\geqslant N-4$ 				& $\mathbf{\frac12 (N+2)(N-1)}$				\\[.3ex]
$\USp(2N)$& $N \geqslant 3$ 		& $k_{4d}\geqslant N+2$				& $\mathbf{\frac12 (2N+1)(2N-2)}$				\\[.3ex]
$G_2$ 	 &					& $k_{4d}\geqslant \frac{10}{3}$  	& $\mathbf{27}$								\\[.3ex]
$F_4$ 	 &					& $k_{4d}\geqslant 5$  				& $\mathbf{324}$							\\[1ex]
$E_6$ 	 &					& $k_{4d}\geqslant 6$           	 	& $\mathbf{650}$							\\[.3ex]
$E_7$ 	 &					& $k_{4d}\geqslant 8$  				& $\mathbf{1539}$							\\[.3ex]
$E_8$ 	 &					& $k_{4d}\geqslant 12$  				& $\mathbf{3875}$							\\[.3ex]
\hline
\end{tabular} 
\caption{Unitarity bounds for the anomaly coefficient $k_{4d}$ arising from positivity of the $\hat\BB_2$ three-point function in non-singlet channels.\label{tab:bounds} }
\end{table} 

The same analysis can be performed for the functions $f_{{\cal R} \neq{\bf 1} }(z_i)$. In these channels there will be no contribution from the stress tensor multiplet, so the resulting bounds make reference only to the anomaly coefficient $k_{4d}$, as in \eqref{singletkbound}. \emph{A priori}, an independent bound may be obtained for each representation ${\cal R}$ appearing in the tensor product of two copies of the adjoint. For example, in the adjoint channel itself, there can be contributions from $\hat\BB_1$ and $\hat\CC_{1(\frac\ell2,\frac\ell2)}$ multiplets with even $\ell$. Unitarity then imposes a bound on $k_{4d}$ that turns out to be equivalent to that of \eqref{singletkbound}. Stronger bounds can be found by considering other choices of ${\cal R}$, the possible values of which will depend on the particular choice of simple Lie algebra we consider. In general, we find that for a given choice of $G_F$, the strongest bound comes from requiring positivity of the contributions of $\hat\BB_2$ multiplets in a single channel. The bounds from other channels are then automatically satisfied when the strongest bound is imposed. These strongest bounds are displayed in Table \ref{tab:bounds}, where we also indicate the representation ${\cal R} \in\otimes^2{\bf adj}$ that leads to the bound in question. It should be noted that for the special case $G_F=SO(8)$, the same strongest bound is obtained from multiple channels. The representation appearing in the third line of Table \ref{tab:bounds} is in fact decomposable as ${\bf 70}={\bf 35}_s\oplus{\bf 35}_{c}$, and the degeneracy in the bounds can be understood as a consequence of $SO(8)$ triality. For $G_F=SU(2)$ one finds no additional bounds to the ones given in \eqref{centralchargebound} and in \eqref{singletkbound}. Finally, we can see that the bound \eqref{singletkbound} arising from positivity of the $\hat\CC_{1(\frac12,\frac12)}$ multiplet in the singlet channel is made obsolete by bounds arising from other channels for all choices of $G_F$ listed in the table.

\subsection{Saturation of unitarity bounds}
\label{subsec:saturation}

\begin{table}
\centering
\renewcommand{\arraystretch}{1.3}
\begin{tabular}{ |c||c|c|c|c|c|c|c|c| }
  \hline
  $G_F$ 	& $A_1$ 	& $A_2$ 	& $D_4$ 	& $E_6$ 		& $E_7$ 		& $E_8$ 		& \cellcolor[gray]{0.9} $F_4$ 		& \cellcolor[gray]{0.9} $G_2$ 			\\ \hline\hline  
  $\dce$ 	& $2$ 		& $3$ 		& $6$ 		& $12$			& $18$			& $30$			& \cellcolor[gray]{0.9} $9$ 		& \cellcolor[gray]{0.9} $4$				\\ \hline
  $k_{4d}$ 	& $\frac83$ & $3$ 		& $4$ 		& $6$ 			& $8$ 			& $12$ 			& \cellcolor[gray]{0.9} $5$ 		& \cellcolor[gray]{0.9} $\frac{10}{3}$ 	\\ \hline
  $c_{4d}$ 	& $\frac12$ & $\frac23$ & $\frac76$ & $\frac{13}{6}$& $\frac{19}{6}$& $\frac{31}{6}$& \cellcolor[gray]{0.9} $\frac53$ 	& \cellcolor[gray]{0.9} $\frac56$		\\ \hline
\end{tabular}
\caption{Central charges for $\NN=2$ SCFTs with Higgs branches given by one-instanton moduli spaces for $G_F$ instantons. Models corresponding to the right-most two columns are not known to exist, but must satisfy these conditions for their central charges if they do.\label{Tab:one_instanton}}
\end{table}

Given the existence of these unitarity bounds, it is incumbent upon us to consider the question of whether the bounds are saturated in any known superconformal models. To understand what sort of theory might saturate the bounds, it helps to identify any physical properties that a theory will necessarily possess if it saturates a bound. When the inequalities in \eqref{centralchargebound} or Table \ref{tab:bounds} are saturated, it means precisely that there is no $\hat\BB_2$ multiplet in the corresponding representation of $G_F$ contributing to the four-point function in question. The absence of such an operator is intimately connected with a well-known feature of theories with $\NN=2$ supersymmetry in four dimensions. Recalling that the Schur operators in the $\hat\BB_R$ multiplets are Higgs branch chiral ring operators, the absence of a $\hat\BB_2$ multiplet contributing to the four-point function of $\hat\BB_1$ multiplets in the ${\cal R}$ channel amounts to a relation in the Higgs branch chiral ring of the form
%%%%%%
\be
\restr{(M \otimes M)}{\cal R}=0~,
\ee
%%%%%%
where $M$ is the moment map operator and the tensor product is taken in the chiral ring.

There exists an interesting set of theories for which precisely such relations are known to hold. These are the superconformal field theories that arise on a single $D3$ brane probing a codimension one singularity in $F$-theory on which the dilaton is constant \cite{Sen:1996vd,Banks:1996nj,Dasgupta:1996ij,Minahan:1996fg,Minahan:1996cj,Aharony:1998xz}. There are seven such singularities, labelled $H_0,H_1,H_2,D_4,E_6,E_7,E_8$, for which the corresponding SCFT has global symmetry given by the corresponding group (with $H_i\rightarrow A_i$). The Higgs branch of each such theory is isomorphic to the minimal nilpotent orbit of the flavor group $G_F$. These minimal nilpotent orbits admit a simple description: they are generated by a complex, adjoint-valued moment map $M$, subject to a set of relations that defined the so-called ``Joseph ideal'' (see \cite{Gaiotto:2008nz} for a nice discussion),
%%%%%%
\be\label{eq:josephideal}
\restr{(M \otimes M)}{\II_2}=0~,\qquad \mathrm{Sym}^2({\bf adj})=(2\,{\bf adj})\oplus\II_2~,
\ee
%%%%%
where $(2\,{\bf adj})$ is the representation with Dynkin indices twice those of the adjoint representation.

This leads to an interesting set of conclusions. For one, these theories must saturate some of the $\hat\BB_2$-type bounds listed above. In particular, this allows us to predict the value of $c_{4d}$ and $k_{4d}$ for these theories as a direct consequence of the Higgs branch relations. These predictions are listed in Table \ref{Tab:one_instanton}. Indeed, these anomaly coefficients have been computed by other means and the results agree \cite{Aharony:2007dj}. On the other hand, an $\NN=2$ superconformal theory with $G_F$ symmetry can have as its Higgs branch the one-instanton moduli space of $G_F$ instantons \emph{only} if the $\hat\BB_2$ bound for all representations in $\II_2$ can be simultaneously saturated. It is not hard to verify that the list of cases for which this can be true includes the cases described above in F-theory, along with $G_F=F_4$ and $G_F=G_2$. Theories with Higgs branches isomorphic to the one-instanton $F_4$ and $G_2$ moduli spaces appear to be absent from the literature, and it is tempting to speculate that such theories should nonetheless exist and have as their central charges the values listed in the right-most two columns of Table \ref{Tab:one_instanton}.

Finally, it is interesting to rephrase the above discussion purely in the language of the chiral algebra $\goodchi[\TT]$. From this perspective, there is a marked difference between the bound \eqref{centralchargebound} for the singlet sector and those of Table \ref{tab:bounds} for non-singlets. In a theory saturating the non-singlet bounds, the coefficient of a conformal block is actually set to zero in the OPE of \ref{eq:4pt_conf_block_2d}. This should be considered in contrast to a theory that saturates the singlet bound, in which case all of the $\mf{sl}(2)$ conformal blocks are present with nonzero coefficients. It follows that saturation of a non-singlet bound is equivalent to the presence of a null state in the chiral algebra. In particular, because the bounds in question appear in the $\hat\BB_1$ four-point function, such null states can be understood entirely in terms of the affine Lie subalgebra of the chiral algebra. This interpretation can be verified directly by studying an affine Lie algebra with the level listed in Table \ref{tab:bounds}.

The bound \eqref{centralchargebound}, on the other hand, does \emph{not} imply the presence of a null state in the chiral algebra. Instead, a theory $\goodchi[\TT]$ that saturates the singlet bound should have the property that the only $\mf{sl}(2)$ primary of dimension two that appears in the OPE of two affine currents is identically equal to the chiral vertex operator that arises from the $\hat\CC_{0(0,0)}$ multiplet in four dimensions, \ie, it should be the two-dimensional stress tensor. We thus identify saturation of the singlet bound with the property that the Sugawara construction gives the true stress tensor of the chiral algebra,
%%%%%%
\be
T_{2d}=\frac{1}{k_{2d}+\dce}(J^aJ^a)~.
\ee
%%%%%%
Sure enough, if the bound \eqref{centralchargebound} is saturated, then we can rewrite the bound as an equation for the central charge
%%%%%%
\be
c_{2d}=\frac{k_{2d}\dim{G_F}}{k_{2d}+\dce}~.
\ee
%%%%%%
This is precisely the central charge associated with the Sugawara construction for the stress tensor of an affine Lie algebra.

Finally, we mention a number of additional theories that saturate some of the unitarity bounds derived here. In particular, though the rank one theory corresponding to the $H_0$ singularity has no flavor symmetry, it will have an extra $SU(2)$ symmetry for rank larger than one (as will all the other rank $\geqslant1$ theories). In particular, for the case of rank two the flavor central charge corresponding to this extra $SU(2)$ is $\frac{17}{5}$ and the central charge is $c_{4d}=\frac{17}{12}$ \cite{Aharony:2007dj}. This theory therefore saturates the bound \eqref{centralchargebound}.
Additionally, we have found a number of theories that saturate bounds appearing in Table~\ref{tab:bounds}. 
In particular, the new rank one SCFTs found in \cite{Argyres:2007tq} with flavor symmetry $USp(10)_7$ and $USp(6)_5\times SU(2)_8$, where $k_{4d}$ is indicated as a subscript for each group, saturate the bounds on $k_{4d}$ for the $USp$ factors. However for these theories the central charge bound is not saturated. The following theories described in \cite{Chacaltana:2010ks} also saturate bounds on $k_{4d}$: $S_5$ with flavor symmetry $SU(10)_{10}$ (but not the rest of the $S_N$ series), the $R_{0,N}$ series with flavor symmetry $SU(2)_6 \times SU(2N)_{2N}$, and the $R_{2,N}$ series with $SO(2N+4)_{2N} \times U(1)$ flavor symmetry.

\subsection{Torus partition function and the superconformal index}
\label{subsec:index}

Just as correlators of the chiral algebra are related to certain supersymmetric correlators of the parent four-dimensional theory, it will not come as a surprise that the torus partition function of the chiral algebra is related to a certain four-dimensional supersymmetric index -- indeed, to the Schur limit of the superconformal index, as foreshadowed in our terminology.
 
We should first identify which quantum numbers can be meaningfully assigned to chiral algebra operators. Of the various Cartan generators of the four-dimensional superconformal algebra, only the holomorphic dimension $L_0$ and the transverse spin $M^\perp =j_1-j_2$ (which is equal to $-r$ for Schur operators) survive as independent conserved charges of the chiral algebra. The torus partition function therefore takes the form\footnote{To avoid clutter, we have omitted the obvious refinement by flavor fugacities. If the theory is invariant under some global symmetry group $G_F$, we may refine the trace formula by $\prod_i a_i^{f_i}$, where the $f_i$ are Cartan generators of $G_F$ and $a_i$ the associated fugacities.}
%%%%%%
\be \label{partition}
Z(x, q)\colonequals\Tr\,x^{M^\perp}\,q^{L_0}~.
\ee
%%%%%%
As usual, the trace is over the Hilbert space in radial quantization, or equivalently over the local operators of the chiral algebra.
 
Specializing to $x=-1$, and noting that by the four-dimensional spin-statistics connection implies $(-1)^{j_1-j_2} = (-1)^F$, where $F$ is the fermion number, we find a weighted Witten index,
%%%%%%
\be \label{schurindex}
\II(q) \colonequals Z( -1, q ) = \Tr\,(-1)^F\,q^{L_0} = \Tr\,(-1)^Fq^{E-R}~.
\ee
%%%%%%
We recognize this as the trace formula that defines the Schur limit of the superconformal index \cite{Gadde:2011uv}, \cf\ Appendix \ref{app:shortening}.\footnote{It was observed in \cite{Razamat:2012uv} that the Schur index has interesting modular properties under the action of $SL(2,\Zb)$ on the superconformal and flavor fugacities. The identification of the Schur index with a two-dimensional index may serve to shed some light on these observations.} We should check that in the two-dimensional and four-dimensional interpretations of this formula the trace can be taken over the same space of states. Strictly speaking, in the four-dimensional interpretation the trace is over the entire Hilbert space of the radially quantized theory. However, the point of the Schur index is that only states obeying the Schur condition can conceivably contribute -- the contributions of all other states cancel pairwise. As the states of the chiral algebra are in one-to-one correspondence with Schur states, the chiral algebra index \eqref{schurindex} is indeed equivalent to the Schur index.
 
The index is a cruder observable than the partition function, but because it is invariant under exactly marginal deformations, it is generally easier to evaluate. In practice, to evaluate the index of a Lagrangian SCFT, one enumerates all gauge-invariant states that can be formed by combining the elementary ``letters'' that obey the Schur condition, see Table \ref{schurTable}. This combinatorial exercise is efficiently solved with the help of a matrix integral, where the integration over the gauge group enforces the projection onto gauge singlets. Examples of this prescription will be seen in the following section. By this procedure, one enumerates all gauge-invariant states that obey the tree-level Schur condition; there will be cancellations in the index corresponding to the recombinations of Schur multiplets into long multiplets that are \emph{a priori} allowed by representation theory.

There is an entirely isomorphic computation in the associated chiral algebra. The ``letters'' obeying the tree-level Schur condition are nothing but the states of the symplectic bosons and the ghost small algebra (in the appropriate representations), and one is again instructed to project onto gauge singlets. To reiterate, to evaluate the index we do not really need to compute the cohomology of $Q^-$, which defines the states of the chiral algebra of the interacting gauge theory, \cf\ \eqref{eq:corrected_chiral_algebra}. We can simply let the trace run over the redundant set of states of the free theory. By contrast, the trace in the partition function \eqref{partition} must be taken over only the states of the chiral algebra for the interacting theory, which are the cohomology classes of $Q^-$.

At the risk of being overly formal, we may point out that the physical state space of the chiral algebra (which for gauge theories is defined by the cohomological problem \eqref{eq:corrected_chiral_algebra}), acts as a \emph{categorification} of the Schur index. Once this vector space and the action of the charges are known, we can perform the more refined counting \eqref{partition}. In physical terms, the categorification contains extra information relative to the Schur index in that it knows about sets of short multiplets that are kinematically allowed to recombine but do not. In addition, there may be multiplets that cannot recombine but nonetheless make accidentally cancelling contributions to the index, and these are also seen in the categorification. Of course, the chiral algebra structure goes well beyond categorification -- it is a rich algebraic system that also encodes the OPE coefficients of the Schur operators, and is subject to non-trivial associativity constraints. 

It should be noted that as a graded vector space, we also have a categorification of the Macdonald limit of the superconformal index. Recall that the states contributing to the Macdonald index are really the same as the states that contribute to the Schur index, but their counting is refined by an extra fugacity $t/q$ associated to the charge $r + R$ (for $t=q$ we recover the Schur index). Since each state in the vector space defined by the chiral algebra corresponds to a Schur operator, the additional grading by $r+R$ is perfectly well-defined. However, there is no obvious chiral algebra interpretation of the Macdonald limit of the superconformal index, because the additional grading is incompatible with the chiral algebra structure. More precisely, while $L_0$ and $r$ are conserved charges for the twisted-translated operators \eqref{displaced}, $r + R$ is not, since away from the origin the operators are linear combinations of operators with different $R$ eigenvalues. In particular $r +R$ is \emph{not} preserved by the OPE.

\section{Examples and conjectures}
\label{sec:lagrangian_examples}

In this section we consider a number of illustrative examples in which the four-dimensional superconformal field theory $\TT$ admits a weakly coupled Lagrangian description. In such cases, the chiral algebra $\chi[\TT] $ can be defined via the BRST procedure of \S\ref{sec:new2d4d}, which at the very least allows for a level-by-level analysis of the physical states/operators in the algebra.

We can also consider the problem of giving an economical description of the chiral algebra in terms of a set of generators and their singular OPEs. A natural question is whether this set is finite, or in other words whether the chiral algebra is a $\WW$-algebra. The results of \S\ref{subsec:notable} suggest a very general ansatz for a possible $\WW$-algebra structure: the generators should be the operators associated to HL chiral ring generators in four dimensions, and possibly in addition the stress tensor. In each of the first three examples, our results are compatible with this guess, and we formulate concrete conjectures for the precise definition of each chiral algebra as a $\WW$-algebra. In the final example, we find a counterexample to this simplistic picture. Namely, we find a theory for which the chiral algebra contains at least one additional generator beyond those included in our basic ansatz.

For the first example, we turn to perhaps the most familiar $\NN=2$ superconformal gauge theory.

\subsection{\texorpdfstring{$SU(2)$}{SU(2)} superconformal QCD}
\label{subsec:so8}

The theory of interest is the $SU(2)$ gauge theory with four fundamental hypermultiplets. Many aspects of this theory that are relevant to the structure of the associated chiral algebra have been analyzed in, \eg, \cite{Argyres:1996eh}. The field content is an $SU(2)$ vector multiplet and four fundamental hypermultiplets. Because the fundamental representation of $SU(2)$ is pseudo-real, the obvious $U(4)$ global symmetry is enhanced to $SO(8)$, with the four fundamental hypermultiplets being reinterpreted as eight half-hypermultiplets. In $\NN=1$ notation we then have an adjoint-valued $\NN=1$ field strength superfield $W_{\a}^A$, an adjoint-valued chiral multiplet $\Phi^B$, and fundamental chiral multiplets $Q^i_{a}$ transforming in the ${\bf 8}_v$ of $SO(8)$.
Here $a,b=1,2$ are vector color indices that can be raised and lowered with epsilon tensors, $A,B=1, 2, 3$ are adjoint color indices, and $i=1, \ldots, 8$ are $SO(8)$ vector indices. By a common abuse of notation, we use the same symbol for the scalar squarks in the matter chiral multiplets as for the superfields, whereas the gauginos in the vector multiplet are denoted $\lambda^A_\a$ and $\tilde\lambda_{A\ad}$. In terms of the $\NN=1$ superfields listed above, the Lagrangian density takes the form
%%%%%%
\be\label{eq:so8_lagrangian_density}
\LL=\text{Im}
\left[
	\tau\int d^2\,\theta d^2\bar\theta\;\Tr\left(
		\Phi^\dagger e^V\Phi+Q_i^\dagger e^V Q^i
	\right)
	+\tau\int d^2\theta\;\left(
		\tfrac12 \Tr\,W_\alpha W^\alpha+\sqrt2 Q_a^i\Phi^a_b Q^{ib}
	\right)
\right]~,
\ee
%%%%%%
Where $\tau=\theta/2\pi+4\pi i/g_{\rm YM}^2$ is the complexified gauge coupling. The central charge of the $SU(2)$ color symmetry acting on the hypermultiplets is $k_{4d}^{SU(2)}=8$, which satisfies condition \eqref{eq:4d_marginal_central_charge} for $\tau$ to be an exactly marginal coupling. The central charge for the $SO(8)$ flavor symmetry and the conformal anomaly $c_{4d}$ can also be read off directly from the field content,
%%%%%%
\be\label{eq:so8_central_charges}
k_{4d}^{SO(8)}=4~,\quad c_{4d}=\frac76~.
\ee
%%%%%%

Although this description is sufficient to set up a BRST cohomology problem that defines the chiral algebra in the manner of \S\ref{sec:new2d4d}, it is useful to first review some of the features of this theory that we expect to see reflected in the two-dimensional analysis. We have seen that a special role is played in the chiral algebra by the HL chiral ring, the elements of which are the superconformal primary operators in $\hat\BB$ and $\DD$-type multiplets. In this example, these are the lowest components of $\NN=1$ chiral superfields that are gauge-invariant polynomials in $Q_a^i$ and $W_\alpha^A$. As this theory is represented by an acyclic quiver diagram, all $\DD$-type multiplets recombine and the HL chiral ring is identically the Higgs chiral ring.

In purely gauge invariant terms, the Higgs branch chiral ring is generated by a single dimension two operator in the adjoint of $SO(8)$,
%%%%%%
\be\label{eq:so8_chiral_ops}
M^{[ij]}=Q_a^iQ^{aj}~.
\ee
%%%%%%
This is the moment map for the action of $SO(8)$ on the Higgs branch.\footnote{It is a special feature of this theory (in contrast to, say, the $N_f=2N_c$ theories with $N_c>2$ that will be considered next) that the generators of the Higgs branch chiral ring all have dimension two. In general, there will be higher-dimensional baryonic generators that are not directly related to the global symmetry currents of the theory.} There are additional relations that make the structure of the Higgs branch more interesting. Already at tree-level, there are relations that follow automatically from the underlying description in terms of squarks. When organized in representations of $SO(8)$, the of generators of these relations are as follows,
%%%%%%
\be\label{eq:so8_higgs_relations_D}
M\otimes M\big|_{{\bf 35}_s}=0~,\qquad M\otimes M\big|_{{\bf 35}_c}=0~.
\ee
%%%%%%
On the other hand, there are $F$-term relations as a consequence of the superpotential in \eqref{eq:so8_lagrangian_density}. They are absent in the theory with strictly zero gauge coupling, and encode the fact that certain operators that are present in the chiral ring of the free theory recombine and are lifted from the protected part of the spectrum when the coupling is turned on. The generators of $F$-term relations, again organized according to $SO(8)$ representation, are as follows,
%%%%%%
\be\label{eq:so8_higgs_relations_F}
M\otimes M\big|_{{\bf 35}_v}=0~,\qquad M\otimes M\big|_{\bf 1}=0~.
\ee
%%%%%%
One immediately recognizes the complete set of relations in \eqref{eq:so8_higgs_relations_D} and \eqref{eq:so8_higgs_relations_F} as defining the $SO(8)$ Joseph ideal described in \S\ref{sec:4d_consequences}. Indeed, for the particular case of $G_F=SO(8)$ we have $\II_2={\bf 1}\oplus{\bf 35}_v\oplus{\bf 35}_s\oplus{\bf 35}_{c}$. The Higgs branch of this theory is known to be isomorphic to the $SO(8)$ one-instanton moduli space, and the central charges \eqref{eq:so8_central_charges} do in fact saturate the appropriate unitarity bounds outlined in \S\ref{sec:4d_consequences}.

As a final comment, let us recall that the gauge coupling appearing in the Lagrangian \eqref{eq:so8_lagrangian_density} is exactly marginal and parameterizes a one-complex-dimensional conformal manifold. $S$-duality acts by $SL(2,\Zb)$ transformations on $\tau$, and the conformal manifold is identified with the familiar fundamental domain of $SL(2,\Zb)$ in the upper half plane. In the various weak-coupling limits the theory can always be described using the same $SU(2)$ gauge theory, but in comparing one such limit to another, the duality transformations act by triality on the $SO(8)$ flavor symmetry. Consequently, though a given Lagrangian description of this theory (and of the chiral algebra in the next subsection) singles out a certain triality frame, the protected spectrum of the theory, and so in particular the chiral algebra, should be triality invariant.

\subsubsection{BRST construction of the associated chiral algebra}

The chiral algebra can now be constructed using the procedure of \S\ref{sec:new2d4d}. We first define the chiral algebra $\goodchi[\,\TT_{\text{free}}\,]$ of the free theory. Each half-hypermultiplet gives rise to a pair of commuting, dimension $1/2$ currents, whose OPE is that of symplectic bosons
%%%%%%
\be\label{eq:sym_boson_OPE}
q_{a}^i(z)\colonequals\goodchi[\,Q^i_a\,]~,
\qquad q_{a}^{i} (z) \, q_{b}^{j}(w) \sim \frac{\delta^{ij}\epsilon_{ab}}{z-w}~.
\ee
%%%%%%
Meanwhile, the vector multiplet contributes a set of adjoint-valued $(b,c)$ ghosts of dimension $(1,0)$ with the standard OPE,
%%%%%%
\be\label{eq:bc_ghost_OPE}
b^A(z)\colonequals\goodchi[\,\tilde \lambda^A]~,\qquad \del c^B(z)\colonequals\goodchi[\,\lambda^B]~,\qquad
b^A(z) c^B(w) \sim \frac{\delta^{AB}}{z-w}~.
\ee
%%%%%%%
The generators of the $SU(2)$ gauge symmetry in the matter sector arise from the moment maps in the free theory, while in the ghost system they take the canonical form described in \S\ref{sec:new2d4d},
%%%%%%
\be\label{eq:lag_currents}
J^A(T^A)_a^b = q_{a}^{i} q^{ib}~,\qquad
J_{\text{gh}}^A = - i f^{ABC}(c^{B}\,b^{C})~.
\ee
%%%%%%
The chiral algebra of the free theory is then given by the gauge-invariant part of the tensor product of the symplectic boson and small algebra Fock spaces,
%%%%%%
\be
\goodchi[\TT_{\rm free}]=\{\psi\in\FF(q_a^i,\rho^A_+,\rho^A_-)\,|\,J^A_{{\rm tot},0}\psi=0\}~.
\ee
%%%%%%

The current algebra generated by the $J_{\text{mat}}^A$ has level $k_{2d}^{SU(2)}=-4=-2\dce$, which ensures the existence of a nilpotent BRST differential. The BRST current and differential are then constructed in terms of these currents,
%%%%%%
\be
J_{\text{BRST}}=c^A\left(J^A+\frac12 J_{\text{gh}}^A\right)~,\qquad Q_{\text{BRST}}=\oint \frac{dz}{2\pi i}\,J_{\text{BRST}}(z)~.
\ee
%%%%%%
The chiral algebra of the interacting theory is now the $BRST$ cohomology
%%%%%%
\be
\goodchi[\,\TT\,]=\HH^*_{\text{BRST}}\left[\goodchi[\TT_{\rm free}]\right]~.
\ee
%%%%%%
We now perform a basic analysis of this cohomology. Already at this rudimentary level, we will find that a substantial amount of four-dimensional physics is packaged elegantly into the chiral algebra framework.

\subsubsection{Enumerating physical states}

It is a straightforward exercise to enumerate the physical operators up to any given dimension and to compute the singular terms in their OPEs. This is made easier with computer assistance -- we have made extensive use of  K.~Thielemans' \texttt{Mathematica} package \cite{Thielemans:1991uw}. We now describe this enumeration in detail for operators of dimension one and two in the chiral algebra. In this example, the material we have reviewed above is already enough to predict the results of this enumeration. We will nevertheless find it instructive to explore in some detail how the inevitable spectrum comes about.

We begin at dimension one. Dimension one currents in the chiral algebra can only originate in $\DD_{0(0,0)}$ and $\hat\BB_1$ multiplets (\cf\ Table \ref{schurTable}). The former contain free vector multiplets, and so are not gauge invariant. Thus the physical spectrum at dimension one should be isomorphic to the spectrum of $\hat\BB_1$ multiplets. Sure enough, the complete list dimension-one operators in $\goodchi[\,\TT_{\rm free}\,]$ is the following,
%%%%%%
\be\label{eq:so8_akm_currents}
J^{[ij]}=q_a^i  q^{ja}~,
\ee
%%%%%%
and these operators are the chiral counterparts of the $SO(8)$ moment maps, \ie,
%%%%%%
\be
J^{[ij]}=\goodchi[M^{[ij]}]~.
\ee
%%%%%%
Direct computation further verifies that these operators exhaust the nontrivial BRST cohomology at dimension one. It is also straightforward to determine the singular terms in the OPEs of these currents,
%%%%%%
\be\label{eq:so8_akm_opes}
J^{[ij]}(z)J^{[kl]}(0)\sim\frac{-2(\delta^{ik}\delta^{jl}-\delta^{il}\delta^{jk})}{z^2}+\frac{i f^{[ij][kl]}_{[mn]}J^{[mn]}(0)}{z}~.
\ee
%%%%%%
This is just an $\mf{so}(8)$ affine Lie algebra at level $k_{2d}=-2$, which confirms the general prediction of \S\ref{sec:new2d4d} that flavor symmetries are affinized in the chiral algebra, subject to the relation $k_{2d}=-\frac12 k_{4d}$.

Moving on, the four-dimensional multiplets that can give rise to two-dimensional quasi-primary currents of dimension two are $\hat\CC_{0(0,0)}$, $\hat\BB_2$, $\DD_{0(0,1)}$, and $\DD_{\frac12(0,\frac12)}$ multiplets (along with the conjugates of the last two). In addition, conformal descendants of dimension two can arise from holomorphic derivatives of the dimension one operators. Since no $\DD$-type multiplets appear in this theory, the only quasi-primaries at dimension two will correspond to Higgs branch operators and the two-dimensional stress tensor.

The latter descends from the four-dimensional $SU(2)_R$ current. That current being bilinear in the free fields of the noninteracting theory, the corresponding two-dimensional operator can be obtained by simply replacing the four-dimensional fields with their chiral counterparts and conformally normal ordering,
%%%%%%
\be\label{eq:so8_T2d}
T_{2d} = \tfrac{1}{2} q_{a}^{i} \del q^{ia}  - b^A \del c_A ~.
\ee
%%%%%%
Alternatively, this is just the canonical stress tensor for the combined system of free symplectic bosons and ghosts. Given the multiplicities of matter and ghost fields, the two-dimensional central charge is easily determined to be $c_{2d}=-14$.

The remaining BRST-invariant currents of dimension two can be constructed as normal ordered products and derivatives of the $\mf{so}(8)$ affine currents,
%%%%%%
\be\label{eq:so8_dim2_ops}
\partial J^{[ij]}~,\quad(J\otimes J)\big|_{\bf 1,35,35,35,300}~.
\ee
%%%%%%
The singlet term in the tensor product above, once appropriately normalized, is the Sugawara stress tensor of the $\mf{so}(8)$ affine Lie algebra,
%%%%%%
\be
T_{\text{sug}}^{\mf{so}(8)} = \tfrac{1}{8}(J^{[ij]}J^{[ij]})~.
\ee
%%%%%%
The Sugawara central charge is determined by the usual formula,
%%%%%%
\be
c_{\text{sug}}=\frac{k_{2d}\dim G_F}{k_{2d}+\dce}=-14~.
\ee
%%%%%%
This matches the value for the canonical stress-tensor. This comes as no surprise, since the central charges of this theory saturate the unitarity bound \eqref{centralchargebound}, which implies that the canonical stress tensor should be equivalent to the Sugawara stress tensor. Indeed, \eqref{eq:so8_T2d} and \eqref{eq:so8_dim2_ops} constitute an overcomplete list, and we in fact have the following relations,
%%%%%%
\begin{subequations}\label{eq:so8_rel_list}
\begin{eqnarray}
\makebox[.6in][l]{$J\otimes J\big|_{\bf 1}		$}	&=&~T_{2d}+\{Q_{\text{BRST}},q_a^iq^{ib}b^a_b\}~,\label{rel1}\\
\makebox[.6in][l]{$J\otimes J\big|_{{\bf 35}_v}	$}	&=&~\{Q_{\text{BRST}},q_a^{(i}q^{j)b}b^a_b\}~,\label{rel2}\\
\makebox[.6in][l]{$J\otimes J\big|_{{\bf 35}_c}	$}	&=&~0~,\label{rel3}\\
\makebox[.6in][l]{$J\otimes J\big|_{{\bf 35}_s}	$}	&=&~0~,\label{rel4}
\end{eqnarray}
\end{subequations}
%%%%%%
The relations appearing here can be traced back to different aspects of the four-dimensional physics. Relations \eqref{rel1} and \eqref{rel2} are the two-dimensional avatars of the $F$-term relations in \eqref{eq:so8_higgs_relations_F}. Note that the first relation appears differently in this two-dimensional context due to the presence of the two-dimensional stress tensor on the right hand side. This is a remnant of the more complicated structure of normal ordering in the chiral algebra as compared to the chiral ring. Relations \eqref{rel3} and \eqref{rel4} are the tree-level relations. In the context of the chiral algebra, they can be seen as a simple consequence of Bose symmetry and normal ordering without making any reference to the BRST differential. This perfectly mirrors of the nature of tree-level relations in four dimensions. 

\subsubsection{A \texorpdfstring{$\WW$}{W}-algebra conjecture}

Although the cohomological description of the chiral algebra is sufficient to compute the physical operators to any given level, it would be ideal to have a characterization entirely in terms of physical operators -- for example, we may hope for a description as a $\WW$ algebra. We have seen that the physical dimension two currents are all generated by the affine currents of dimension one, \ie, the physical states enumerated so far all lie in the vacuum module of the $\mf{so}(8)$ affine Lie algebra at level $k=-2$. What's more, these operators exhaust the list of operators that are guaranteed to be generators of the chiral algebra according to \S\ref{sec:new2d4d}. We are thus led to a natural conjecture:

\newtheorem{conj}{Conjecture}

\begin{conj}\label{conj_so8}
When $\TT$ is $\NN=2$ $SU(2)$ SQCD with four fundamental flavors, then $\goodchi[\,\TT\,]$ is isomorphic to the $\mf{so}(8)$ affine Lie algebra at level $k_{2d}=-2$.
\end{conj}

This is a mathematically well-posed conjecture, since the cohomological characterization of the chiral algebra is entirely concrete. It seems plausible that a more sophisticated approach to the cohomological problem could lead to a proof of the conjecture. We will be satisfied in the present work to test it indirectly.

\subsubsection{The superconformal index and affine characters}

Conjecture \ref{conj_so8} can be tested at the level of the indices of these theories. In particular, we have the following conjectural relationship
%%%%%%
\be
\II_{\mathrm{Schur}}(q;\vec{a})=\Tr_{\goodchi[\TT_{\rm free}]}(-1)^Fq^{L_0}\prod_{i=1}^4a_i^{\mu_i}=\Tr_{\mf{so}(8)_{-2}}(-1)^Fq^{L_0}\prod_{i=1}^4a_i^{\mu_i}~.
\ee
%%%%%%
The shorthand $\vec{a}=(a_1,a_2,a_3,a_4)$ denotes the $SO(8)$ fugacities. Of course, the affine Lie algebra has only bosonic states, so the factor of $(-1)^F$ is immaterial. In particular this observation implies that if Conjecture \ref{conj_so8} is correct, then all possible recombinations of tree-level Schur operators occur already at one loop.

On the one hand, the Schur limit of the superconformal index for this theory can be computed directly to fairly high orders in the $q$ expansion by starting with the defining matrix integral,
%%%%%%
\be
\II_{\mathrm{Schur}}(q;\vec{a}) = \oint[db] \text{P.E.}\left[ \left(\frac{\sqrt{q}}{1-q} \right)\chi_{SO(8)}^{\mathbf{8}}(\vec{a})\chi_{SU(2)}^{\mathbf{2}}(b) + \left(\frac{-2q}{1-q} \right) \chi_{SU(2)}^{\mathbf{3}}(b)\right]~,
\ee
%%%%%%
and expanding the exponential. Here $\oint[db]$ denotes integration over the fugacity for the gauge group with the Haar measure.  

\begin{table}
\centering
\begin{tabular}{cl}
\hline
\hline
level & $SO(8)$ representations and their multiplicities\\
\hline
$0$ & ${\bf 1}$ \\
$1$ & ${\bf 28}$\\
$2$ & ${\bf 1},~{\bf 28},~{\bf 300}$ \\
$3$ & ${\bf 1},~2\times{\bf 28},~{\bf 300},~{\bf 350},~{\bf 1925}$ \\
$4$ & $2\times{\bf 1},~3\times{\bf 28},~{\bf 35}_{v},~{\bf 35}_s,~{\bf 35}_c,~3\times{\bf 300},~{\bf 350},~{\bf 1925},~{\bf 4096},~{\bf 8918}$\\
$5$ & $2\times{\bf 1},~6\times{\bf 28},~{\bf 35}_{v},~{\bf 35}_s,~{\bf 35}_c,~4\times{\bf 300},~3\times{\bf 350},~{\bf 567}_v,~{\bf 567}_s,~{\bf 567}_c,~3\times{\bf 1925}$,\\
    & $2\times{\bf 4096},~{\bf 8918},~{\bf 25725},~{\bf 32928^\prime}$\\
\hline
\end{tabular}
\caption{\label{tab:so8_index}The operator content of the chiral algebra up to level 5.}
\end{table}

On the other hand, the vacuum character of the $\mf{so}(8)$ affine Lie algebra at level $k=-2$ can be computed once the spectrum of null primaries is known. Said spectrum can be determined with the aid of the Kazhdan-Lusztig polynomials, as we review in Appendix \ref{app:KL}. Ultimately, both the character and the index are expanded in the form
$$
1+\sum_{i=1}^\infty q^n \left(\sum_{R}d_{\RR}\chi^{\RR}(\vec{a})\right)~,
$$
where the $d_{\RR}$ are positive integer multiplicities. At a given power of $q$, there are only a finite number of non-zero $d_{\RR}$. Up to $O(q^5)$, the resulting degeneracies have been computed in both manners and agree. They are displayed in Table \ref{tab:so8_index}.

\subsection{\texorpdfstring{$SU(N)$}{SU(N)} superconformal QCD with \texorpdfstring{$N\geqslant3$}{N>=3}}
\label{sec:Nfeq2Nc}

We next consider the generalization of the previous example to the case of $SU(N)$ superconformal QCD with $N\geqslant3$. In these theories, the Higgs branch has generators of dimension greater than two, thus guaranteeing the existence of nonlinear $\WW$-symmetry generators in the chiral algebra. The cohomological construction of the corresponding chiral algebra is analogous to the $SU(2)$ case, \emph{mutatis mutandi}. We will not repeat the description here in any detail. We first provide a brief outline of the relevant four-dimensional physics of these models, and then perform a systematic analysis of the physical operators of low dimension in the associated chiral algebra.

As in the $SU(2)$ theory, there is a Lagrangian description of these models in terms of the $\NN=1$ chiral superfields
%%%%%%
\be
W_{\a}^A~,\quad \Phi^B~,\quad Q_{a}^{i}~,\quad \wt Q^{b}_{j}~,
\ee
%%%%%%
where $a,b=1,\ldots,N$ are vector color indices, $A,B=1,\ldots,N^2-1$ are adjoint color indices, and $i,j=1, \ldots, N_f$ with $N_f=2N$ are vector flavor indices. The central charge is fixed by the field content to $c_{4d}=\frac{2N^2-1}{6}$.

For our purposes, the principal difference between the $N\geqslant3$ theories and the $N=2$ case is in the structure of the Higgs branch chiral ring. In the higher rank theories, the hypermultiplets transform in a complex representation of the gauge group, so the global symmetry is not enhanced and we have $G_F=SU(N_f)\times U(1)$. The moment map operators for the global symmetry reside in mesonic $\hat\BB_1$ multiplets, which can be separated into $SU(N_f)$ and $U(1)$ parts,
%%%%%%
\be
M_j^{\ i} \colonequals \wt Q_j^{a} Q^i_{a} \quad\Longrightarrow\qquad \mu \colonequals M_i^{\ i}~,\quad \mu_j^{\ i} \colonequals M_j^{\ i} - \frac{1}{N_f} \mu\,\delta_j^{\ i}~.
\ee
%%%%%%
The level of the non-Abelian part of the global symmetry is $k^{SU(N_f)}_{4d}=2N$. The baryons are of dimension $N$ and no longer generate any additional global symmetries. Rather, they transform in the $N$-fold antisymmetric tensor representations of the flavor symmetry:
%%%%%%
\be\label{eq:baryons}
\begin{split}
B^{i_1 \dots i_N} &\colonequals Q^{i_1}_{a_1}\cdots Q^{i_N}_{a_N}\epsilon^{a_1\ldots a_N}~,\\
\wt B_{i_1 \dots i_N} &\colonequals \wt Q_{i_1}^{a_1}\cdots \wt Q_{i_N}^{a_N}\epsilon_{a_1\ldots a_N}~.
\end{split}
\ee
%%%%%%
The mesons and baryons satisfy a set of polynomial relations. Following \cite{Argyres:1996eh}, we introduce notation where ``$\cdot$'' denotes contraction of an upper and a lower index and ``$*$'' denotes the contraction of flavor indices with the completely antisymmetric tensor in $N_f$ indices. The relations are then given by 
%%%%%%
\be\label{eq:nf2ncHiggsrelations}
\begin{alignedat}{3}
(*B)\wt B 			&~=~& *(M^{N})				&~,& 		\qquad M \cdot *B 		&~=~ M\cdot * \wt B = 0~,\\
M^\prime 	\cdot B 	&~=~& \wt B \cdot M^\prime ~=	&& 0~,	\qquad M \cdot M^\prime		&~=~ 0~,
\end{alignedat}
\ee
%%%%%%
where $(M^\prime)_i^{\ j} \colonequals M_i^{\ j} - \frac{1}{N}\mu \delta_i^{j} = \mu_i^{\ j} - \frac{1}{2N}\mu \delta_i^{j}$. Additionally, all quantities antisymmetrized in more than $N$ flavor indices must vanish.

This completes the description of the Hall-Littlewood chiral ring, since again this theory admits a linear quiver description, so there are no $\DD$-type multiplets after turning on interactions. The final representation of canonical interest is the $\hat\CC_{0(0,0)}$ multiplet, which again contributes an important Schur operator in the form of the $R=1$ component of the $SU(2)_R$ current:
%%%%%%
\be\label{Nf2NcRcurrent}
\JJ_{+\dot+}^{R=1}\sim\frac12\left(Q_a^i\partial_{+\dot+}\wt Q_i^a-\wt Q_i^a\partial_{+\dot+}Q_a^i\right)+\lambda^A_+\tilde\lambda_{\dot+A}~.
\ee
%%%%%%

Like the $SU(2)$ theory, these models all have one-complex-dimensional conformal manifolds with interesting behaviors at the boundary points, where $S$-dual descriptions become appropriate. In contrast to the $SU(2)$ theory, these $S$-dual descriptions are not the same as the original description, and rather involve intrinsically strongly-coupled non-Lagrangian sectors. While such dualities imply interesting structures for the associated chiral algebras, their dependence on non-Lagrangian theories takes us outside the scope of the current examples. This is discussed in much greater detail in \cite{WIP_Class_S}.

\subsubsection{Physical operators of low dimension}

The nontrivial BRST cohomology classes of the chiral algebra can be computed by hand for small values of the dimension. The physical operators of dimension one again correspond to the moment map operators of the global symmetry, which in this case includes only the mesonic chiral ring operators,
%%%%%%
\bea\label{eq:lag_AKM}
J_i^j &\colonequals& q_{a i} \tilde q^{a j} - \frac{1}{N_f} \delta_i^j q_{a k} \tilde q^{a k} = \goodchi[\mu_i^j]~,\\
J     &\colonequals& q_{a k} \tilde q^{a k} = \goodchi[\mu]~.
\eea
%%%%%%
The singular OPEs of these currents are given by
%%%%%%
\be \label{JJOPE}
\begin{alignedat}{3}
&&	J_i^j(z) J_k^l(0) 	&~~~\sim~~~& 	&-\frac{N (\delta_i^l \delta^j_k - \text{trace})}{z^2} ~+~ \frac{\delta_i^l J_k^j(z) - \delta_k^j J_i^l(z)}{z}~,\\
&&	J(z) J(0) 			&~~~\sim~~~& 	&-\frac{2N^2}{z^2}~.
\end{alignedat}
\ee
%%%%%%
This is an $\mf{su}(N_f)\times \mf{u}(1)$ affine Lie algebra at level $k_{2d} = -N$.

At dimension two, we first consider the operators that are invariant under the flavor symmetry. As expected, there is a canonical stress tensor,
%%%%%%
\be
T \colonequals \frac{1}{2}\left( q_{a i} \del \tilde q^{a i} -  \tilde q^{a i}\del q_{a i} \right) - b^a_b \del c_a^b = \goodchi[\JJ_{+\dot+}^1]~,
\ee
%%%%%%
whose self-OPE fixes the two-dimensional central charge,
%%%%%%
\be
\label{c2dnfeq2nc}
c_{2d} = 2-4N^2~.
\ee
%%%%%%

Additionally, the algebra generated by the affine $\mf{su}(N_f)\times\mf{u}(1)$ currents \eqref{eq:lag_AKM} contains a dimension two singlet that is the Sugawara stress tensor of the current algebra,
%%%%%%
\be
T_{\text{sug}} \colonequals \frac{1}{N_f} \left(J_i^j J_j^i - \frac{1}{N_f}JJ\right)~.
\ee
%%%%%%
The corresponding Sugawara central charge is also equal to $2-4N^2$, which suggests that the two stress tensors $T$ and $T^{\text{sug}}$ may be equivalent operators as they were in the $N=2$ theory. Indeed, we expect this to be the case since the central charges in this theory again saturate the unitarity bound \eqref{centralchargebound}. A short computation verifies that their difference is BRST exact,
%%%%%%
\be
\label{sugawararelation}
T - T_{\text{sug}} =  \frac{1}{N_f} \{ Q_{\rm BRST},  q_{a i} \tilde q^{b j} b^a_b \}~.
\ee
%%%%%%
A complete basis for the physical flavor singlets of dimension two is given by $T$, $JJ$, and $\del J$.

The remaining physical operators of dimension two are charged under $U(N_f)$. An overcomplete basis of such operators is given by flavored current bilinears $J_i^j J_k^l$ and $J_i^j J$, in addition to derivatives of the currents $\del J_i^j$. These operators are not all independent. For example, the usual rules of conformal normal ordering imply that
%%%%%%
\be
\label{contracedJJantisym}
J_i^j J_k^l - J_k^l J_i^j =  \delta_i^l \del J_k^j - \delta_k^j \del J_i^l ~,
\ee
%%%%%%
so the antisymmetric normal ordered product of two $SU(N_f)$ currents is a combination of descendants. For the symmetrized normal ordered product there exists another relation:
%%%%%%
\be
\label{contractedJJsym}
\hf(J_i^k J_k^j + J_k^j J_i^k) =  \delta_i^j \left(\frac{1}{N_f^2} JJ + T\right) - \{Q_{\rm BRST},  q_{\a i} \tilde q^{\b j} b_\b^\a\}\,.
\ee
%%%%%%
In group-theoretic terms, the relations amount to the statement that the parts of the symmetric product of two currents that transform in the singlet and adjoint representations do not correspond to independent operators.

It is worth jumping ahead to the case of dimension $N/2$, where we find operators that correspond to the baryonic chiral ring generators \eqref{eq:baryons}:
%%%%%%
\be\label{eq:chiral_baryons}
\begin{split}
b_{i_1 i_2 \ldots i_{N_c}} &\colonequals \varepsilon^{\a_1 \a_2 \ldots \a_{N_c}} q_{\a_1 i_1} q_{\a_2 i_2} \ldots q_{\a_{N_c} i_{N_c}}=\goodchi[B_{i_1i_2\cdots i_N}]~,\\
\tilde b^{i_1 i_2 \ldots i_{N_c}} &\colonequals \varepsilon_{\a_1 \a_2 \ldots \a_{N_c}} \tilde q^{\a_1 i_1} \tilde q^{\a_2 i_2} \ldots \tilde q^{\a_{N_c} i_{N_c}}=\goodchi[\tilde B^{i_1i_2\cdots i_N}]~.
\end{split}
\ee
%%%%%%
These are Virasoro primaries of dimension $N_f / 4$. The only non-trivial OPE that is not entirely fixed by symmetry is the $b \times \tilde b$ OPE. For $N_c = 3$, for example, it is given by
%%%%%%
\be\label{eq:baryonOPE}
b_{i_1 i_2 i_3}(z) \tilde b^{j_1 j_2 j_3}(0) \sim 
\frac{36\, 	\delta_{[i_1}^{[j_1} \delta_{\ph{[}\!i_2}^{\ph{[}\!j_2} \delta_{i_3]}^{j_3]}}{z^3} - 
\frac{36\,  \delta_{[i_1}^{[j_1} \delta_{\ph{[}\!i_2}^{\ph{[}\!j_2} 	 J_{i_3]}^{j_3]}(0)}{z^2} + 
\frac{18\, 	\delta_{[i_1}^{[j_1} 	  J_{\ph{[}\!i_2}^{\ph{[}\!j_2} 	 J_{i_3]}^{j_3]}(0) - 
	  18\, 	\delta_{[i_1}^{[j_1} \delta_{\ph{[}\!i_2}^{\ph{[}\!j_2} \del J_{i_3]}^{j_3]}(0)}{z}~,
\ee
%%%%%%
where square brackets denote antisymmetrization with weight one. 

\subsubsection{Relation to the Higgs branch chiral ring}

Again, certain features of the Higgs branch chiral ring arise organically from the chiral algebra. According to the general discussion in \S\ref{subsec:notable}, the dimension two operators in the chiral algebra should in particular contain the image of the Schur operators in $\hat\BB_2$ multiplets, which in the theories under consideration simply correspond to the product of two of the mesonic operators $\mu$ and $\mu_i^j$ subject to the final relation in \eqref{eq:nf2ncHiggsrelations}. Furthermore, these Schur operators necessarily become Virasoro primary operators in the chiral algebra.

From amongst the BRST cohomology classes at level two -- spanned by $T$, $JJ$, $J_i^j J$, the symmetrized combination $J_i^j J_k^l + J_k^l J_i^j$ modulo relation \eqref{contractedJJsym}, and derivatives of level one currents -- we find exactly three Virasoro primary operators:
%%%%%%
\be \label{Xdefs}
\begin{split}
\XX 			&\colonequals JJ - \frac{N_f^2}{N_f^2 - 2} T\,,\\
\XX_i^j 		&\colonequals J_i^j J\,,\\
\XX_{ik}^{jl} 	&\colonequals \hf(J_i^j J_k^l + J_k^l J_i^j) - \frac{N_f}{N_f^2 -2} \left( \delta_i^l \delta_k^j - \frac{1}{N_f} \delta_i^j \delta_k^l \right) T\,,
\end{split}
\ee
%%%%%%
which are subject to the additional constraints,
%%%%%%
\be \label{Xreln}
\XX_{ik}^{jl} = \XX_{ki}^{lj}\,, \qquad \qquad \XX_{ik}^{il} = 0\,, \qquad \qquad \XX_{i j}^{j k} = \frac{1}{N_f^2} \delta_i^k \XX + \{Q_{\rm BRST}, \ldots \}~.
\ee
%%%%%%
We see that we should identify $\XX = \goodchi[\, \mu \mu \,]$, $\XX_i^j = \goodchi[\, \mu \mu_i^j\,]$ and $\XX_{ik}^{jl} = \goodchi[\, \mu_i^j \mu_k^l \,]$. The first two relations in \eqref{Xreln} then reflect the natural symmetry properties of the original Schur operator, whilst the last equation precisely reproduces the final equation in \eqref{eq:nf2ncHiggsrelations}.

We note that the definitions \eqref{Xdefs} somewhat obscure the relationship to four-dimensional physics because of the conformal normal ordering used to define the products of interacting fields. The same dimension two operators take a completely natural form in terms of creation/annihilation normal ordered products of symplectic bosons,
%%%%%%
\be
\begin{split}
\XX 			&= ~~:q_{\alpha i}\tilde q^{\alpha i}q_{\beta j}\tilde q^{\beta j}:~,\\
\XX_i^j 		&= ~~:q_{\alpha i}\tilde q^{\alpha j}q_{\beta k}\tilde q^{\beta k}:~,\\
\XX_{ik}^{jl} 	&= ~~:q_{\alpha i}\tilde q^{\alpha j}q_{\beta k}\tilde q^{\beta l}:~,
\end{split}
\ee
%%%%%%
and this description also nicely illustrates the commutative diagram of \S\ref{subsec:freetheories}.

Finally, at the level of Virasoro representations, the OPEs of the dimension one currents can now be summarized by the following fusion rules,
%%%%%%
\be
\label{virOPEs}
\begin{alignedat}{6}
&J_i^j & \;\times\; & J_k^l & \quad\rightarrow\quad & - N (\delta_i^l \delta_k^j - \text{trace}) \mathds{1} + (\delta_i^l J_k^j- \delta_k^j J_i^l) + \XX_{ik}^{jl} + \ldots~,\\
&J_i^j & \;\times\; & J     & \quad\rightarrow\quad & \ph{-} \XX_i^j + \ldots~,\\
&J     & \;\times\; & J     & \quad\rightarrow\quad & - 2N^2 \mathds{1} + \XX + \ldots~,\\
\end{alignedat}
\ee
%%%%%%
where we have omitted operators of dimension higher than two. We see that the product structure of the Higgs branch chiral ring is reproduced precisely by the $O(1)$ terms in these fusion rules.\footnote{We may similarly speculate that the Poisson bracket is encoded in the terms of the OPE that correspond to simple poles, but we have not checked this in detail.}

\subsubsection{A $\WW$-algebra conjecture}\label{subsubsec:Nf2Nc_W_algebra}

The chiral algebra is not as simple in this case as it was for the $SU(2)$ theory, since the generators $b$ and $\tilde b$ are higher-spin $\WW$-symmetry generators rather than simple affine currents. Nevertheless, there is a natural guess as to how to describe this more involved theory as a $\WW$ algebra. It is useful to think of the operator content of the algebra in terms of representations of the affine $\mf{u}(N_f)$ current algebra. From the analysis of levels one and two, we know that there is the vacuum representation -- which in particular contains the affine currents and the stress tensor -- and the ``baryonic'' representations, for which the highest weight state is given by the baryon or anti-baryon of \eqref{eq:chiral_baryons}. Other representations of the affine Lie algebra can only come from multi-baryon states or from new generators of dimension greater than two, where we have not performed a detailed analysis of the cohomology.

In four dimensions the mesons and the baryons are the complete set of generators for the Hall-Littlewood chiral ring. The most obvious conjecture is then that the corresponding two-dimensional operators generate the entire $\WW$-algebra: 
%%%%%%
\begin{conj}\label{conj_Nf2Nc}
When $\TT$ is $\NN=2$ $SU(N)$ superconformal QCD for with $2N$ flavors for $N>2$, then $\goodchi[\TT]$ is isomorphic to the $\WW$ algebra generated by affine $\mf{u}(N_f)$ currents at level $k_{\mf{su}(N_f)} = - N$ along with baryonic generators $b$ and $\tilde b$ with the OPE \eqref{eq:baryonOPE} (or its generalizations to $N\geqslant4$).
\end{conj}
%%%%%%
Because no additional generators make an appearance in the singular OPEs of the affine currents and baryons, it is guaranteed to be the case that the $\WW$ algebra we have just described forms a chiral subalgebra of $\goodchi[\TT]$. Our conjecture is that this is in fact the whole thing. If true, this conjecture would imply that the Schur index for the $N_f=2N$ theories decomposes into characters of affine $\mf{u}(2N)_{-N}$ with highest weights given by the vacuum or by one or more baryons.

\subsubsection{Superconformal Index}

We can provide support for this conjecture by comparing with the superconformal index. The Schur index of the theory is given by the following contour integral,
%%%%%%
\begin{align}
\II_{\mathrm{Schur}}(q;c,\vec{a}) = \int[d\vec{b}] P.E.&\left[ \frac{\sqrt{q}}{1-q} \left( c\  \chi_{SU(N_f)}^{\mathbf{N_f}}(\vec{a}) \chi_{SU(N)}^{\mathbf{N}}(\vec{b})    + c^{-1}\  \chi_{SU(N_f)}^{\mathbf{N_f}}(\vec{a}^{-1}) \chi_{SU(N)}^{\mathbf{N}}(\vec{b}^{-1})  \right)  \right. \notag \\
&\left. \qquad +\left(\frac{-2q}{1-q}\right) \chi_{SU(N)}^{\mathbf{N^2-1}}(\vec{b})  \right],
\end{align}
%%%%%%
where $c$ is the $U(1)$ fugacity and $\vec{a}=(a_1,a_2,\ldots,a_{N_f-1})$ denotes $SU(N_f)$ fugacities. For $N=3,$ the first few orders are given by
%%%%%%
\begin{align}
\II_{\mathrm{Schur}}(q;c,\vec{a}) = &1+\left(1+  \chi_{SU(6)}^{\mathbf{35}}(\vec{a}) \right) q + (c^3+ c^{-3}) \chi_{SU(6)}^{\mathbf{20}}(\vec{a})   q^{3/2} \notag \\
&+\left( \left(\chi_{SU(6)}^{sym^2(\mathbf{35})}(\vec{a}) - \chi_{SU(6)}^{\mathbf{35}}(\vec{a})\right) + 2\chi_{SU(6)}^{\mathbf{35}}(\vec{a}) +2\right) q^2 \notag \\
&+ (c^3+ c^{-3} ) \left( 2\chi_{SU(6)}^{\mathbf{20}}(\vec{a})  + \left(\chi_{SU(6)}^{\mathbf{35}\otimes \mathbf{20}}(\vec{a}) - \chi_{SU(6)}^{\mathbf{20}}(\vec{a}) - \chi_{SU(6)}^{\mathbf{70}}(\vec{a})-\chi_{SU(6)}^{\overline{\mathbf{70}}}(\vec{a})\right) \right)q^{5/2} \notag \\
&+ \ldots,
\end{align}
%%%%%%
where we have explicitly indicated the presence of relations by listing them with a minus sign. The dimension two relations in the chiral algebra were elaborated upon in the previous subsection. At level $5/2$, we can similarly determine the Virasoro primaries
%%%%%%
\begin{align}
Y_{ijk}= J b_{ijk} + \partial b_{ijk}, \qquad \wt Y^{ijk}= J \wt b^{ijk} - \partial \wt b^{ijk} \\
Y_{i,klm}^{j}=\frac{1}{2}\left( J_i^{\ j} b_{klm} +  b_{klm}  J_i^{\ j} -\frac{1}{6} \delta_i^j \partial b_{klm}  + \delta_{[k}^j \partial b_{|i|lm]} \right) \\
\wt Y^{j,klm}_{i}=\frac{1}{2}\left( J_i^{\ j} \wt b^{klm} +  \wt b^{klm}  J_i^{\ j} +\frac{1}{6} \delta_i^j \partial \wt b^{klm}  - \delta_i^{[k} \partial \wt b^{|j|lm]} \right) \, ,
\end{align}
%%%%%%
subject to the constraints
%%%%%%
\begin{align}
\epsilon^{iklmnp}\left( Y_{i,mnp}^{j} + \frac{1}{6}\delta_i^j Y_{mnp}\right) = 0, \qquad Y_{i,jlm}^{j} - \frac{1}{6}Y_{ilm} = \{Q_{\rm BRST},\ldots\} \\
\epsilon_{jklmnp}\left( \wt Y^{j,mnp}_{i} + \frac{1}{6}\delta_i^j \wt Y^{mnp}\right) = 0, \qquad \wt Y^{j,kli}_{i} - \frac{1}{6}\wt Y^{jkl} = \{Q_{\rm BRST},\ldots\}\, ,
\end{align}
%%%%%%
which again encode precisely the Higgs branch relations.

At level three, we have checked agreement between the Schur index and the cohomology generated by the $SU(6)\times U(1)$ currents and the baryons by explicitly computing the null states.

\subsection{\texorpdfstring{$\NN=4$}{N=4} supersymmetric Yang-Mills theory}

The theories considered in the previous two subsections all shared the special quality of admitting descriptions as linear quiver gauge theories, which  meant that $\DD$-type multiplets played no role in the analysis. We now turn to a case where this simplification no longer holds, and so there will necessarily be generators outside of the Higgs chiral ring. The theory in question is $\NN=4$ supersymmetric Yang-Mills theory with gauge group $SU(N)$. For our purposes, this is an $\NN=2$ theory with an $SU(N)$ vector multiplet and a single adjoint-valued hypermultiplet. In $\NN=1$ notation, we have the following chiral superfields,
%%%%%%
\be
W^A_\alpha~,\qquad \Phi^A~,\qquad Q_{i}^A~,
\ee
%%%%%%
where $A=1,\ldots N^2-1$ an $SU(N)$ adjoint index and $i=1,2$ is an $SU(2)_F$ vector index. The flavor symmetry $SU(2)_F$ is the commutant of $SU(2)_R\times U(1)_r\subset SU(4)_R$, and so is an $R$-symmetry with respect to the full superalgebra. The central charges of the theory are given by
%%%%%%
\be 
k_{4d}^{SU(2)} = N^2 -1~,  \quad   c_{4d} = \frac{(N^2 - 1)}{4}~.
\ee
%%%%%%

The Higgs branch chiral ring has $N-1$ generators. In terms of the $N\times N$ matrices $Q_i\colonequals Q_{i}^{A}t^A$, these are given by
%%%%%%
\be\label{eq:n4_higgs_ops}
\Tr \,Q_{(i_1}\cdots Q_{i_k)}~,\qquad k=1,\ldots,N-1~,
\ee
%%%%%%
subject to trace relations. In this theory, the Hall-Littlewood chiral ring contains additional $\DD$-type multiplets that are not described by the Higgs chiral ring. More specifically, for $SU(N)$ gauge group there are an additional $N-1$ HL generators given by 
%%%%%%
\be\label{eq:n4_D_ops}
\Tr \,Q_{(i_1}\cdots Q_{i_k)}\tilde\lambda_{\dot+}^1~,\qquad k=1,\ldots,N-1~.
\ee
%%%%%%
There are corresponding generators of the HL anti-chiral ring that lie in $\overline{\DD}$ multiplets and take the same form with $\tilde\lambda_{\dot+}^1$ replaced by $\lambda_{+}^1$. Finally, the Schur component of the $SU(2)_R$ current, which will give rise to the stress tensor in two-dimensions, is given in terms of four-dimensional fields by
%%%%%%
\be 
\JJ_{+\dot{+}}^{R=1} \sim \frac{1}{2}\Tr\,Q_{i} \partial_{+\dot{+}} Q_{j}\varepsilon^{ij} -\Tr\,\tilde{\lambda}_{\dot{+}} \lambda_{+}~.
\ee
%%%%%%

\subsubsection{Cohomological description of the associated chiral algebra}

The free chiral algebra follows the same pattern as the previous examples. The two dimensional counterparts of the hypermultiplet scalars and gauginos can be introduced as usual,
%%%%%%
\be 
q_{i}^A(z)\colonequals\goodchi[Q_{i}^A]~,\qquad b^A(z)\colonequals\goodchi[\tilde \lambda^A]~,\qquad \partial c^A(z)\colonequals\goodchi[\lambda^A]~.
\ee
%%%%%%
The free chiral algebra has the free OPEs,
%%%%%%
\be
q_{i}^{A}(z) q_{j}^{B}(0) \sim  \frac{\varepsilon_{ij}\delta^{AB}}{z}~, \qquad 
b^A (z) c^B(0) \sim \frac{\delta^{AB}}{z}~.\nn 
\ee
%%%%%%
The stress tensor is given by the usual canonical expression
%%%%%%
\begin{align}
T = \frac{1}{2} q_{i}^A \partial q_{j}^B \    \varepsilon^{ij} - b^A\partial c^A~,
\end{align}
%%%%%%
which has a central charge of $c_{2d} = -3 (N^2 - 1)$. The $SU(2)_F$ currents are given by 
%%%%%%
\begin{align}
J_{ij} =-\frac{1}{2}q_{i}^{A}q_{j}^{A}\ \,,
\end{align}
%%%%%%
and satisfy a current algebra at level $k_{2d} = -\frac{N^2 -1}{2}$. The current algebra contains a Sugawara stress tensor of the usual form,
%%%%%%
\begin{align}
T_{\text{Sug}} (z)= \frac{1}{N^2-5}\ J_{ij} J_{kl}\ \varepsilon^{ik}\varepsilon^{jl}\,,
\end{align}
%%%%%%
with central charge equal to $\frac{3(N^2-1)}{N^2-5}$. Note that precisely for $N=2$ and for no other value of $N$, the Sugawara central charge matches with the true central charge. As we will see, this is again a consequence of the two stress tensors being equivalent in BRST cohomology.

The $SU(N)$ currents for the matter and ghost sectors are given by
%%%%%%
\begin{align}
J^A = \frac{i}{2} f^{ABC} q_{i}^Bq_{j}^C\ \varepsilon^{ij}\,,\qquad J_{\rm{gh}}^A =- if^{ABC}c^B b^C\,.
\end{align}
%%%%%%
The levels for the corresponding current algebras are $-2N$ and $2N$, respectively. The BRST current is constructed as usual,
%%%%%%
\begin{align}
J_{\text{BRST}} = c^A\left(J^A_{SU(N)} + \frac{1}{2} J_{\rm{gh}}^A\right)~,
\end{align}
%%%%%%
and its zero mode defines the nilpotent BRST operator $Q_{\text{BRST}}$.

\subsubsection{Low-lying physical states}

Let us first consider the case of $SU(2)$ gauge group. In this case the difference between the Sugawara stress tensor and the canonical stress tensor is BRST exact,
%%%%%
\begin{align}
T - T_{\text{Sug}} ~~\sim~~ \{ Q_{\text{BRST}}~,~f^{ABC}q_{i}^A q_{j}^B b^C \ \varepsilon^{ij} \}~.
\end{align}
%%%%%%
Based on the description of the HL chiral ring generators, we expect that amongst the physical states should be an $SU(2)_F$ triplet of affine currents and an $SU(2)_F$ doublet of dimension $3/2$ fermionic generators. Up to dimension two, the cohomology is generated by precisely these operators,
%%%%%%
\be
\begin{alignedat}{2}
J_{ij} 		&=	 		 & -\frac12(q_i^Aq_j^A) 		 &= \goodchi[\Tr\,Q_iQ_j]~,\\
G_{i} 		&\colonequals& \sqrt{2}(q_{i}^Ab^A) 		 &= \goodchi[\Tr\,Q_i\tilde\lambda_+]~,\\ 
\tilde G_{i}&\colonequals& -\sqrt{2}(q_{i}^A\partial c^A) &= \goodchi[\Tr\,Q_i\lambda_+]~.
\end{alignedat}
\ee
%%%%%%
The OPEs of these generators can be computed directly,
%%%%%%
\begin{align}\label{eq:n4_chiral_opes}
J_{ij}(z)J_{kl}(w) &\sim - \frac{N^2-1}{2}\frac{\varepsilon_{l (i}\varepsilon_{j)k}}{(z-w)^2} + \frac{2\varepsilon_{(k(i}J_{j)l)}}{z-w} \,, \\
J_{ij}(z) G_k(w) &\sim\frac{ \frac{1}{2}(\varepsilon_{ki} G_j(w) + \varepsilon_{kj} G_i(w) ) }{z-w} \, , \\
J_{ij}(z) \tilde G_k(w) &\sim\frac{ \frac{1}{2}(\varepsilon_{ki} \tilde G_j(w) + \varepsilon_{kj} \tilde G_i(w) ) }{z-w} \,, \\
G_{i}(z)G_{j}(w) &\sim 0\,, \\
\tilde G_{i}(z)\tilde G_{j}(w) &\sim 0 \,, \\
G_{i}(z)\tilde G_{j}(w) &\sim -\frac{2(N^2-1)\varepsilon_{ij}}{(z-w)^3} + \frac{4J_{ij}(w)}{(z-w)^2} + \frac{2\varepsilon_{ij} T(w) + 2\partial J_{ij}(w)}{z-w}\,,
\label{OPEGenN4SCA}
\end{align}
%%%%%%
where $N=2$ and the symmetrization in the indices $i,j$ and $k,l$ has weight one. The value of $N$ has been left unspecified in \eqref{eq:n4_chiral_opes} because the OPEs will continue to hold for higher rank gauge groups. For the same reason, $T(z)$ has been included separately, though for $N=2$ it not a distinct generator, but rather is identified with the Sugawara stress tensor.

The operator product algebra in \eqref{eq:n4_chiral_opes} can be immediately recognized to be the ``small'' $\NN=4$ superconformal algebra with central charge $c_{2d} = -3 (N^2 - 1)$ \cite{Ademollo:1976pp}. It is natural that there should be supersymmetry acting in the chiral algebra, since the holomorphic $\mf{sl}(2)$ that commutes with the supercharges $\qq\,_i$ is in enhanced to a holomorphic $\mf{sl}(2\,|\,2)$ when the four-dimensional theory is $\NN=4$ supersymmetric. However, like the case of the global conformal algebra being generated not by the four-dimensional stress tensor but by the chiral operator associated to the $SU(2)_R$ current, here the enhanced supersymmetry in the chiral algebra is generated {\it not} by the four-dimensional supercurrents, but by the Schur operators that lie in the same $\DD_{\frac12(0,0)}$  and $\overline{\DD}_{\frac12(0,0)}$ multiplets with them. Those are the Schur operators that are transmuted into the two-dimensional supercurrents $G_i$ and $\tilde G_i$.

In $SU(3)$ theory there are additional generators arising from the additional HL generators. Sure enough, direct computation produces the following list of new generators of dimension less than or equal to $5/2$:
%%%%%%
\be\label{eq:n4_su3_gens}
\begin{alignedat}{3}
&B_{ijk}&			&~\colonequals~& \makebox[1.3in][l]{$\ph{3}\Tr\,q_{i}q_{j}q_{k}$}		&~=~\goodchi[\Tr\,Q_iQ_jQ_k]~,\\
&B_{ij}&			&~\colonequals~& \makebox[1.3in][l]{$\ph{3}\Tr\,q_{i}q_{j}b$}			&~=~\goodchi[\Tr\,Q_iQ_j\tilde\lambda_{+}]~,\\
&\tilde B_{ij}&		&~\colonequals~& \makebox[1.3in][l]{$\ph{3}\Tr\,q_{i}q_{j}\partial c$} 	&~=~\goodchi[\Tr\,Q_iQ_j\lambda_{\dot+}]~,\\
&B_i&				&~\colonequals~& \makebox[1.3in][l]{$3\Tr\,q_{i}b\del c + \Tr\,\partial q_{j}q^{j}q_{i}$} &  ~=~ \goodchi[3\Tr\,Q_i\tilde\lambda_+\lambda_{\dot+}+\Tr\,\partial_{+\dot+}Q_jQ^jQ_i]~.
\end{alignedat}
\ee
%%%%%%
Precisely for the $SU(3)$ case, the operator $B_i$ is in fact equivalent to a composite operator,
%%%%%%
\be
B_i\sim \ve^{jj^\pr}\ve^{kk^\pr}J_{jk}B_{ij^\pr k^\pr}~.
\ee
%%%%%%
This is a consequence of a chiral ring relation for this value of $N$ which sets $\ve^{jj^\pr}\ve^{kk^\pr}\Tr\,Q_jQ_k\Tr\,Q_iQ_{j^\pr}Q_{k^\pr}$ to zero. This will not be the case for higher rank gauge groups, and $B_i$ will be an authentic generator of the algebra.

\subsubsection{A super $\WW$-algebra conjecture}

Because the chiral algebras of $\NN=4$ SYM theories are supersymmetric, we can introduce a more restrictive notion of generators for these algebras. More precisely, we would like to identify those operators that generate the chiral algebra under the operations of normal ordered products and \emph{super}derivatives, or the action of $\mf{sl}(2\,|\,2)$. In other words, we allow not just $L{_1}$ descendants, but also $G_{i,-\frac12}$ and $\tilde G_{i,-\frac12}$ descendants.

The last three generators in \eqref{eq:n4_su3_gens} are superdescendants of $B_{ijk}$, so we have really only found one additional super-generator in the $SU(3)$ theory. In general, HL operators will be grouped by $\NN=4$ supersymmetry into multiplets comprised of a single $\hat\BB$-type operator, an $SU(2)_F$ doublet of $\DD$-type operators, and an $SU(2)_F$ doublet worth of $\bar\DD$-type operators.

For a general simple gauge group, the natural guess is that the chiral algebra is generated by the small $\NN=4$ superconformal algebra along with additional chiral primary operators arising from the Higgs chiral ring generators. Our conjecture is then the following:
%%%%%%
\begin{conj}\label{conj_Neq4}
The chiral algebra for $\NN=4$ SYM theory with gauge group $G$ is isomorphic to an $\NN=4$ super $\WW$-algebra with $\mathrm{rank }(G)$ generators given by chiral primaries of dimensions $\frac{d_i}{2},$ where $d_i$ are the degrees of the Casimir invariants of $G$.
\end{conj}
%%%%%%
We now perform some tests of this conjecture at the level of the superconformal index.

\subsubsection{The superconformal index}

Conjecture \ref{conj_Neq4} can be tested up to any given level by comparing the index of the chiral algebra defined in the conjecture with the superconformal index of $\NN=4$ SYM in the Schur limit. For gauge group $SU(N)$, the Schur index is given by a contour integral,
%%%%%%
\begin{align}
\II_{\mathrm{Schur}}(q;a)=\oint[d\vec{b}] \text{P.E.}\left[ \left(\frac{\sqrt{q}}{1-q} \right) \chi^{\mathbf{2}}(a) \chi^{\mathbf{N^2-1}}(\vec{b}) +  \left(\frac{-2q}{1-q} \right) \chi^{\mathbf{N^2-1}}(\vec{b}) \right]\,,
\end{align}
%%%%%%
where $a$ is an $SU(2)_F$ flavor fugacity. For $SU(2)$ gauge group, expanding the integrand in powers of $q$ and integrating gives the following result up to $O(q^4)$, where we have collected terms into $SU(2)_F$ characters $\chi^{\bf R}(a)$,
%%%%%%
\begin{align}
\II_{\mathrm{Schur}}(q;a) =& 1 +  \chi^{\mathbf{3}}(a) q - 2  \chi^{\mathbf{2}}(a)q^{3/2} + \left( \chi^{\mathbf{1}}(a) +\chi^{\mathbf{3}}(a) +\chi^{\mathbf{5}}(a) \right)q^2 \notag \\
&-2\left(\chi^{\mathbf{2}}(a) +\chi^{\mathbf{4}}(a)  \right)q^{5/2} + \left(\chi^{\mathbf{1}}(a) + 3 \chi^{\mathbf{3}}(a) +\chi^{\mathbf{5}}(a) +\chi^{\mathbf{7}}(a)  \right)q^{3}\notag \\
&-\left(4\chi^{\mathbf{2}}(a) +4\chi^{\mathbf{4}}(a) +2\chi^{\mathbf{6}}(a)  \right)q^{7/2}\notag \\
&+\left(3\chi^{\mathbf{1}}(a) +7\chi^{\mathbf{3}}(a) +4\chi^{\mathbf{5}}(a) +\chi^{\mathbf{7}}(a) +\chi^{\mathbf{9}}(a)  \right)q^4 + \ldots~.
%%%%%%
\end{align}
We can compare this result with the index of the $\WW$-algebra appearing in the conjecture (in this case, just the small superconformal algebra with the appropriate value of the central charge) by enumerating the states of the chiral algebra and then finding and subtracting the null states at each level. We have checked up to level four, and the results match exactly.

The same comparison can be done for the $SU(3)$ case, where the Schur index to $O(q^3)$ is given by
%%%%%%
\begin{align}
\mathcal{I}_{\mathrm{Schur}}(q;a) =& 1 +  \chi^{\mathbf{3}}(a) q +\left(\chi^{\mathbf{4}}(a)- 2  \chi^{\mathbf{2}}(a)\right)q^{3/2} + \left( 2\chi^{\mathbf{1}}(a) +\chi^{\mathbf{5}}(a) - \chi^{\mathbf{3}}(a) \right)q^2 \notag \\
&+\left(\chi^{\mathbf{6}}(a)-3\chi^{\mathbf{2}}(a) \right)q^{5/2} + \left(5\chi^{\mathbf{1}}(a) +  \chi^{\mathbf{3}}(a)  +2\chi^{\mathbf{7}}(a) -3\chi^{\mathbf{5}}(a) \right)q^{3} + \ldots\,.
\end{align}
%%%%%%
Up to level three the nulls were computed and they agree with the index. Note that in this case there are cancellations in the index of the chiral algebra, since there are bosonic and fermionic states appearing at the same level.

\subsection{Class \texorpdfstring{$\SS$}{S} at genus two}
\label{subsec:genustwo}

At this point, the reader may be starting to get the impression that the chiral algebra of any four-dimensional theory be entirely determined by the structure of its various chiral rings. The purpose of this next example is to show that such a simplistic picture is untenable.

\begin{figure}[t]
\label{fig:genus2_quivers}
  \centering
  $\vcenter{\hbox{\includegraphics[width=2.7in]{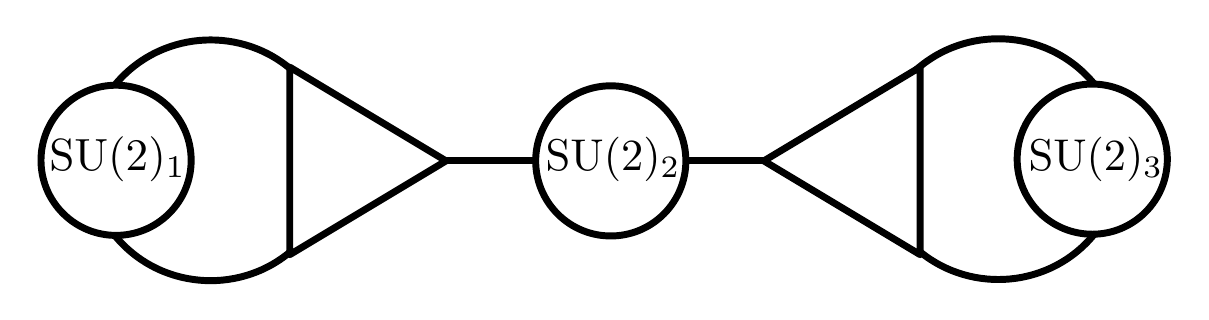}}}$
  \hspace*{.2in}
  $\vcenter{\hbox{\includegraphics[width=1.84in]{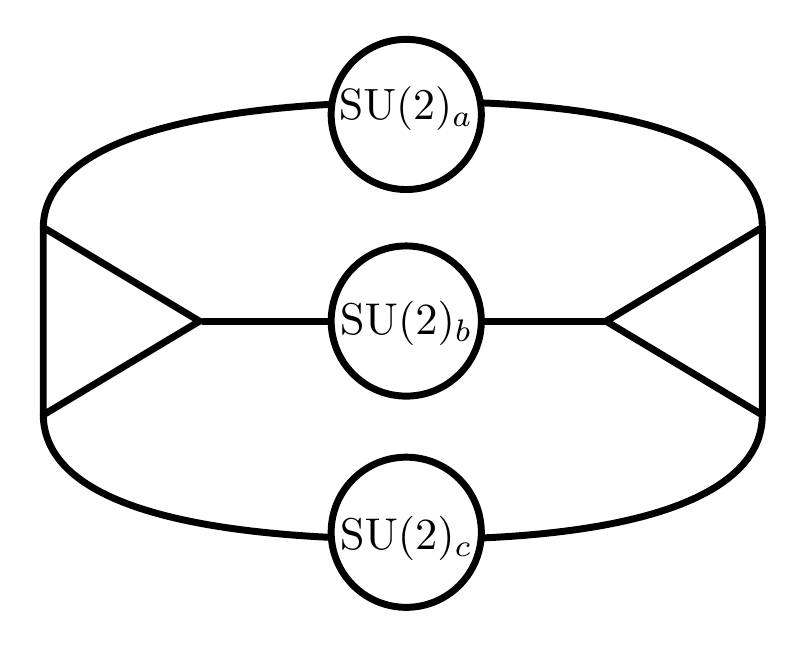}}}$
  \caption{Weak coupling limits of the genus two class $\SS$ theory.}
\end{figure}

Our example is the rank one class $\SS$ theory associated to an unpunctured genus two Riemann surface \cite{Gaiotto:2009we,Gaiotto:2009hg}. The theory admits two inequivalent weak-coupling limits, or $S$-duality frames, corresponding to the two generalized quiver constructions illustrated in Fig. \ref{fig:genus2_quivers}. We will focus on the first case, which is sometimes called the dumbbell quiver. The gauge groups are denoted $SU(2)_1$ for the left loop, $SU(2)_2$ for central line, and $SU(2)_3$ for the right loop. The fields of the theory are two sets of half-hypermultiplets transforming in the trifundamental representation of $SU(2)^3$ and three $SU(2)$ vector multiplets. In $\NN=1$ notation, we denote these by
%%%%%%
\begin{align}
Q_{a_1b_1a_2}~,\qquad S_{a_3b_3a_2}~, \qquad W^{(\nu)}_{\alpha \ A_\nu}~, \qquad \Phi^{(\nu)}_{B_\nu}~,
\end{align}
%%%%%%
where $\nu=1,2,3$ indexes the three $SU(2)$ gauge groups, $a_\nu, b_\nu$ are fundamental indices of $SU(2)_\nu$, and $A_\nu, B_\nu$ are adjoint indices of $SU(2)_\nu$. It is convenient to rearrange the fields $Q_{a_1b_1a_2}$ and $S_{a_3b_3a_2}$ in terms of irreducible representations of the gauge groups.  
In particular, we can define 
%%%%%%
\be
\begin{alignedat}{2}
Q_{A_1a_2}&\colonequals& -i Q_{a_1b_1a_2} (T_J)^{a_1b_1}~,\quad 
Q_{a_2}&\colonequals \frac{1}{\sqrt{2}}\varepsilon^{a_1b_1}Q_{a_1b_1a_2}~,\\
\quad S_{A_3a_2}&\colonequals& -i S_{a_3b_3a_2} (T_J)^{a_3b_3}~,\quad 
S_{a_2}&\colonequals \frac{1}{\sqrt{2}}\varepsilon^{a_3b_3}S_{a_3b_3a_2}~.
\end{alignedat}
\ee
%%%%%
Finally, we introduce the fields 
%%%%%%
\be
\phi_{a_2} =\frac{1}{\sqrt{2}}( Q_{a_2} + i  S_{a_2})~,\qquad \bar\phi_{a_2} =\frac{1}{\sqrt{2}}( Q_{a_2} - i  S_{a_2})~.
\ee
%%%%%%
The theory has a $U(1)_F$ flavor symmetry that is not completely obvious given the usual structure of flavor symmetries in class $\SS$ theories. The fields $\phi$ and $\bar\phi$ have charges $+1$ and $-1$ respectively under the flavor symmetry, and the remaining fields are neutral.

The BRST cohomology problem for this theory can be set up as in the previous sections. In fact, the analysis may be somewhat simplified by leveraging the $\NN=4$ analysis of the previous section. In particular, each loop in the quiver corresponds to a small $\NN=4$ superconformal algebra along with a decoupled $SU(2)$ doublet of symplectic bosons. The genus two theory is obtained by gauging the diagonal subgroup of the $SU(2)$ flavor symmetries for each side. Nevertheless, the resulting cohomology problem is substantially more intricate than those of the previous examples, and we will not describe the level-by-level analysis.

Instead, we will take an indirect approach to understand the spectrum of \emph{generators} of this chiral algebra at low levels. In particular, by analyzing various superconformal indices of this theory and comparing with the structure of the HL chiral ring, we will be able to prove that the full chiral algebra must have generators in addition to those related to HL chiral ring generators and the stress tensor. More precisely, by studying the spectrum up to dimension three, we find that there must be additional generators that arise from $\hat{\CC}_{1 (0,0)}$ multiplets in four dimensions.

\medskip

The Higgs branch chiral ring for this theory has been analyzed in \cite{Hanany:2010qu}. It has three generators: a $U(1)_F$ neutral operator of dimension two, which is actually the moment map for $U(1)_F$,
%%%%%%
\begin{align}
M = - \epsilon^{a_2 a_2'} \phi_{a_2}\bar\phi_{a_2'}~,
\end{align}
%%%%%%
and two operators of dimension four,
%%%%%%
\begin{align}
\OO_1 &= 2\ \phi_{a_2}\phi_{a_2'}\ \epsilon^{a_2 b_2}\epsilon^{a_2' b_2'}\  Q_{A_1 b_2}Q_{B_1 b_2'}\  \delta^{A_1B_1}~.\\
\OO_2 &=2\  \bar\phi_{a_2}\bar\phi_{a_2'}\ \epsilon^{a_2 b_2}\epsilon^{a_2' b_2'}\  Q_{A_1 b_2}Q_{B_1 b_2'}\  \delta^{A_1B_1}~,
\end{align}
%%%%%%
that have charges $+2$ and $-2$ under the flavor symmetry. These generators satisfy a flavor neutral relation of dimension eight:
%%%%%%
\begin{align}
\OO_1\OO_2 =  M^4~.
\end{align}
%%%%%%
It will be helpful for us to write down the Hilbert series \cite{Hanany:2010qu} for this theory, refined by the $U(1)_F$ flavor symmetry:
%%%%%%
\be
\label{Hilberseries}
g(\tau,a)=\frac{1-t^4}{(1-t)(1-a^2t^2)(1-a^{-2}t^2)}= 1 + t + \left( a^2 + a^{-2} + 1 \right) t^2 + \left( a^2 + a^{-2} + 1 \right)  t^3 + \ldots~,
\ee
where $a$ is the $U(1)_F$ fugacity, and $t$ is the fugacity for the dimension of the operator.\\
%%%%%%
\begin{table}[t] 
\centering
\renewcommand{\arraystretch}{1.5}
\begin{tabular}{|c|c|c|c|c|}
\hline
Multiplet 							& Index contribution 		&~~$h$~~&~~$U(1)_R$~~&~~$U(1)_F$~~\\
\hline\hline
$\hat{\BB}_1$						& $\frac{t}{1-q}$ 			& $1$	& $0$ 				& $ 0$ 				\\
$\hat{\BB}_2$						& $\frac{t^2 a^2}{1-q}$		& $2$	& $0$				& $+2$				\\
$\hat{\BB}_2$						& $\frac{t^2/a^2}{1-q}$		& $2$	& $0$ 				& $-2$				\\
\hline
$2\times\DD_{1\,(0,0)}$				& $-2\frac{t^2 a}{1-q}$		& $2$	& $ \frac12$		& $+1$				\\
$2\times\bar{\DD}_{1\,(0,0)}$		& $-2\frac{t q a}{1-q}$		& $2$	& $-\frac12$		& $+1$				\\
$2\times\DD_{1\,(0,0)}$				& $-2\frac{t^2/a}{1-q}$		& $2$	& $ \frac12$		& $-1$				\\
$2\times\bar{\DD}_{1\,(0,0)}$		& $-2\frac{t q/a}{1-q}$		& $2$	& $-\frac12$		& $-1$				\\
$\DD_{\frac32 \,(0,\frac12)}$		& $\frac{t^3}{1-q}$			& $3$	& $ 1$				& $ 0$				\\
$\bar{\DD}_{\frac32 \,(\frac12,0)}$	& $\frac{t q^2}{1-q}$		& $3$	& $-1$				& $ 0$				\\
\hline
$\hat{\CC}_{0(0,0)}$				& $\frac{t q}{1-q}$			& $2$	& $0$				& $ 0$				\\
\hline
$3\times\hat{\CC}_{1(0,0)}$			& $3\frac{t^2 q}{1-q}$		& $3$	& $0$				& $ 0$				\\
\hline
\end{tabular} 
\caption{Chiral algebra generators for the genus two theory with $h\leqslant 3$. The first columns lists the name and multiplicity of the four dimensional multiplets giving rise to the generators. The second column lists the contribution of each multiplet to the Macdonald superconformal index, including the flavor fugacity. The  last columns list the two-dimensional quantum numbers of the generators. The first block of the table consists of Higgs chiral ring generators, the second the remaining HL chiral and anti-chiral ring generators, the third the two-dimensional stress tensor, and the last block the extra generators deduced from the superconformal index.\label{tab:genus2gens} }
\end{table} 
%%%%%%
The generalized quiver for this theory has closed loops, so there will be additional elements of the HL chiral ring coming from $\DD$-type multiplets. The HL index for this theory can be computed by standard methods, and is given by
%%%%%%
\begin{align}
\II_{\mathrm{HL}}(t;a)&= 1 + t + (a^2 + a^{-2} -2a - 2a^{-1} +1 )t^2 +  (a^2 + a^{-2} -2a - 2a^{-1} +2 )t^3  +\ldots~.
\end{align}
%%%%%%
By subtracting off the contributions of the Higgs chiral ring operators (obtained from \eqref{Hilberseries}), we can find the contributions of just the $\DD$-type multiplets. In turn, we can extract the structure of the $\DD$-type generators.\footnote{We have checked by a computation of the HL cohomology that the HL index captures faithfully the complete spectrum of ${\DD}$-type multiplets up to dimension three.} All told, at dimension two there are two $\DD_{1(0,0)}$ multiplets with $U(1)_F$ charge $+1$ and two with charge $-1$, and at dimension three there is a single $\DD_{\frac{3}{2}(0,\frac{1}{2})}$ multiplet that is $U(1)_F$ neutral. The two-dimensional counterparts of these operators can be defined in an explicit calculation of the BRST cohomology.\\

Up to dimension three, we have now determined all of the generators of the HL chiral ring. The question is whether these operators (along with the conjugates of the $\DD$-type operators), in addition to the two-dimensional stress tensor, are sufficient to explain the full spectrum of the chiral algebra (up to dimension three). The generators are listed in the three blocks of Table~\ref{tab:genus2gens}, together with their contribution to the Macdonald index and the quantum numbers of the corresponding Schur operators.

The Macdonald limit of the superconformal index of this theory is obtained from the following contour integral,
%%%%%%
\begin{align}
\II_{\mathrm{MD}}(q,t;a) = &\oint[db_1][db_2][db_3]\,P.E.\left[ \frac{\sqrt{t}}{1-q} \left[\left(\chi^{\mathbf{3}}(b_1) \chi^{\mathbf{2}}(b_3) + \chi^{\mathbf{3}}(b_2) \chi^{\mathbf{2}}(b_3)\right)  + (a + a^{-1})\chi^{\mathbf{2}}(b_3) \right]\right. \notag \\ &\left.+  \left(\frac{-t-q}{1-q} \right) \left( \chi^{\mathbf{3}}(b_1) +\chi^{\mathbf{3}}(b_2)+\chi^{\mathbf{3}}(b_3)\right) \right]~,
\end{align}
%%%%%%
and the expansion including all operators up to dimension three is as follows,
\begin{align}
\II_{\mathrm{MD}}(q,t;&a)= 1 + t + (a^2 + a^{-2} -2a - 2a^{-1} +1 )t^2 + ( -2a - 2a^{-1} + 2 )qt  + \\
 + & ( a^2 + a^{-2} - 2 ( a + a^{-1} ) + 2 ) t^3 + 
(3 - 2 ( a + a^{-1} ) ) q^2 t + ( a^2 + a^{-2} - 4 (a + a^{-1}) + 5 )t^2 q + \ldots~.\nn
\end{align}
%%%%%%
We find that not all of the terms in this expansion can be accounted for by enumerating normal ordered products of generators and their descendants. In particular, from the list of known generators, the only operators that could contribute as $t^2 q$ to the index (with no flavor fugacity) are the normal-ordered product of a $\hat{\BB}_1$ and a $\hat{\CC}_{0(0,0)}$ and the derivative of the normal-ordered product of two $\hat{\BB}_1$ operators. This leaves a contribution of $3 t^2 q$ remains to be explained. We can thus conclude that there are at least three new operators, and they must all must correspond to $\hat{\CC}_{1,(0,0)}$ multiplets that are uncharged under the flavor symmetry. We have included these as the last entry in Table~\ref{tab:genus2gens}. The argument presented above shows that at least these three multiplets must be present, however it does not take into account possible cancellations in the index, which could hide even more additional generators.
%%%%%%

\section{Beyond Lagrangian theories}
\label{sec:classS}

Although the discussion of the previous section focused on theories admitting Lagrangian descriptions, the correspondence between $\NN=2$ SCFTs and chiral algebras is of course much more general. In particular, the vast landscape of superconformal theories of class $\SS$, most of which are non-Lagrangian in nature, will be mapped to an intricate and interesting class of chiral algebras. The purpose of this section is to draw a sketch of the class of chiral algebras defined by this map. Most of the features discussed here follow from the general structure of class $\SS$ and the correspondence with chiral algebras. We do however include a few specific claims that will be left unsubstantiated here, but which are explained in the more complete analysis of \cite{WIP_Class_S}. To begin, we offer a quick reminder of the salient features of $\NN=2$ SCFTs of class $\SS$.

\subsection{A review of class \texorpdfstring{$\SS$}{S} in four dimensions}
\label{subsec:class_S_review}

Class $\SS$ theories \cite{Gaiotto:2009we,Gaiotto:2009hg} are those that arise from compactification of any of the $\NN=(2,0)$ six-dimensional superconformal theories on a Riemann surface $\CC$, known as the \emph{UV curve}, possibly with the inclusion of real codimension two defect operators at points of $\CC$.\footnote{We restrict to the case of \emph{regular} defects in all that follows. These are defects that are specified by an embedding $\rho:\mf{su}(2)\rightarrow \mf{g}$, where $\mf{g}$ is the simply laced Lie algebra that labels the six-dimensional theory. Such a defect supports a flavor symmetry equal to the centralizer of the embedded $\mf{su}(2)$ subalgebra of $\mf{g}$.} We will be interested in the case of \emph{superconformal} theories of class $\SS$, which means that the mass parameters associated to defect operators will all be set to zero. The conformal manifold of a theory of class $\SS$ is equal to the complex structure moduli space of the UV curve, with boundaries at which the curve degenerates corresponding to physical limits in which a gauge coupling goes to zero and a free vector multiplet decouples from the rest of the spectrum.

For our purposes, the most useful way to think about these theories is in terms of a set of four-dimensional ``building block'' theories associated to three-punctured spheres, or \emph{trinions} \cite{Gaiotto:2009we}. Such a theory can be denoted $T_{\mf{g}}^{(\rho_1,\rho_2,\rho_3)}$, where $\mf{g}$ is the lie algebra of the underlying six-dimensional theory, and the $\rho_i$ label the defects at the three punctures. When all three embeddings are trivial, the theory is sometimes simply denoted $T_{\mf{g}}$ (or $T_N$ for the case that $\mf{g}=A_{N-1}$). These building blocks can be assembled into more complex theories in a manner that is represented by a generalized quiver diagram such as those displayed in \S\ref{subsec:genustwo}. The shape of the generalized quiver is necessarily a tropical limit of the corresponding UV curve, with different tropical limits corresponding to different $S$-duality frames of the same theory.

A number of known features of the building block theories can be used to predict the structure of the associated chiral algebras. In the interest of  simplifying the discussion, we shall henceforth restrict to the case where $\mf{g}=A_{N-1}$. The maximal building block (that is, the one with the largest flavor symmetry group) is then the $T_N$ theory mentioned above. We begin by reviewing its properties. 

Generically, $T_N$ has $SU(N)_1\times SU(N)_2\times SU(N)_3$ flavor symmetry, and central charges \cite{Gaiotto:2009gz,Benini:2009gi}
%%%%%%
\be
c_{4d}=\frac{N^3}{6}-\frac{N^2}{4}-\frac{N}{12}+\frac{1}{6}~,\qquad k^{SU(N)}_{4d}=2N=2\dce~.
\ee
%%%%%%
When $N=2$, this is just the theory of free trifundamental half-hypermultiplets that appeared in the example of \S\ref{subsec:genustwo}, so the associated chiral algebra is already known. In the special case of the $T_3$ theory, the global symmetry is enhanced to $E_6$ and this is the classic theory of \cite{Minahan:1996fg}. In that case, the four-dimensional level for the $E_6$ symmetry is $k_{4d}^{E_6}=6$.

The generators of the Higgs branch chiral ring are known for these theories. There are always dimension two moment maps $\mu_{i=1,2,3}$ that transform in the adjoint of $SU(N)_i$ and obey the relation
%%%%%%
\be
\Tr\,\mu_1^k=\Tr\,\mu_2^k=\Tr\,\mu_3^k~,\quad k=2,\ldots,N~.
\ee
%%%%%%
These are supplemented by generators $Q_{(k)}$ of dimension $k(N-k)$ for $k=1,\ldots,N-1$, which transform in the $(\wedge^k,\wedge^k,\wedge^k)$ representation of $SU(N)_1\times SU(N)_2\times SU(N)_3$, where $\wedge^k$ denotes the $k$-fold antisymmetric tensor representation. For $N=2$ the only operator of this type is $Q_{(1)}$, which is the free hypermultiplet itself. The moment maps are actually composites of this basic operator. For the $N=3$ case the operators are $Q_{(1)}$ and $Q_{(2)}$, which are the additional moment maps of $E_6$. For higher values of $N$, these are genuine new generators of the Higgs branch chiral ring, all with dimension greater than two. Some of the relations amongst these higher generators and the moment maps have been derived in \cite{Maruyoshi:2013hja}, though we do not list them here. In the case of the $E_6$ theory, the full set of Higgs branch relations are precisely those that define the Joseph ideal for the $E_6$ one-instanton moduli space.

The trinion theories with reduced punctures (\ie, with nontrivial defining embeddings $\rho_i$) can be thought of as arising by coupling the theory with a maximal puncture to a certain superconformal tail and then turning on specific Higgs branch vacuum expectation values \cite{Gaiotto:2012uq,Gaiotto:2012xa}. Though we do not write down the explicit formulae, the central charges for these theories can be computed for any choice of defining representations \cite{Chacaltana:2010ks}. Important special cases are the trinions for which the theory that results from reducing the punctures of the non-Lagrangian $T_N$ theory is described in terms of free fields. A canonical example is the theory where $\rho_1$ and $\rho_2$ are trivial, but $\rho_3$ is the subregular embedding of $\mf{su}(2)$ into $\mf{su}(N)$. In this case puncture three is known as a \emph{minimal} punctures, and the resulting trinion theory is that of $N$ free hypermultiplets.

Finally, we mention that index considerations suggest that there are no $\DD$-type multiplets for these theories, in which case the HL chiral ring is just the Higgs chiral ring \cite{Gadde:2011uv,Maruyoshi:2013hja}.

\subsection{An outline of class \texorpdfstring{$\SS$}{S} chiral algebras}
\label{subsec:class_S_chiral_algebras}

We now turn to the class of chiral algebras that form the image of the class $\SS$ SCFTs under the map $\goodchi$. In parallel with the full four-dimensional story, there will be a set of basic building block chiral algebras corresponding to the sphere with three maximal punctures. These will be the chiral algebras $\goodchi[\,T_N\,]$. General aspects of the chiral algebra correspondence allow us to predict a number of properties of these theories. The two-dimensional central charge is fixed by the usual proportionality with the four-dimensional conformal anomaly,
%%%%%%
\be
c_{2d}=-2N^3+3N^2+N-2~.
\ee
%%%%%%
Additionally, these chiral algebras have $\widehat{\mf{su}}(n)_k^3$ affine symmetry with
%%%%%%
\be
k_{2d}=-\dce~.
\ee
%%%%%%
It is interesting to note that this is precisely the level that is relevant for the connection between two-dimensional vertex algebras and the geometric Langlands program (see, \eg, \cite{Frenkel:2005pa}). In addition to the generating currents of the affine flavor symmetry, the chiral algebra will have additional generators $\goodchi[Q_{(k)}]$ of holomorphic dimension $h=\frac12 k(N-k)$ transforming in the appropriate representations of the flavor symmetries.

For the case of the $T_3$ theory, the Higgs chiral ring generators are just the $E_6$ moment maps. The relations are generated by the $E_6$ Joseph ideal, and correspondingly the central charges of this theory saturate the appropriate unitarity bounds of \S\ref{subsec:unitarity_bounds}. In particular, this means that the stress tensor is not an independent generator, but rather is equivalent to the Sugawara stress tensor of the $E_6$ current algebra (see \S\ref{subsec:saturation}). Given our prior experience in \S\ref{subsec:so8}, it is natural to make a preliminary conjecture concerning the description of the $T_3$ chiral algebra:
%%%%%%
\begin{conj}\label{conj_class_S}
The chiral algebra for the rank one $E_6$ theory, also known as $T_3$, is isomorphic to the $E_6$ affine Lie algebra at level $k_{2d}=-3$.
\end{conj}
%%%%%%
It is difficult to directly address this conjecture, since we do not have the free-field realization of this chiral algebra that was present for Lagrangian theories. Nevertheless, a variety of indirect checks have been performed and are presented in \cite{WIP_Class_S}.

The chiral algebras associated to more general punctured Riemann surfaces can be realized in a procedure that parallels the gluing construction in four dimensions. In particular, for a given generalized quiver construction we start with a number of copies of $\goodchi[T_N]$ along with $SU(N)$ ghost small algebras, and then perform the BRST reduction associated to four-dimensional gauging to define the chiral algebra. Because the chiral algebra that is associated to a given four-dimensional theory is independent of the exactly marginal couplings, the chiral algebras associated to a given UV curve will not depend on the complex structure moduli of the curve, and in particular will not depend on the choice of generalized quiver within a given topological class. Thus, there will be a generalized topological quantum field theory that associates a chiral algebra to any appropriately decorated Riemann surface. This is very much in the spirit of \cite{Gadde:2009kb} and \cite{Moore:2011ee}, where the superconformal index and the symplectic holomorphic variety of the Higgs branch, respectively, were used to define a generalized TQFT via class $\SS$. Associativity of the gluing imposes highly nontrivial requirements on the chiral algebra of the elementary $T_N$ building block. There are three {\it a priori} inequivalent gauging procedures of two  $T_N$  theories that must lead to the unique theory associated to the four-punctured sphere. From the $2d$ perspective, the BRST complexes associated to the different gaugings  must give the same cohomology. In the simple case of $T_2$, this follows at once from Conjecture \ref{conj_so8}, as the $\widehat {\mf{so}}(8)$ current algebra is manifestly independent of the choice of gluing. 
   
Having focused thus far on the case of maximal punctures, we should also consider chiral algebras $\goodchi[T_N^{(\rho_1,\rho_2,\rho_3)}]$ associated to the non-maximal theories. The task of reducing the rank of a puncture can be accomplished directly within the two-dimensional chiral algebra setting. We propose that the chiral algebra for the theory $T_N^{(\rho_1,\rho_2,\rho_3)}$ is determined by quantum Drinfeld-Sokolov (DS) reduction of the $T_N$ theory with respect to the three embeddings. In the canonical setting, quantum DS reduction is an operation that is performed on an affine Lie algebra in order to produce a different $\WW$-algebra as the cohomology of an appropriate BRST operator. In the present setting, the reduction is performed on a theory with an affine Lie subalgebra, so one may think of this as quantum DS reduction with modules. The generalization is  conceptually straightforward, but somewhat involved technically. This proposal passes several checks, most notably that the central charges of the reduced theory precisely reproduce the expected answers. The behavior of the class $\SS$ chiral algebras under the reduction of punctures imposes additional powerful constraints on the form of these two-dimensional theories. In particular, complete reduction of a puncture (corresponding to choosing a maximal embedding $\rho$) must lead to the chiral algebra for the theory with one fewer puncture. Similarly, reducing one maximal puncture in $\goodchi[T_N]$ to a minimal punctures must lead to the free hypermultiplet chiral algebra. A detailed discussion will be presented in \cite{WIP_Class_S}.

The connection between reducing the rank of a puncture and quantum DS reduction has made previous appearances in the context of the AGT correspondence \cite{Alday:2009aq,Wyllard:2010rp}, and the fact that the same procedure is used here suggests a deeper connection between the chiral algebras defined here and those that appear in the AGT correspondence.

\section{Open questions}
\label{sec:conclusions}

We have outlined the main features of a new surprising correspondence between the four-dimensional $\NN=2$ superconformal field theories and chiral algebras. It should be apparent that there is a great deal more to learn about this rich structure. There are many aspects that should be clarified further, and many natural directions in which the construction could be generalized. We will simply provide a concise list of what we consider to be the most salient open questions, some of which are currently under investigation. 

\begin{itemize}

\item[$\bullet$]
For the Lagrangian examples considered in \S\ref{sec:lagrangian_examples}, as well as the class $\SS$ examples
sketched in \S\ref{sec:classS}, we have made specific conjectures for the description of the resulting chiral algebras as $\WW$-algebras. We hope that some of these conjectures can be proved by more advanced homological-algebraic techniques.

\item[$\bullet$]
A detailed analysis of the $\hat\BB_1$ four-point function that compared $4d$ and $2d$ perspectives led to powerful new unitarity bounds that must be obeyed in any interacting $\NN=2 $ SCFT with flavor symmetry. It is likely that applying the same methods to more general correlators will lead to further unitarity constraints.

\item[$\bullet$]
A better understanding of the implications of four-dimensional unitarity may help clarify what sort of chiral algebra can be associated to a four-dimensional theory. A sharp characterization of the class of chiral algebras that descend from four-dimensional SCFTs could prove invaluable, both as a source of structural insights and as a possible first step towards a classification program for $\NN=2$ SCFTs.

\item[$\bullet$]
We have seen that the four-dimensional operators that play a role in the chiral algebra are closely related to the ones that contribute to the Schur and Macdonald limits of the superconformal index. While the Schur limit has been interpreted in \S\ref{subsec:index} as an index of the chiral algebra, the additional grading that appears in the Macdonald index is not natural in the framework that we have developed. It would be interesting if the additional refinement of the Macdonald index could be captured by a deformation of the chiral algebra structure, perhaps along the lines of \cite{deformed_chiral}.

\item[$\bullet$]
It seems inevitable that extended operators will ultimately find a place in our construction. We expect that codimension-two defects orthogonal to the chiral algebra plane will play the role of vertex operators transforming as non-trivial modules of the chiral algebra. One could also apply the tools developed here to study protected operators that live on conformal defects that fill the chiral algebra plane.

\item[$\bullet$] As it was presented here, the definition of a protected chiral algebra appears to use extended superconformal symmetry in an essential way. Nevertheless, one wonders whether some aspects of this structure may survive away from conformality, perhaps after putting the theory on a nontrivial geometry.

\item[$\bullet$] A related question is whether some aspects of our construction for Lagrangian theories may be accessible to the techniques of supersymmetric localization. The chiral algebra itself may emerge after an appropriate localization of the four-dimensional path integral.

\item[$\bullet$] In many examples, the structure of the $4d$ Higgs branch appears to play a dominant role in determining the structure of the associated chiral algebra. It is an interesting question whether there is a sense in which the chiral algebra is an intrinsic property of the Higgs branch, possibly with some additional structure added as decoration.

\item[$\bullet$] The structure that we have utilized in this article does not admit a direct generalization to odd space-time dimensions. However, a philosophically similar approach leads to a correspondence between three-dimensional $\NN=4$ superconformal field theories and one-dimensional topological field theories. The topological field theory captures twisted correlators of three-dimensional BPS operators whose positions are constrained to a line. We hope to return to investigate this structure in the future.

\item[$\bullet$]
The cohomological approach to chiral algebras that was successfully pursued in this article can be repeated in two-dimensional theories with at least $\NN=(0,4)$ superconformal symmetry and six-dimensional theories with $\NN=(2,0)$ superconformal symmetry \cite{WIP_6d}. As it was in the four-dimensional case, correlation functions of the six-dimensional chiral algebra should provide the jumping off point for a numerical bootstrap analysis of the elusive $(2,0)$ theories.

\item[$\bullet$]
Combining the extension of this story to six dimensions with the inclusion of defect operators has the potential to provide a direct explanation for the AGT relation between conformal field theory in two-dimensions and $\NN=2$ supersymmetric field theories in four dimensions.

\end{itemize}

\bigskip

\acknowledgments
The authors have benefited from discussions with N.~Arkani-Hamed, N.~Bobev, T.~Dimofte, N.~Mekareeya, J.~Maldacena, S.~Minwalla, G.~Moore, H.~Ooguri, S.~Razamat, N.~Seiberg, D.~Simmons-Duffin, E.~Sokatchev, Y.~Tachikawa, A.J.~Tolland, E.~Witten, and A. Zhiboedov. 

C.B. gratefully acknowledges the Aspen Center for Physics and NSF Grant \#1066293 for a stimulating environment during early stages of this work.
L.R. is grateful to the IAS for providing a wonderful scientific home during his sabbatical leave.
C.B., L.R., and B.v.R. would also like to thank the Kavli IPMU for warm hospitality and a stimulating environment during the workshop on Gauge and String Theory, supported in part by the EU under the Marie Curie UNIFY agreement.
The work of C.B. is supported in part by NSF grant PHY-1314311.
L. R. gratefully acknowledges the generous support of the Simons Foundation and of the Solomon Guggenheim Foundation.
The work of M.L. is supported in part by FCT - Portugal through grant SFRH/BD/70614/2010.
The work of P.L. is supported in part by SFB 647 ``Raum-Zeit-Materie. Analytische und Geometrische Strukturen''.
The work of  M.L., P.L., W.P., and L.R. is supported in part by NSF Grant  PHY-0969919.

\appendix

\section{Superconformal algebras}
\label{app:SCAs}

This appendix lists useful superconformal algebras that are used in the body of this paper. We adopt the convention of working in terms of the complexified version of symmetry algebras. We adopt bases for the complexified algebras such that the restriction to the real form that is relevant for physics in Lorentzian signature is the most natural. In general, the structures described in this paper are insensitive to the spacetime signature of the four-dimensional theory, with the caveat that we will assume that the theories in question, when Wick rotated to Lorentzian signature, are unitary.

\subsection{The four-dimensional superconformal algebra}

The spacetime symmetry algebra for $\NN=2$ superconformal field theories in four dimensions is the superalgebra $\mf{sl}(4\,|\,2)$. The maximal bosonic subalgebra is $\mf{so}(6,\Cb)\times\mf{sl}(2)_R\times\Cb^*$. The $\mf{so}(6,\Cb)$ conformal algebra, in a spinorial basis with $\aa,\aad=1,2$, is given by
%%%%%%
\be
\begin{alignedat}{4}
&[\MM_{\aa}^{~\bb},\MM_{\gg}^{\ph{\gg}\delta}]	&~=~&	\delta_{\gg}^{~\bb}\MM_{\aa}^{~\delta}-\delta_{\aa}^{~\delta}\MM_{\gg}^{~\bb}~,\\
&[\MM^{\aad}_{~\bbd},\MM^{\ggd}_{~\ddd}]			&~=~&	\delta^{\aad}_{~\dd}\MM^{\ggd}_{~\bbd}-\delta^{\ggd}_{~\bbd}\MM^{\aad}_{~\ddd}~,\\
&[\MM_{\aa}^{~\bb},\PP_{\gg\ggd}]				&~=~&	\delta_{\gg}^{~\bb}\PP_{\aa\ggd}-\tfrac12\delta_{\aa}^{\ph{\aa}\bb}\PP_{\gg\ggd}~,\\
&[\MM^{\aad}_{~\bbd},\PP_{\gg\ggd}]				&~=~&	\delta^{\aad}_{~\ggd}\PP_{\gg\bbd}-\tfrac12\delta^{\aad}_{\ph{\aad}\bbd}\PP_{\gg\ggd}~,\\
&[\MM_{\aa}^{~\bb},\KK^{\ggd\gg}]				&~=~&	-\delta_{\aa}^{~\gg}\KK^{\ggd\bb}+\tfrac12\delta_{\aa}^{\ph{\aa}\bb}\KK^{\ggd\gg}~,\\
&[\MM^{\aad}_{~\bbd},\KK^{\ggd\gg}]				&~=~&	-\delta^{\ggd}_{~\bbd}\KK^{\aad\gg}+\tfrac12\delta^{\aad}_{\ph{\aad}\bbd}\KK^{\ggd\gg}~,\\
&[\HH,\PP_{\aa\aad}]							&~=~&	\PP_{\aa\aad}~,\\
&[\HH,\KK^{\aad\aa}]							&~=~&	- \KK^{\aad\aa}~,\\
&[\KK^{\aad\aa},\PP_{\bb\bbd}]					&~=~&	\delta_{\bb}^{\ph{\bb}\aa}\delta^{\aad}_{\ph{\aad}\bbd}\HH+\delta_{\bb}^{\ph{\bb}\aa}\MM^{\aad}_{\ph{\aad}\bbd}+\delta^{\aad}_{\ph{\aad}\bbd}\MM_{\bb}^{\ph{\bb}\aa}~.
\end{alignedat}
\ee
%%%%%%
The $\mf{sl}(2)_R$ algebra has a Chevalley basis of generators $\RR^{\pm}$ and $\RR$, where
%%%%%%
\be
[\RR^+,\RR^-]=2\RR~,\qquad [\RR,\RR^{\pm}]=\pm\RR^{\pm}~.
\ee
%%%%%%
In Lorentz signature where the appropriate real form of this algebra is $\mf{su}(2)_R$, these generators will obey hermiticity conditions $(\RR^+)^{\dagger}=\RR^{-}$, $\RR^\dagger=\RR$.
The generator of the Abelian factor $\Cb^*$ is denoted by $r$ and is central in the bosonic part of the algebra. It is also convenient to introduce the basis $\RR^\II_{\ph1\JJ}$, with
%%%%%%
\be\label{eq:R-sym_change_of_basis}
\RR^1_{\ph{1}2}=\RR^+~,\qquad\RR^2_{\ph{2}1}=\RR^-~,\qquad\RR^1_{\ph{1}1}=\frac12 r+\RR~,\qquad\RR^2_{\ph{1}2}=\frac12 r-\RR~,
\ee
%%%%%%
where we follow the conventions of \cite{Dolan:2002zh} for $r$, and which obey the commutation relations
%%%%%%
\be
[\RR^\II_{\ph{\II}\JJ},\RR^{\KK}_{\ph{\KK}\LL}]=\delta^\KK_{\ph{\KK}\JJ}\RR^\II_{\ph{\II}\LL}-\delta^\II_{\ph{\II}\LL}\RR^\KK_{\ph{\KK}\JJ}~.
\ee
%%%%%%

There are sixteen fermionic generators in this superconformal algebra -- eight Poincar\'e supercharges and eight conformal supercharges -- denoted $\{\QQ^{\II}_{\aa},\,\wt\QQ_{\II\aad},\,\SS_{\JJ}^{\aa},\,\wt\SS^{\JJ\aad}\}$. The nonvanishing commutators amongst them are as follows,
%%%%%%
\be\label{eq:4dSCA}
\begin{alignedat}{4}
&\{\QQ_{\aa}^\II,\,\wt\QQ_{\JJ\aad}\}  			&~=~~&	\delta^\II_{\ph{\II}\JJ} \PP_{\aa\aad}~,\\
&\{\wt\SS^{\II\aad},\,\SS_{\JJ}^{\ph{\aa}\aa}\} &~=~~&	\delta^\II_{\ph{\II}\JJ} \KK^{\aad\aa}~,\\
&\{\QQ_{\aa}^\II,\,\SS^{\ph{\aa}\bb}_\JJ\}     	&~=~~&	\tfrac12 \delta^\II_{\ph{\II}\JJ}\delta_{\aa}^{\ph{\aa}\bb}\HH   + \delta^\II_{\ph{\II}\JJ} \MM_{\aa}^{\ph{\aa}\bb}-\delta_\aa^{\ph{\aa}\bb} \RR^\II_{\ph{\II}\JJ}~,\\
&\{\til\SS^{\II\aad},\,\til\QQ_{\JJ\bbd}\}		&~=~~&	\tfrac12 \delta^\II_{\ph{\II}\JJ}\delta^{\aad}_{\ph{\aad}\bbd}\HH + \delta^\II_{\ph{\II}\JJ} \MM^{\aad}_{\ph{\aad}\bbd}+\delta^{\aad}_{\ph{\aad}\bbd} \RR^\II_{\ph{\II}\JJ}~.
\end{alignedat}
\ee
%%%%%%
Finally, the commutators of the supercharges with the bosonic symmetry generators are the following:
%%%%%%
\be
\begin{alignedat}{4}
&[\MM_{\aa}^{~\bb},\QQ_{\gg}^\II]	&~=~&	\delta_{\gg}^{~\bb} \QQ_{\aa}^\II -\tfrac12\delta_{\aa}^{\ph{\aa}\bb} \QQ_{\gg}^\II~,\\
&[\MM^{\aad}_{~\bbd},\wt\QQ_{\II \ddd}]			&~=~& \delta^{\aad}_{~\ddd}\wt\QQ_{\II \bbd} -\tfrac12\delta^{\aad}_{\ph{\aad}\bbd}\wt\QQ_{\II \ddd}~,\\
&[\MM_{\aa}^{~\bb},\SS_{\II}^{\ph{\aa}\gg}]				&~=~&	-\delta_{\aa}^{~\gg}\SS_{\II}^{\ph{\aa}\bb}+\tfrac12\delta_{\aa}^{\ph{\aa}\bb} \SS_{\II}^{\ph{\aa}\gg}~,\\
&[\MM^{\aad}_{~\bbd},\wt\SS^{\II\ggd}]				&~=~&	-\delta^{\ggd}_{~\bbd}\wt\SS^{\II\aad}+\tfrac12\delta^{\aad}_{\ph{\aad}\bbd}\wt\SS^{\II\ggd}~,\\
&[\HH,\QQ_{\aa}^\II]							&~=~& 	\tfrac12 \QQ_{\aa}^\II~,\\
&[\HH,\wt\QQ_{\II \aad}]							&~=~& 	\tfrac12 \wt\QQ_{\II \aad}~,\\
&[\HH, \SS_{\II}^{\ph{\aa}\aa}]							&~=~&	-\tfrac12  \SS_{\II}^{\ph{\aa}\aa}~,\\
&[\HH, \til\SS^{\II\aad} ]							&~=~&	-\tfrac12 \til\SS^{\II\aad} ~,\\
&[\RR^\II_{\ph{\II}\JJ},\QQ_{\aa}^\KK]	&~=~&	\delta_{\JJ}^{~\KK} \QQ_{\aa}^\II -\frac{1}{4} \delta_{\JJ}^{\II} \QQ_{\aa}^\KK~,\\
&[\RR^\II_{\ph{\II}\JJ},\wt\QQ_{\KK \aad}]	&~=~&	-\delta_{\KK}^{~\II} \wt\QQ_{\JJ \aad} +\frac{1}{4} \delta_{\JJ}^{\II} \wt\QQ_{\KK \aad}~,\\
&[\KK^{\aad\aa},\QQ_{\bb}^\II]					&~=~&	\delta_{\bb}^{\ph{\bb}\aa}\wt\SS^{\II\aad}~,\\
&[\KK^{\aad\aa},\wt\QQ_{\II \bbd}]					&~=~&	\delta_{\bbd}^{\ph{\bbd}\aad} \SS_{\II}^{\ph{\aa}\aa}~,\\
&[\PP_{\aa\aad},\SS_{\II}^{\ph{\aa}\bb}]					&~=~&	- \delta_{\aa}^{\ph{\aa}\bb}\wt\QQ_{\II \aad}~,\\
&[\PP_{\aa\aad},\til\SS^{\II\bbd} ]					&~=~&	-\delta_{\aad}^{\ph{\aad}\bbd} \QQ_{\aa}^\II~.
\end{alignedat}
\ee
%%%%%%

\subsection{The two-dimensional superconformal algebra}

The second superalgebra of interest is $\mf{sl}(2|2)$, which corresponds to the right-moving part of the global superconformal algebra in $\NN=(0,4)$ SCFTs in two dimensions. The maximal bosonic subgroup is $\mf{sl}(2)\times \mf{sl}(2)_R$, with generators $\{L_0,L_{\pm1}\}$ for $\mf{sl}(2)$ and $\{\RR^{\pm},\RR\}$ for $\mf{sl}(2)_R$. The non-vanishing bosonic commutation relations are given by
\bea
&[\RR, \RR^\pm ]= \pm \RR^\pm~, \quad       &[\RR^+, \RR^- ] = 2 \RR~,\nonumber\\
&[\Lt_0, \Lt_{\pm 1} ] = \mp \Lt_{\pm1}~,\quad  &[\Lt_{1}, \Lt_{-1} ] = 2 \Lt_0~.\nonumber 
\eea
There are additionally right-moving Poincar\'e supercharges $\QQ^\II,\;\tilde\QQ_\JJ$ and right-moving superconformal charges $\SS_\JJ,\;\tilde\SS^\II$. The commutation relations amongst the fermionic generators are given by
%%%%%%
\bea
&\{\QQ^\II,\tilde\QQ_{\JJ}\}  &=~\delta^\II_\JJ \Lt_{-1}~,\nonumber\\
&\{\tilde\SS^\II,\SS_\JJ\}    &=~\delta^\II_\JJ \Lt_{+1}~,\nonumber\\
&\{\QQ^\II,\SS_\JJ\}      &=~\delta^\II_\JJ \Lt_0-\RR^\II_\JJ - \frac{1}{2} \delta^\II_\JJ\ZZ~,\nonumber\\
&\{\tilde\QQ_\JJ,\tilde\SS^\II\}&=~\delta^\II_\JJ \Lt_0+\RR^\II_\JJ+ \frac{1}{2} \delta^\II_\JJ\ZZ~,\nonumber
\eea
%%%%%%
where $\RR^\II_{\ph{1}\JJ}$ are defined as in \eqref{eq:R-sym_change_of_basis}, but with $r$ set to zero. Here $\ZZ$ is a central element, the removal of which gives the algebra $\mf{psl}(2|2)$. The additional commutators of bosonic symmetry generators with the supercharges are given by
\bea \label{SU(1,1|2)comm}
\begin{aligned}
\,[\Lt_{-1} \, , \til{\SS}^\II ]     	& =  -\QQ^\II ~, \\
\,[\Lt_{-1} \, , \SS_\II       ]     	& =  -\til{\QQ}_\II~,   \\
\,[\Lt_{+1} \, , \til{\QQ}_\II ]     	& =   \SS_\II~,   \\
\,[\Lt_{+1} \, , \QQ^\II       ]     	& =   \til{\SS}^\II~,\\
\,[\Lt_{0\ph{+}} \, , \til{\SS}^\II] 	& =  -\tfrac12\til{\SS}^\II~,\\
\,[\Lt_{0\ph{+}} \, , \SS_\II]	   	& =  -\tfrac12{\SS}_\II~,\\
\,[\Lt_{0\ph{+}} \, , \til{\QQ}_\II] 	& =  \tfrac12\til{\QQ}_\II~,\\
\,[\Lt_{0\ph{+}} \, , \QQ^\II 	 ] 	& =  \tfrac12{\QQ}^\II~.
\end{aligned}
\eea

\section{Shortening conditions and indices of \texorpdfstring{$\mf{su}(2,2\,|\,2)$}{su(2,2|2)}}
\label{app:shortening}
The classification of short representations of the four-dimensional $\NN=2$ superconformal algebra \cite{Dobrev:1985qv,Dolan:2002zh,Kinney:2005ej} plays a major role in the structure of the chiral algebras described in this paper. This appendix provides a review of the classification, as well as of the various indices that can be defined on any representation of the algebra that are insensitive to the recombination of collections of short multiplets into generic long multiplets. 

Short representations occur when the norm of a superconformal descendant state in what would otherwise be a long representation is rendered null by a conspiracy of quantum numbers. The unitarity bounds for a superconformal primary operator are given by
%%%%%%
\be\label{eq:unit_bounds}
\begin{alignedat}{3}
E	&\geqslant E_i~,&\qquad				&&				 j_i&\neq0~,\\
E	&=	E_i-&2~~\mbox{~or~}~~E&\geqslant& E_i~,	\qquad j_i&=0~,\\
\end{alignedat}
\ee
%%%%%%
where we have defined
%%%%%%
\be
E_1=2+2j_1+2R+ r~, \qquad E_2=2+2j_2+2R- r~,
\ee
%%%%%%
and short representations occur when one or more of these bounds are saturated. The different ways in which this can happen correspond to different combinations of Poincar\'e supercharges that will annihilate the superconformal primary state in the representation. There are two types of shortening conditions, each of which has four incarnations corresponding to an $SU(2)_R$ doublet's worth of conditions for each supercharge chirality:
%%%
\bea\label{eq:constitutent_shortening_conditions}
\BB^\II&:&\qquad \QQ^\II_{\alpha}|\psi\rangle=0~,\quad\alpha=1,2\\
{\bar \BB}_\II&:&\qquad \wt\QQ_{\II\dot\alpha}|\psi\rangle=0~,\quad\dot\alpha=1,2\\
\CC^\II&:&\qquad
\begin{cases} \epsilon^{\a\b}\QQ^\II_{\alpha}&|\psi\rangle_\beta=0~,\quad j_1\neq0\\
\epsilon^{\a\b}\QQ^\II_{\alpha}\QQ^\II_{\beta}&|\psi\rangle=0~,\quad j_1=0
\end{cases}~,\\
{\bar\CC}_\II&:&\qquad
\begin{cases} \epsilon^{\ad\bd}\wt\QQ_{\II\ad}&|\psi\rangle_\beta=0~,\quad j_2\neq0\\
\epsilon^{\ad\bd}\wt\QQ_{\II\ad}\wt\QQ_{\II\bd}&|\psi\rangle=0~,\quad j_2=0
\end{cases}~,
\eea
%%%
The different admissible combinations of shortening conditions that can be simultaneously realized by a single unitary representation are summarized in Table \ref{Tab:shortening}, where the reader can also find the precise relations that must be satisfied by the quantum numbers $(E,j_1,j_2,r,R)$ of the superconformal primary operator, as well as the notations used to designate the different representations in \cite{Dolan:2002zh} (DO) and \cite{Kinney:2005ej} (KMMR).\footnote{We follow the R-charge conventions of DO.}

\begin{table}[t]
\begin{centering}
\renewcommand{\arraystretch}{1.3}
\begin{tabular}{|l|l|l|l|}
\hline
Shortening & Quantum Number Relations & DO & KMMR \tabularnewline
\hline
\hline 
$\varnothing$						& \makebox[3.8cm][l]{$E\geqslant2R+r$}									  	& $\AA^\Delta_{R,r(j_1,j_2)}$ & ${\bf aa}_{\Delta,j_1,j_2,r,R}$ 	\tabularnewline
\hline 
$\BB^1$ 							& \makebox[3.8cm][l]{$E=2R+r$}			\makebox[3cm][l]{$j_1=0$}	  	& $\BB_{R,r(0,j_2)}$ 		& ${\bf ba}_{0,j_2,r,R}$ 		\tabularnewline
\hline 
$\bar\BB_2$							& \makebox[3.8cm][l]{$E=2R-r$}			\makebox[3cm][l]{$j_2=0$}   	& $\bar{\BB}_{R,r(j_1,0)}$ 	& ${\bf ab}_{j_1,0,r,R}$ 		\tabularnewline
\hline 
$\BB^1\cap\BB^2$  					& \makebox[3.8cm][l]{$E=r$}  			\makebox[3cm][l]{$R=0$}  		& $\EE_{r(0,j_2)}$ 			& ${\bf ba}_{0,j_2,r,0}$ 		\tabularnewline
\hline
$\bar\BB_1\cap\bar\BB_2$  			& \makebox[3.8cm][l]{$E=-r$}  			\makebox[3cm][l]{$R=0$}  		& $\bar \EE_{r(j_1,0)}$ 	& ${\bf ab}_{j_1,0,r,0}$ 		\tabularnewline
\hline 
$\BB^1\cap\bar\BB_{2}$  			& \makebox[3.8cm][l]{$E=2R$}  			\makebox[3cm][l]{$j_1=j_2=r=0$}	& $\hat{\BB}_{R}$ 			& ${\bf bb}_{0,0,0,R}$ 			\tabularnewline
\hline\hline 
$\CC^1$ 							& \makebox[3.8cm][l]{$E=2+2j_1+2R+r$}  									& $\CC_{R,r(j_1,j_2)}$ 		& ${\bf ca}_{j_1,j_2,r,R}$ 		\tabularnewline
\hline 
$\bar\CC_2$  						& \makebox[3.8cm][l]{$E=2+2 j_2+2R-r$}  								& $\bar\CC_{R,r(j_1,j_2)}$ 	& ${\bf ac}_{j_1,j_2,r,R}$		\tabularnewline
\hline
$\CC^1\cap\CC^2$  					& \makebox[3.8cm][l]{$E=2+2j_1+r$}  	\makebox[3cm][l]{$R=0$}  		& $\CC_{0,r(j_1,j_2)}$ 		& ${\bf ca}_{j_1,j_2,r,0}$ 		\tabularnewline
\hline 
$\bar\CC_1\cap\bar\CC_2$			& \makebox[3.8cm][l]{$E=2+2 j_2-r$} 	\makebox[3cm][l]{$R=0$}  	 	& $\bar\CC_{0,r(j_1,j_2)}$ 	& ${\bf ac}_{j_1,j_2,r,0}$ 		\tabularnewline
\hline 
$\CC^1\cap\bar\CC_2$  				& \makebox[3.8cm][l]{$E=2+2R+j_1+j_2$}	\makebox[3cm][l]{$r=j_2-j_1$}   & $\hat{\CC}_{R(j_1,j_2)}$ 	& ${\bf cc}_{j_1,j_2,j_2-j_1,R}$\tabularnewline
\hline\hline 
$\BB^1\cap\bar\CC_2$  				& \makebox[3.8cm][l]{$E=1+2R+j_2$}  	\makebox[3cm][l]{$r=j_2+1$}   	& $\DD_{R(0,j_2)}$ 			& ${\bf bc}_{0,j_2,j_2+1,R}$	\tabularnewline
\hline 
$\bar\BB_2\cap\CC^1$  				& \makebox[3.8cm][l]{$E=1+2R+j_1$}		\makebox[3cm][l]{$-r=j_1+1$}    & $\bar\DD_{R(j_1,0)}$ 		& ${\bf cb}_{j_1,0,-j_1-1,R}$ 	\tabularnewline
\hline 
$\BB^1\cap\BB^2\cap\bar\CC_2$  		& \makebox[3.8cm][l]{$E=r=1+j_2$} 		\makebox[2.5cm][l]{$r=j_2+1$} 	\makebox[1.5cm][l]{$R=0$}	& $\DD_{0(0,j_2)}$ 		& ${\bf bc}_{0,j_2,j_2+1,0}$  \tabularnewline
\hline
$\CC^1\cap\bar\BB_1\cap\bar\BB_2$  	& \makebox[3.8cm][l]{$E=-r=1+j_1$}  	\makebox[2.5cm][l]{$-r=j_1+1$} 	\makebox[1.5cm][l]{$R=0$}	& $\bar\DD_{0(j_1,0)}$ 	& ${\bf cb}_{j_1,0,-j_1-1,0}$ \tabularnewline
\hline
\end{tabular}
\par\end{centering}
\caption{\label{Tab:shortening}Unitary irreducible representations of the $\NN=2$ superconformal algebra.}
\end{table}
%%%%%%

At the level of group theory, it is possible for a collection of short representations to recombine into a generic long representation whose dimension is equal to one of the unitarity bounds of \eqref{eq:unit_bounds}. In the DO notation, the generic recombinations are as follows:
%%%%%%
\bea\label{eq:recombination}
\AA_{R,r(j_1,j_2)}^{2R+r+2+2j_1}&\simeq& \CC_{R,r(j_1,j_2)}\oplus \CC_{R+\hf,r+\hf(j_1-\hf,j_2)}~,\\
\AA_{R,r(j_1,j_2)}^{2R-r+2+2j_2}&\simeq& \bar\CC_{R,r(j_1,j_2)}\oplus \bar\CC_{R+\hf,r-\hf(j_1,j_2-\hf)}~,\\
\AA_{R,j_1-j_2(j_1,j_2)}^{2R+j_1+j_2+2}&\simeq& \hat\CC_{R(j_1,j_2)}\oplus \hat\CC_{R+\hf(j_1-\hf,j_2)}\oplus\hat\CC_{R+\hf(j_1,j_2-\hf)}\oplus\hat\CC_{R+1(j_1-\hf,j_2-\hf)}~.
\eea
%%%%%%
There are special cases when the quantum numbers of the long multiplet at threshold are such that some Lorentz quantum numbers in \eqref{eq:recombination} would be negative and unphysical:
%%%%%%
\bea\label{eq:special_recombination}
\AA_{R,r(0,j_2)}^{2R+r+2} 			& \simeq & 			\CC_{R,r(0,j_2)} 		\oplus 		\BB_{R+1,r+\hf(0,j_2)}~,\\
\AA_{R,r(j_1,0)}^{2R-r+2} 			& \simeq & 			\bar\CC_{R,r(j_1,0)} 	\oplus 		\bar\BB_{R+1,r-\hf(j_1,0)}~,\\
\AA_{R,-j_2(0,j_2)}^{2R+j_2+2} 			& \simeq & 			\hat\CC_{R(0,j_2)} 	\oplus 		\DD_{R+1(0,j_2)} \oplus 		\hat\CC_{R+\hf(0,j_2-\hf)} 		\oplus 		\DD_{R+\frac{3}{2}(0,j_2-\hf)}~,\\
\AA_{R,j_1(j_1,0)}^{2R+j_1+2} 			& \simeq & 			\hat\CC_{R(j_1,0)} 	\oplus 		\hat\CC_{R+\hf(j_1-\hf,0)} \oplus 		\bar\DD_{R+1(j_1,0)} 	\oplus 		\bar\DD_{R+\frac{3}{2}(j_1-\hf,0)}~,\\
\AA_{R,0(0,0)}^{2R+2} 			& \simeq & 			\hat\CC_{R(0,0)} 	\oplus 		\DD_{R+1(0,0)} 	\oplus 		\bar\DD_{R+1(0,0)} 	\oplus 		\hat\BB_{R+2}~.
\eea
%%%%%%
The last three recombinations involve multiplets that make an appearance in the associated chiral algebra described in this work. Note that the $\EE$, $\bar\EE$, $\hat  {\cal B}_{\frac{1}{2}}$, 
$\hat  {\cal B}_{1}$, $\hat  {\cal B}_{\frac{3}{2}}$, $\DD_0$,  $\bar\DD_0$, $\DD_{\frac{1}{2}}$ and $\bar\DD_{\frac{1}{2}}$ multiplets can never recombine, along with $\BB_{\frac12,r(0,j_2)}$ and $\bar\BB_{\frac12,r(j_1,0)}$.

There exist a variety of trace formulas \cite{Kinney:2005ej,Gadde:2011uv} that can be defined on the Hilbert space of an $\NN=2$ SCFT such that the result receives contributions only from states that lie in short representations of the superconformal algebra, with the contributions being such that the indices are insensitive to recombinations. The indices are defined and named as follows:
%%%%%%
\bea
\mbox{Superconformal Index}~&:&\qquad \Tr_{\HH}(-1)^F p^{\frac12(E+2j_1-2R-r)}q^{\frac12(E-2j_1-2R-r)}t^{R+r}   ,\\
\mbox{Macdonald}~&:&\qquad \Tr_{\HH_{\rm M}}(-1)^F q^{\frac12(E-2j_1-2R-r)}t^{R+r}~,\\
\mbox{Schur}~&:&\qquad \Tr_{\HH}(-1)^Fq^{E-R}~,\\
\mbox{Hall-Littlewood}~&:&\qquad \Tr_{\HH_{\rm HL}}(-1)^F\tau^{2E-2R}~,\\
\mbox{Coulomb}~&:&\qquad \Tr_{\HH_{\rm C}}(-1)^F\sigma^{\frac12(E+2j_1-2R-r)}\rho^{\frac12(E-2j_1-2R-r)}~.
\eea
%%%%%%
The specialized Hilbert spaces appearing in the trace formulas above are defined as follows,
%%%%%%
\bea
\HH_{\rm M}	&\colonequals&\{\psi\in\HH~\big|~ E+2j_1-2R-r=0\}~,\\
\HH_{\rm HL}&\colonequals&\{\psi\in\HH~\big|~ E-2R-r=0~,j_1=0\}~,\\
\HH_{\rm C}	&\colonequals&\{\psi\in\HH~\big|~ E+2j_1+r=0\}~.
\eea
%%%%%%
The different indices are sensitive to different superconformal multiplets. In particular, the Coulomb index counts only $\EE$ and $\DD_0$ type multiplets. These can be thought of as $\NN=1$ chiral ring operators that are $SU(2)_R$ singlets. Similarly, the Hall-Littlewood index counts only $\hat\BB_R$ and $\DD_R$ multiplets, which can be thought of as the consistent truncation of the $\NN=1$ chiral ring to operators that are neutral under $U(1)_r$. The Schur and Macdonald indices count only the operators that are involved in the chiral algebras of this paper: $\hat\BB_R$, $\hat\CC_R$, $\DD$, and $\bar\DD$ multiplets. The full index receives contributions from all of the multiplets appearing in Table \ref{Tab:shortening}.

\section{Kazhdan-Lusztig polynomials and affine characters}
\label{app:KL}

Computing the characters of irreducible modules of an affine Lie algebra at a negative integer level is a nontrivial task. For low levels, the multiplicity and norms of states can be found by hand using the mode expansion of the affine currents $J^A (z)$, but this computation quickly becomes rather involved. Fortunately there exists another method to compute these characters, based on the work of Kazhdan and Lusztig \cite{kazhdan1979}, which (with the aid of a computer) can produce results to very high order. In this appendix we give a brief introduction to this method. The interested reader is referred to, \eg, \cite{Fuchs:1997jv,DeVos:1995an} for more details. 

A generic method to obtain an irreducible representation of any (affine) Lie algebra is to start with the Verma module $M$ built on a certain highest weight state $\psi_{h.w.}$, and then to subtract away all the null states in this module with the correct multiplicities. Let us recall that according to the Poincar\'e-Birkhoff-Witt theorem, the Verma module is spanned by all the states of the form
%%%%%%
\be
(E^{-\alpha_1,1})^{n_{1,1}} (E^{-\alpha_1,2})^{n_{1,2}} \ldots (E^{-\alpha_1,m_1})^{n_{1,m_1}} \ldots (E^{-\alpha_2,1})^{n_{2,1}}\ldots (E^{-\alpha_N,m_N})^{n_{N,m_N}} \psi_{h.w.}~,
\ee
%%%%%%
with nonnegative integer coefficients $n_{i,j}$. Here the $E^{-\alpha,k_\alpha}$ are the negative roots with weight $-\alpha$, and the auxiliary index $k_\alpha \in \{1,\ldots,m_\alpha\}$ is only necessary when the multiplicity $m_\alpha$ of the given weight is greater than one. The ordering of the roots in the above equation is arbitrary but fixed. If the highest weight state $\psi_{h.w.}$ has weight $\mu$ then the state defined as above has weight
%%%%%%
\be
\mu - \alpha_1 (n_{1,1} + n_{1,2} + \ldots + n_{1,m_1}) - \alpha_2 (n_{2,1} + \ldots) - \ldots - \alpha_N( \ldots + n_{N,m_N})~,
\ee
%%%%%%
and with a moment's thought one sees that the character $M_{\mu}$ of the Verma module is given by
%%%%%%
\be\label{Kostantpar}
\text{char} M_\mu = e^{\mu} \prod_{\a > 0} (1 - e^{-\alpha})^{-\text{mult}(\alpha)}~.
\ee
%%%%%%
This is the \emph{Kostant partition function}. The product is taken over the set of all the positive roots, which is infinite for an affine Lie algebra.

For a given affine Lie algebra there are special values of the highest weights for which the Verma module becomes reducible due to the existence of null states. We need to subtract all these null states to recover the irreducible module. Since any descendant of a null state is also null, the null states are themselves organized into Verma modules and we can subtract away entire modules at a time. This procedure is further complicated by the existence of ``nulls of nulls'', \ie, null states inside the Verma module that we are subtracting. In general, this leads to a rather intricate pattern of subtractions. It follows that the character of the irreducible module with highest weight $\lambda$, which we denote as $L_\lambda$, can be obtained from a possibly infinite sum of the form
%%%%%%
\be
\label{decompirredchar}
\text{char} L_\lambda = \sum_{\mu \leqslant \lambda} m_{\lambda,\mu} \text{char} M_\mu~,
\ee
%%%%%%
where the integers $m_{\lambda,\mu}$ are not of definite sign and reflect the aforementioned pattern of null states. Of course $m_{\lambda,\lambda} = 1$. The vectors labeled by $\mu$ in the above sum are called the \emph{primitive null vectors} of the Verma module $M_\lambda$.

This leaves us with the task of determining the weights $\mu$ that appear in \eqref{decompirredchar} along with their associated multiplicities $m_{\lambda,\mu}$. The first task is accomplished by noting that these weights are necessarily annihilated by all raising operators, and therefore must be highest weight states in themselves. The quadratic Casimir operator of an affine Lie algebra acts simply on highest weight states with weight $\mu$ as multiplication by $|\mu + \rho|^2$, where $\rho$ is the Weyl vector with unit Dynkin labels. On the other hand, the eigenvalue should be an invariant of the full representation, which means that the only states $\mu$ that can appear in \eqref{decompirredchar} have to satisfy
%%%%%%
\be \label{Casimirequality}
|\mu + \rho|^2 = |\lambda + \rho|^2~.
\ee
%%%%%%
Notice that so far we have made no distinction between unitary representations, where the highest weight $\lambda$ is dominant integral (\ie, its Dynkin labels are nonnegative integers), and non-unitary representations like the ones in which we are interested. This distinction becomes crucial in the computation of the multiplicities $m_{\lambda,\mu}$.

For the irreducible representations associated to dominant integral weights, the weight multiplicities are invariant under the action of the Weyl group, and correspondingly $\text{char} L_{\lambda}$ is invariant under the action of the Weyl group on the fugacities. On the other hand, the Kostant partition function is essentially \emph{odd} under this action (\cf\ \cite{Fuchs:1997jv}),
%%%%%%
\be
w( e^{-\rho - \mu} \text{char} M_{\mu}) = \text{sign}(w) e^{-\rho - \mu} \text{char} M_{\mu}~,
\ee
%%%%%%
where the sign of an element $w$ in the Weyl group is simply given by $-1$ raised to the power of the number of generators used to express $w$. One can easily convince oneself that the multiplicities $m_{\lambda,\mu}$ therefore necessarily satisfy
%%%%%%
\be
m_{\lambda,\mu} = \text{sign}(w) m_{\lambda,w \cdot \mu}~,
\ee
%%%%%%
where $w\cdot\mu\colonequals w(\mu+\rho)-\rho$ is the shifted action of the Weyl group on the weight $\mu$. All the multiplicities $m_{\lambda,\mu}$ for weights $\mu$ on the same shifted Weyl orbit are therefore related by factors of $\text{sign}(w)$, and it suffices to know only one multiplicity on each orbit. Happily, if the highest weight $\lambda$ is dominant integral, then it lies on the shifted Weyl orbit of \emph{any} primitive null vector. This essentially follows from the fact that there is a unique dominant integral weight on every shifted Weyl orbit, and from \eqref{Casimirequality} it can be shown that this has to be $\lambda$. So, using that $m_{\lambda,\lambda} = 1$, we find that all the weights appearing in \eqref{decompirredchar} are given by the shifted Weyl orbit of $\lambda$ and have multiplicities equal to $\text{sign}(w)$. In summary, then,
%%%%%%
\be
\text{char} L_\lambda = \frac{\sum_{w \in W} \text{sign}(w) e^{w(\rho + \lambda) - \rho}}{\prod_{\a > 0} (1 - e^{-\alpha})^{\text{mult}(\alpha)}}~,
\ee
%%%%%%
which is the famous Weyl-Kac character formula.

Let us return to the case where the $\lambda$ is not dominant integral. This is the case that interests us: indeed, for $\mf{so}(8)_{-2}$ the vacuum representation has Dynkin labels $[-2\,0\,0\,0\,0]$ and the zeroth Dynkin label is not positive.\footnote{Recall that the zeroth Dynkin label for a weight vector in an affine Lie algebra $\hat{\mf{g}}$ is given by $k - (\lambda,\theta)$ with $\lambda$ the part of the weight vector corresponding to the original Lie algebra $\mf{g}$ and $\theta$ the highest root of $\mf{g}$.} For non-dominant integral weights the above derivation already fails at the very first step: the weight multiplicities in the irreducible representation are not invariant under the action of the Weyl group. This is most easily seen by taking the infinite irreducible representation of $\mf{su}(2)$ whose highest weight is negative. In this case the single Weyl reflection maps the highest weight, which of course has multiplicity one, to a positive weight, which has multiplicity zero. The derivation of the coefficients $m_{\lambda,\mu}$ now becomes considerably more involved. Since we find qualitative differences depending on the sign of $k + h^\vee$, we will in the remainder of this appendix focus on the relevant case $k + h^\vee > 0$.

For the non-unitary representations considered here it is still true that all the primitive null vectors lie on the shifted Weyl orbit of the highest weight $\lambda$, and for $k+h^\vee > 0$ there is still a unique \emph{dominant} weight $\Lambda$ on the same orbit such that $\Lambda + \rho$ has nonnegative Dynkin labels. For example, for the vacuum module of $\mf{so}(8)_{-2}$ the dominant weight has Dynkin labels $[0\,0-\!1\,0\,0]$ which happens to be related to $[-2\,0\,0\,0\,0]$ by a single elementary reflection. All the weights in \eqref{decompirredchar}, including $\lambda$ itself, can thus be written as $\mu = w \cdot \Lambda$ for some Weyl element $w$. We can therefore alternatively try to label these weights with the corresponding element of the Weyl group $w$ instead of $\mu$. We will see that such a relabeling has great benefits, but first we need to mention two important subtleties.

The first subtlety concerns the fact that we may restrict ourselves to elementary reflections of the Weyl group for which the corresponding Dynkin label in $\Lambda$ is integral, since it is only in those cases that null states can possibly appear. These reflections generate a subgroup of the Weyl group that we will denote as $W_\Lambda$. In the case of $\mf{so}(8)_{-2}$ the weights are all integral and $W_\Lambda = W$. The second subtlety is the possibility of the existence of a subgroup $W^0_\Lambda$ of $W_\Lambda$ that leaves $\Lambda$ invariant. This happens precisely when some of the Dynkin labels of $\Lambda + \rho$ are zero - in our case there is a single such zero. It is clear that the weights $\mu$ can then at best be uniquely labeled by elements of the coset $W_\Lambda / W^0_\Lambda$.

It is now a deep result that the multiplicities $m_{\lambda,\mu}$ depend on the dominant integral weight $\Lambda$ \emph{only} through the corresponding elements $w$ and $w'$ of the coset $W_\Lambda / W^0_\Lambda$. We may therefore replace
%%%%%%
\be
m_{\lambda,\mu} \to m_{w,w'}~,
\ee
%%%%%%
where $\lambda = w \cdot \Lambda$, $\mu = w' \cdot \Lambda$ and $w$ and $w'$ are elements of the coset. The celebrated \emph{Kazhdan-Lusztig conjecture} tells us that the precise form of these multiplicities is given by
%%%%%%
\be\label{KLmults}
m_{w,w'} = \tilde Q_{w,w'} (1)~.
\ee
%%%%%%
where the Kazhdan-Lusztig polynomial $\tilde Q_{w,w'}(q)$ is a single-variable polynomial depending on two elements $w$ and $w'$ of the coset $W_\Lambda/ W^0_\Lambda$. These polynomials are determined via rather intricate recursion relations that are explained in detail in \cite{DeVos:1995an}. For $k+ h^\vee > 0$ and integral weights, which is the case that interests us here, the Kazhdan-Lusztig conjecture was proven in \cite{kashiwara1990kazhdan,casian1990kazhdan}.

For the computations mentioned in the main text, we have implemented the recursive definitions of the Kazhdan-Lusztig polynomials on cosets given in \cite{DeVos:1995an} in \texttt{Mathematica}. Equations \eqref{Kostantpar}, \eqref{decompirredchar}, and \eqref{KLmults} then allow us to compute all the states in the irreducible vacuum character of $\mf{so}(8)_{-2}$ up to level five. The results are shown in Table \ref{tab:so8_index}.

\providecommand{\href}[2]{#2}

\begingroup
\raggedright

\endgroup


\begin{thebibliography}{10}

\bibitem{Polyakov:1974gs}
A.~Polyakov, {\it {Nonhamiltonian approach to conformal quantum field theory}},
   {\em Zh.Eksp.Teor.Fiz.} {\bf 66} (1974) 23--42.

\bibitem{Ferrara:1973yt}
S.~Ferrara, A.~Grillo, and R.~Gatto, {\it {Tensor representations of conformal
  algebra and conformally covariant operator product expansion}},  {\em Annals
  Phys.} {\bf 76} (1973) 161--188.

\bibitem{Rattazzi:2008pe}
R.~Rattazzi, V.~S. Rychkov, E.~Tonni, and A.~Vichi, {\it {Bounding scalar
  operator dimensions in 4D CFT}},  {\em JHEP} {\bf 0812} (2008) 031,
  [\href{http://xxx.lanl.gov/abs/0807.0004}{{\tt arXiv:0807.0004}}].

\bibitem{ElShowk:2012ht}
S.~El-Showk, M.~F. Paulos, D.~Poland, S.~Rychkov, D.~Simmons-Duffin, {\em
  et.~al.}, {\it {Solving the 3D Ising Model with the Conformal Bootstrap}},
  {\em Phys.Rev.} {\bf D86} (2012) 025022,
  [\href{http://xxx.lanl.gov/abs/1203.6064}{{\tt arXiv:1203.6064}}].

\bibitem{Poland:2010wg}
D.~Poland and D.~Simmons-Duffin, {\it {Bounds on 4D Conformal and
  Superconformal Field Theories}},  {\em JHEP} {\bf 1105} (2011) 017,
  [\href{http://xxx.lanl.gov/abs/1009.2087}{{\tt arXiv:1009.2087}}].

\bibitem{Poland:2011ey}
D.~Poland, D.~Simmons-Duffin, and A.~Vichi, {\it {Carving Out the Space of 4D
  CFTs}},  {\em JHEP} {\bf 1205} (2012) 110,
  [\href{http://xxx.lanl.gov/abs/1109.5176}{{\tt arXiv:1109.5176}}].

\bibitem{Beem:2013qxa}
C.~Beem, L.~Rastelli, and B.~C. van Rees, {\it {The N=4 Superconformal
  Bootstrap}},  {\em Phys.Rev.Lett.} {\bf 111} (2013) 071601,
  [\href{http://xxx.lanl.gov/abs/1304.1803}{{\tt arXiv:1304.1803}}].

\bibitem{Alday:2013opa}
L.~F. Alday and A.~Bissi, {\it {The superconformal bootstrap for structure
  constants}},  \href{http://xxx.lanl.gov/abs/1310.3757}{{\tt
  arXiv:1310.3757}}.

\bibitem{WIP_6d}
C.~Beem, L.~Rastelli, and B.~C. van Rees, {\it {W Symmetry in six dimensions}},
   \href{http://xxx.lanl.gov/abs/1404.1079}{{\tt arXiv:1404.1079}}.

\bibitem{Dolan:2006ec}
F.~Dolan, M.~Nirschl, and H.~Osborn, {\it {Conjectures for large N
  superconformal N=4 chiral primary four point functions}},  {\em Nucl.Phys.}
  {\bf B749} (2006) 109--152,
  [\href{http://xxx.lanl.gov/abs/hep-th/0601148}{{\tt hep-th/0601148}}].

\bibitem{Eden:2000bk}
B.~Eden, A.~C. Petkou, C.~Schubert, and E.~Sokatchev, {\it {Partial
  nonrenormalization of the stress tensor four point function in N=4 SYM and
  AdS / CFT}},  {\em Nucl.Phys.} {\bf B607} (2001) 191--212,
  [\href{http://xxx.lanl.gov/abs/hep-th/0009106}{{\tt hep-th/0009106}}].

\bibitem{Eden:2001ec}
B.~Eden and E.~Sokatchev, {\it {On the OPE of 1/2 BPS short operators in N=4
  SCFT(4)}},  {\em Nucl.Phys.} {\bf B618} (2001) 259--276,
  [\href{http://xxx.lanl.gov/abs/hep-th/0106249}{{\tt hep-th/0106249}}].

\bibitem{Dolan:2001tt}
F.~Dolan and H.~Osborn, {\it {Superconformal symmetry, correlation functions
  and the operator product expansion}},  {\em Nucl.Phys.} {\bf B629} (2002)
  3--73, [\href{http://xxx.lanl.gov/abs/hep-th/0112251}{{\tt hep-th/0112251}}].

\bibitem{Heslop:2002hp}
P.~Heslop and P.~Howe, {\it {Four point functions in N=4 SYM}},  {\em JHEP}
  {\bf 0301} (2003) 043, [\href{http://xxx.lanl.gov/abs/hep-th/0211252}{{\tt
  hep-th/0211252}}].

\bibitem{Dolan:2004mu}
F.~A. Dolan, L.~Gallot, and E.~Sokatchev, {\it {On four-point functions of
  1/2-BPS operators in general dimensions}},  {\em JHEP} {\bf 0409} (2004) 056,
  [\href{http://xxx.lanl.gov/abs/hep-th/0405180}{{\tt hep-th/0405180}}].

\bibitem{Nirschl:2004pa}
M.~Nirschl and H.~Osborn, {\it {Superconformal Ward identities and their
  solution}},  {\em Nucl.Phys.} {\bf B711} (2005) 409--479,
  [\href{http://xxx.lanl.gov/abs/hep-th/0407060}{{\tt hep-th/0407060}}].

\bibitem{Drukker:2009sf}
N.~Drukker and J.~Plefka, {\it {Superprotected n-point correlation functions of
  local operators in N=4 super Yang-Mills}},  {\em JHEP} {\bf 0904} (2009) 052,
  [\href{http://xxx.lanl.gov/abs/0901.3653}{{\tt arXiv:0901.3653}}].

\bibitem{deMedeiros:2001kx}
P.~de~Medeiros, C.~M. Hull, B.~J. Spence, and J.~M. Figueroa-O'Farrill, {\it
  {Conformal topological Yang-Mills theory and de Sitter holography}},  {\em
  JHEP} {\bf 0208} (2002) 055,
  [\href{http://xxx.lanl.gov/abs/hep-th/0111190}{{\tt hep-th/0111190}}].

\bibitem{Kinney:2005ej}
J.~Kinney, J.~M. Maldacena, S.~Minwalla, and S.~Raju, {\it {An Index for 4
  dimensional super conformal theories}},  {\em Commun.Math.Phys.} {\bf 275}
  (2007) 209--254, [\href{http://xxx.lanl.gov/abs/hep-th/0510251}{{\tt
  hep-th/0510251}}].

\bibitem{Gadde:2011ik}
A.~Gadde, L.~Rastelli, S.~S. Razamat, and W.~Yan, {\it {The 4d Superconformal
  Index from q-deformed 2d Yang-Mills}},  {\em Phys.Rev.Lett.} {\bf 106} (2011)
  241602, [\href{http://xxx.lanl.gov/abs/1104.3850}{{\tt arXiv:1104.3850}}].

\bibitem{Gadde:2011uv}
A.~Gadde, L.~Rastelli, S.~S. Razamat, and W.~Yan, {\it {Gauge Theories and
  Macdonald Polynomials}},  {\em Commun.Math.Phys.} {\bf 319} (2013) 147--193,
  [\href{http://xxx.lanl.gov/abs/1110.3740}{{\tt arXiv:1110.3740}}].

\bibitem{WIP_N2_Numerics}
C.~Beem, M.~Lemos, P.~Liendo, L.~Rastelli, and B.~van Rees, {\it {To Appear}}.

\bibitem{Cardy:1988cwa}
J.~L. Cardy, {\it {Is There a c Theorem in Four-Dimensions?}},  {\em
  Phys.Lett.} {\bf B215} (1988) 749--752.

\bibitem{Komargodski:2011vj}
Z.~Komargodski and A.~Schwimmer, {\it {On Renormalization Group Flows in Four
  Dimensions}},  {\em JHEP} {\bf 1112} (2011) 099,
  [\href{http://xxx.lanl.gov/abs/1107.3987}{{\tt arXiv:1107.3987}}].

\bibitem{Bouwknegt:1992wg}
P.~Bouwknegt and K.~Schoutens, {\it {W symmetry in conformal field theory}},
  {\em Phys.Rept.} {\bf 223} (1993) 183--276,
  [\href{http://xxx.lanl.gov/abs/hep-th/9210010}{{\tt hep-th/9210010}}].

\bibitem{Johansen:1994ud}
A.~Johansen, {\it {Infinite conformal algebras in supersymmetric theories on
  four manifolds}},  {\em Nucl.Phys.} {\bf B436} (1995) 291--341,
  [\href{http://xxx.lanl.gov/abs/hep-th/9407109}{{\tt hep-th/9407109}}].

\bibitem{Kapustin:2006hi}
A.~Kapustin, {\it {Holomorphic reduction of N=2 gauge theories, Wilson-'t Hooft
  operators, and S-duality}},
  \href{http://xxx.lanl.gov/abs/hep-th/0612119}{{\tt hep-th/0612119}}.

\bibitem{Dolan:2002zh}
F.~Dolan and H.~Osborn, {\it {On short and semi-short representations for
  four-dimensional superconformal symmetry}},  {\em Annals Phys.} {\bf 307}
  (2003) 41--89, [\href{http://xxx.lanl.gov/abs/hep-th/0209056}{{\tt
  hep-th/0209056}}].

\bibitem{Maldacena:2011jn}
J.~Maldacena and A.~Zhiboedov, {\it {Constraining Conformal Field Theories with
  A Higher Spin Symmetry}},  {\em J.Phys.} {\bf A46} (2013) 214011,
  [\href{http://xxx.lanl.gov/abs/1112.1016}{{\tt arXiv:1112.1016}}].

\bibitem{Argyres:2007cn}
P.~C. Argyres and N.~Seiberg, {\it {S-duality in N=2 supersymmetric gauge
  theories}},  {\em JHEP} {\bf 0712} (2007) 088,
  [\href{http://xxx.lanl.gov/abs/0711.0054}{{\tt arXiv:0711.0054}}].

\bibitem{Thielemans:1991uw}
K.~Thielemans, {\it {A Mathematica package for computing operator product
  expansions}},  {\em Int.J.Mod.Phys.} {\bf C2} (1991) 787--798.

\bibitem{Goddard:1987td}
P.~Goddard, D.~I. Olive, and G.~Waterson, {\it {Superalgebras, Symplectic
  Bosons and the Sugawara Construction}},  {\em Commun.Math.Phys.} {\bf 112}
  (1987) 591.

\bibitem{Friedan:1985ge}
D.~Friedan, E.~J. Martinec, and S.~H. Shenker, {\it {Conformal Invariance,
  Supersymmetry and String Theory}},  {\em Nucl.Phys.} {\bf B271} (1986) 93.

\bibitem{Kausch:1995py}
H.~G. Kausch, {\it {Curiosities at c = -2}},
  \href{http://xxx.lanl.gov/abs/hep-th/9510149}{{\tt hep-th/9510149}}.

\bibitem{Bhardwaj:2013qia}
L.~Bhardwaj and Y.~Tachikawa, {\it {Classification of 4d N=2 gauge theories}},
  \href{http://xxx.lanl.gov/abs/1309.5160}{{\tt arXiv:1309.5160}}.

\bibitem{Karabali:1989dk}
D.~Karabali and H.~J. Schnitzer, {\it {BRST Quantization of the Gauged WZW
  Action and Coset Conformal Field Theories}},  {\em Nucl.Phys.} {\bf B329}
  (1990) 649.

\bibitem{Liendo:2011xb}
P.~Liendo, E.~Pomoni, and L.~Rastelli, {\it {The Complete One-Loop Dilation
  Operator of N=2 SuperConformal QCD}},  {\em JHEP} {\bf 1207} (2012) 003,
  [\href{http://xxx.lanl.gov/abs/1105.3972}{{\tt arXiv:1105.3972}}].

\bibitem{Baggio:2012rr}
M.~Baggio, J.~de~Boer, and K.~Papadodimas, {\it {A non-renormalization theorem
  for chiral primary 3-point functions}},  {\em JHEP} {\bf 1207} (2012) 137,
  [\href{http://xxx.lanl.gov/abs/1203.1036}{{\tt arXiv:1203.1036}}].

\bibitem{Gawedzki:1988nj}
K.~Gawedzki and A.~Kupiainen, {\it {Coset Construction from Functional
  Integrals}},  {\em Nucl.Phys.} {\bf B320} (1989) 625.

\bibitem{Cvitanovic:2008zz}
P.~Cvitanovic, {\it {Group theory: Birdtracks, Lie's and exceptional groups}}.

\bibitem{Sen:1996vd}
A.~Sen, {\it {F theory and orientifolds}},  {\em Nucl.Phys.} {\bf B475} (1996)
  562--578, [\href{http://xxx.lanl.gov/abs/hep-th/9605150}{{\tt
  hep-th/9605150}}].

\bibitem{Banks:1996nj}
T.~Banks, M.~R. Douglas, and N.~Seiberg, {\it {Probing F theory with branes}},
  {\em Phys.Lett.} {\bf B387} (1996) 278--281,
  [\href{http://xxx.lanl.gov/abs/hep-th/9605199}{{\tt hep-th/9605199}}].

\bibitem{Dasgupta:1996ij}
K.~Dasgupta and S.~Mukhi, {\it {F theory at constant coupling}},  {\em
  Phys.Lett.} {\bf B385} (1996) 125--131,
  [\href{http://xxx.lanl.gov/abs/hep-th/9606044}{{\tt hep-th/9606044}}].

\bibitem{Minahan:1996fg}
J.~A. Minahan and D.~Nemeschansky, {\it {An N=2 superconformal fixed point with
  E(6) global symmetry}},  {\em Nucl.Phys.} {\bf B482} (1996) 142--152,
  [\href{http://xxx.lanl.gov/abs/hep-th/9608047}{{\tt hep-th/9608047}}].

\bibitem{Minahan:1996cj}
J.~A. Minahan and D.~Nemeschansky, {\it {Superconformal fixed points with E(n)
  global symmetry}},  {\em Nucl.Phys.} {\bf B489} (1997) 24--46,
  [\href{http://xxx.lanl.gov/abs/hep-th/9610076}{{\tt hep-th/9610076}}].

\bibitem{Aharony:1998xz}
O.~Aharony, A.~Fayyazuddin, and J.~M. Maldacena, {\it {The Large N limit of
  N=2, N=1 field theories from three-branes in F theory}},  {\em JHEP} {\bf
  9807} (1998) 013, [\href{http://xxx.lanl.gov/abs/hep-th/9806159}{{\tt
  hep-th/9806159}}].

\bibitem{Gaiotto:2008nz}
D.~Gaiotto, A.~Neitzke, and Y.~Tachikawa, {\it {Argyres-Seiberg duality and the
  Higgs branch}},  {\em Commun.Math.Phys.} {\bf 294} (2010) 389--410,
  [\href{http://xxx.lanl.gov/abs/0810.4541}{{\tt arXiv:0810.4541}}].

\bibitem{Aharony:2007dj}
O.~Aharony and Y.~Tachikawa, {\it {A Holographic computation of the central
  charges of d=4, N=2 SCFTs}},  {\em JHEP} {\bf 0801} (2008) 037,
  [\href{http://xxx.lanl.gov/abs/0711.4532}{{\tt arXiv:0711.4532}}].

\bibitem{Argyres:2007tq}
P.~C. Argyres and J.~R. Wittig, {\it {Infinite coupling duals of N=2 gauge
  theories and new rank 1 superconformal field theories}},  {\em JHEP} {\bf
  0801} (2008) 074, [\href{http://xxx.lanl.gov/abs/0712.2028}{{\tt
  arXiv:0712.2028}}].

\bibitem{Chacaltana:2010ks}
O.~Chacaltana and J.~Distler, {\it {Tinkertoys for Gaiotto Duality}},  {\em
  JHEP} {\bf 1011} (2010) 099, [\href{http://xxx.lanl.gov/abs/1008.5203}{{\tt
  arXiv:1008.5203}}].

\bibitem{Razamat:2012uv}
S.~S. Razamat, {\it {On a modular property of N=2 superconformal theories in
  four dimensions}},  {\em JHEP} {\bf 1210} (2012) 191,
  [\href{http://xxx.lanl.gov/abs/1208.5056}{{\tt arXiv:1208.5056}}].

\bibitem{Argyres:1996eh}
P.~C. Argyres, M.~R. Plesser, and N.~Seiberg, {\it {The Moduli space of vacua
  of N=2 SUSY QCD and duality in N=1 SUSY QCD}},  {\em Nucl.Phys.} {\bf B471}
  (1996) 159--194, [\href{http://xxx.lanl.gov/abs/hep-th/9603042}{{\tt
  hep-th/9603042}}].

\bibitem{WIP_Class_S}
C.~Beem, W.~Peelaers, L.~Rastelli, and B.~C. van Rees, {\it {Chiral algebras
  of class ${\mathcal S}$}}, \href{http://xxx.lanl.gov/abs/1408.6522}{{\tt
  arXiv:1408.6522}}.

\bibitem{Ademollo:1976pp}
M.~Ademollo, L.~Brink, A.~D'Adda, R.~D'Auria, E.~Napolitano, {\em et.~al.},
  {\it {Dual String with U(1) Color Symmetry}},  {\em Nucl.Phys.} {\bf B111}
  (1976) 77--110.

\bibitem{Gaiotto:2009we}
D.~Gaiotto, {\it {N=2 dualities}},  {\em JHEP} {\bf 1208} (2012) 034,
  [\href{http://xxx.lanl.gov/abs/0904.2715}{{\tt arXiv:0904.2715}}].

\bibitem{Gaiotto:2009hg}
D.~Gaiotto, G.~W. Moore, and A.~Neitzke, {\it {Wall-crossing, Hitchin Systems,
  and the WKB Approximation}},  \href{http://xxx.lanl.gov/abs/0907.3987}{{\tt
  arXiv:0907.3987}}.

\bibitem{Hanany:2010qu}
A.~Hanany and N.~Mekareeya, {\it {Tri-vertices and SU(2)'s}},  {\em JHEP} {\bf
  1102} (2011) 069, [\href{http://xxx.lanl.gov/abs/1012.2119}{{\tt
  arXiv:1012.2119}}].

\bibitem{Gaiotto:2009gz}
D.~Gaiotto and J.~Maldacena, {\it {The Gravity duals of N=2 superconformal
  field theories}},  {\em JHEP} {\bf 1210} (2012) 189,
  [\href{http://xxx.lanl.gov/abs/0904.4466}{{\tt arXiv:0904.4466}}].

\bibitem{Benini:2009gi}
F.~Benini, S.~Benvenuti, and Y.~Tachikawa, {\it {Webs of five-branes and N=2
  superconformal field theories}},  {\em JHEP} {\bf 0909} (2009) 052,
  [\href{http://xxx.lanl.gov/abs/0906.0359}{{\tt arXiv:0906.0359}}].

\bibitem{Maruyoshi:2013hja}
K.~Maruyoshi, Y.~Tachikawa, W.~Yan, and K.~Yonekura, {\it {N=1 dynamics with
  $T_{N}$ theory}},  {\em JHEP} {\bf 1310} (2013) 010,
  [\href{http://xxx.lanl.gov/abs/1305.5250}{{\tt arXiv:1305.5250}}].

\bibitem{Gaiotto:2012uq}
D.~Gaiotto and S.~S. Razamat, {\it {Exceptional Indices}},  {\em JHEP} {\bf
  1205} (2012) 145, [\href{http://xxx.lanl.gov/abs/1203.5517}{{\tt
  arXiv:1203.5517}}].

\bibitem{Gaiotto:2012xa}
D.~Gaiotto, L.~Rastelli, and S.~S. Razamat, {\it {Bootstrapping the
  superconformal index with surface defects}},
  \href{http://xxx.lanl.gov/abs/1207.3577}{{\tt arXiv:1207.3577}}.

\bibitem{Frenkel:2005pa}
E.~Frenkel, {\it {Lectures on the Langlands program and conformal field
  theory}},  \href{http://xxx.lanl.gov/abs/hep-th/0512172}{{\tt
  hep-th/0512172}}.

\bibitem{Gadde:2009kb}
A.~Gadde, E.~Pomoni, L.~Rastelli, and S.~S. Razamat, {\it {S-duality and 2d
  Topological QFT}},  {\em JHEP} {\bf 1003} (2010) 032,
  [\href{http://xxx.lanl.gov/abs/0910.2225}{{\tt arXiv:0910.2225}}].

\bibitem{Moore:2011ee}
G.~W. Moore and Y.~Tachikawa, {\it {On 2d TQFTs whose values are holomorphic
  symplectic varieties}},  \href{http://xxx.lanl.gov/abs/1106.5698}{{\tt
  arXiv:1106.5698}}.

\bibitem{Alday:2009aq}
L.~F. Alday, D.~Gaiotto, and Y.~Tachikawa, {\it {Liouville Correlation
  Functions from Four-dimensional Gauge Theories}},  {\em Lett.Math.Phys.} {\bf
  91} (2010) 167--197, [\href{http://xxx.lanl.gov/abs/0906.3219}{{\tt
  arXiv:0906.3219}}].

\bibitem{Wyllard:2010rp}
N.~Wyllard, {\it {W-algebras and surface operators in N=2 gauge theories}},
  {\em J.Phys.} {\bf A44} (2011) 155401,
  [\href{http://xxx.lanl.gov/abs/1011.0289}{{\tt arXiv:1011.0289}}].

\bibitem{deformed_chiral}
E.~{Frenkel} and N.~{Reshetikhin}, {\it {Towards Deformed Chiral Algebras}},
  in {\em eprint arXiv:q-alg/9706023}, p.~6023, 1997.

\bibitem{Dobrev:1985qv}
V.~Dobrev and V.~Petkova, {\it {All Positive Energy Unitary Irreducible
  Representations of Extended Conformal Supersymmetry}},  {\em Phys.Lett.} {\bf
  B162} (1985) 127--132.

\bibitem{kazhdan1979}
D.~Kazhdan and G.~Lusztig, {\it Representations of coxeter groups and hecke
  algebras},  {\em Inventiones mathematicae} {\bf 53} (1979), no.~2 165--184.

\bibitem{Fuchs:1997jv}
J.~Fuchs and C.~Schweigert, {\em {Symmetries, Lie algebras and representations:
  a graduate course for physicists}}.
\newblock {Cambridge University Press}, 1997.

\bibitem{DeVos:1995an}
K.~De~Vos and P.~Van~Driel, {\it {The Kazhdan-Lusztig conjecture for W
  algebras}},  \href{http://xxx.lanl.gov/abs/hep-th/9508020}{{\tt
  hep-th/9508020}}.

\bibitem{kashiwara1990kazhdan}
{Kashiwara, Masaki and Tanisaki, Toshiyuki}, {\it {Kazhdan-Lusztig conjecture
  for symmetrizable Kac-Moody Lie algebras II}},  {\em {Operator algebras,
  unitary representations, enveloping algebras, and invariant theory}} {\bf 2}
  (1990) 159--195.

\bibitem{casian1990kazhdan}
{Casian, Luis}, {\it {Kazhdan-Lusztig multiplicity formulas for Kac-Moody
  algebras}},  {\em {Comptes Rendus de l'Academie des Sciences Serie
  I-Mathematique}} {\bf 310} (1990), no.~6 333--337.

\end{thebibliography}
\end{document}